%% file: ms.tex
\documentclass[twocolumn]{aastex62}

\usepackage{graphicx}
\usepackage{amsmath}
\usepackage{amssymb}
\usepackage{enumitem}

\defcitealias{lehner15}{LHW15}
\defcitealias{lehner20}{L20}
\defcitealias{kim24}{KZP24}

\newcommand{\dvrange}{$-700 \le \vlsr\ \le -150$ \km}
\newcommand{\dvobs}{$-700 \le \vlsr\ \le -150$ \km}

\def\nodata{ ~$\cdots$~ }

\newcommand{\zem}{\ensuremath{z_{\rm em}}}

\newcommand{\cmm}{cm$^{-2}$}

\newcommand{\lya}{Ly$\alpha$}

\newcommand{\mlnovi}{\log N_{\rm O\,VI}}

\newcommand{\km}{${\rm km\,s}^{-1}$}
\newcommand{\kms}{${\rm km\,s}^{-1}$}
\newcommand{\hst}{{HST}}
\newcommand{\fuse}{{FUSE}}
\newcommand{\vlsr}{\ensuremath{v_{\rm LSR}}}

\newcommand{\lms}{\ensuremath{l_{\rm MS}}}
\newcommand{\bms}{\ensuremath{b_{\rm MS}}}

\newcommand{\rvir}{\ensuremath{R_{\rm vir}}}
\newcommand{\Mvir}{\ensuremath{M_{\rm vir}}}

\newcommand{\hi}{\ion{H}{1}}

\newcommand{\alii}{\ion{Al}{2}}

\newcommand{\cii}{\ion{C}{2}}

\newcommand{\civ}{\ion{C}{4}}
\newcommand{\nni}{\ion{N}{1}}

\newcommand{\nv}{\ion{N}{5}}
\newcommand{\oi}{\ion{O}{1}}

\newcommand{\ovi}{\ion{O}{6}}

\newcommand{\sii}{\ion{S}{2}}

\newcommand{\siii}{\ion{Si}{2}}
\newcommand{\siiii}{\ion{Si}{3}}
\newcommand{\siiv}{\ion{Si}{4}}

\newcommand{\feii}{\ion{Fe}{2}}

\shortauthors{Lehner et al.}
\shorttitle{Project AMIGA: The Inner CGM of M31---Spatial Structure and Mass}
\begin{document}

\title{Project AMIGA: The Inner Circumgalactic Medium of Andromeda from Thick Disk to Halo \footnote{Based on observations made with the NASA/ESA Hubble Space Telescope, obtained from the data archive at the Space Telescope Science Institute. STScI is operated by the Association of Universities for Research in Astronomy, Inc. under NASA contract NAS 5-26555.}}

\author[0000-0001-9158-0829]{Nicolas Lehner}
\affiliation{Department of Physics and Astronomy, University of Notre Dame, Notre Dame, IN 46556, USA}

\author[0000-0002-2591-3792]{J. Christopher Howk}
\affiliation{Department of Physics and Astronomy, University of Notre Dame, Notre Dame, IN 46556, USA}

\author[0009-0008-9243-1428]{Lucy Collins}
\affiliation{Department of Physics and Astronomy, University of Notre Dame, Notre Dame, IN 46556, USA}

\author[0000-0001-9966-6790]{Sameer}
\affiliation{Department of Physics and Astronomy, University of Notre Dame, Notre Dame, IN 46556, USA}

\author[0000-0002-0507-7096]{Bart P. Wakker}
\affiliation{Eureka Scientific, 2452 Delmer Street, Suite 100, Oakland, CA 94602-3017, USA}

\author[0000-0001-7472-3824]{Ramona Augustin}
\affiliation{Leibniz-Institut f{\"u}r Astrophysik Potsdam (AIP), An der Sternwarte 16, 14482 Potsdam, Germany}

\author[0000-0001-5817-0932]{Kathleen A. Barger}
\affiliation{Department of Physics \& Astronomy, Texas Christian University, Fort Worth, TX 76109, USA}

\author[0000-0002-8518-6638]{Michelle A. Berg}
\affiliation{Department of Physics \& Astronomy, Texas Christian University, Fort Worth, TX 76109, USA}

\author[0000-0002-3120-7173]{Rongmon Bordoloi}
\affiliation{Department of Physics, North Carolina State University, Raleigh, 27695, North Carolina, USA}

\author[0000-0002-1793-9968]{Thomas M. Brown}
\affiliation{Space Telescope Science Institute, 3700 San Martin Dr., Baltimore, MD 21218, USA}
\affiliation{The William H. Miller III Department of Physics \& Astronomy, Bloomberg Center for Physics and Astronomy, Johns Hopkins University, 3400 N. Charles Street, Baltimore, MD 21218, USA}

\author[0000-0003-4237-3553]{Frances H. Cashman}
\affiliation{Department of Physics, Presbyterian College, Clinton, SC 29325, USA}

\author[0000-0002-4900-6628]{Claude-Andr\'e Faucher-Gigu\`ere}
\affiliation{CIERA and Department of Physics and Astronomy, Northwestern University, 1800 Sherman Ave, Evanston, IL 60201, USA}

\author[0000-0003-0724-4115]{Andrew J. Fox}
\affiliation{AURA for ESA, Space Telescope Science Institute, 3700 San Martin Drive, Baltimore, MD 21218, USA}

\author[0000-0003-3681-0016]{David M. French}
\affiliation{Space Telescope Science Institute, 3700 San Martin Dr., Baltimore, MD 21218, USA}

\author[0000-0003-0394-8377]{Karoline M. Gilbert}
\affiliation{Space Telescope Science Institute, 3700 San Martin Dr., Baltimore, MD 21218, USA}
\affiliation{The William H. Miller III Department of Physics \& Astronomy, Bloomberg Center for Physics and Astronomy, Johns Hopkins University, 3400 N. Charles Street, Baltimore, MD 21218, USA}

\author[0000-0001-8867-4234]{Puragra Guhathakurta}
\affiliation{Department of Astronomy \& Astrophysics, University of California Santa Cruz, 1156 High Street, Santa Cruz, CA 95064, USA}

\author[0000-0002-7893-1054]{John M. O'Meara}
\affiliation{W.M. Keck Observatory 65-1120 Mamalahoa Highway Kamuela, HI 96743, USA}

\author[0000-0002-2786-0348]{Brian W. O'Shea}
\affiliation{Department of Computational Mathematics, Science, and Engineering, Michigan State University, East Lansing, MI, USA}
\affiliation{Department of Physics and Astronomy, Michigan State University, East Lansing, MI, USA}
\affiliation{Facility for Rare Isotope Beams, Michigan State University, East Lansing, MI 48824, USA}

\author[0000-0003-1455-8788]{Molly S.\ Peeples}
\affiliation{Space Telescope Science Institute, 3700 San Martin Dr., Baltimore, MD 21218, USA}
\affiliation{The William H. Miller III Department of Physics \& Astronomy, Bloomberg Center for Physics and Astronomy, Johns Hopkins University, 3400 N. Charles Street, Baltimore, MD 21218, USA}

\author[0000-0001-7996-7860]{D.J. Pisano}
\affiliation{University of Cape Town, Private Bag X3, Rondebosch, 7701, Republic of South Africa}

\author[0000-0002-7738-6875]{J. Xavier Prochaska}
\affiliation{UCO/Lick Observatory, Department of Astronomy \& Astrophysics, University of California Santa Cruz, 1156 High Street, Santa Cruz, CA 95064}
\affiliation{Kavli IPMU (WPI), UTIAS, The University of Tokyo, Kashiwa, Chiba 277-8583, Japan}
\affiliation{Division of Science, National Astronomical Observatory of Japan, 2-21-1 Osawa, Mitaka, Tokyo 181-8588, Japan}

\author[0000-0002-7541-9565]{Jonathan Stern}
\affiliation{School of Physics and Astronomy, Tel Aviv University, Tel Aviv 69978, Israel}

\author[0000-0002-7982-412X]{Jason Tumlinson}
\affiliation{Space Telescope Science Institute, 3700 San Martin Dr., Baltimore, MD 21218, USA}
\affiliation{The William H. Miller III Department of Physics \& Astronomy, Bloomberg Center for Physics and Astronomy, Johns Hopkins University, 3400 N. Charles Street, Baltimore, MD 21218, USA}

\author[0000-0002-0355-0134]{Jessica K. Werk}
\affiliation{Department of Astronomy, University of Washington, Seattle, WA, 98195, USA}

\author[0000-0002-7502-0597]{Benjamin F. Williams}
\affiliation{Department of Astronomy, University of Washington, Seattle, WA, 98195, USA}


\begin{abstract}
The inner circumgalactic medium (CGM) of galaxies, where disk and halo processes intersect, remains poorly characterized despite its critical role in regulating galaxy evolution. We present results from Project AMIGA Insider, mapping Andromeda's (M31) inner CGM within 0.25 \rvir\ ($\sim$75 kpc) using 11 QSO sightlines, bringing our total sample to 54 sightlines from the disk to 2\rvir. We detect a clear transition between M31's thick disk and CGM at $R \la 30$ kpc, where low/intermediate ions show thick-disk corotating components with higher column densities than the CGM ones, while high ions exhibit similar column densities in both the CGM and thick disk. Beyond this region, all ion column densities decrease with impact parameter, with steeper gradients for low ions than high ions. The inner CGM ($R \la 100$ kpc) shows more complex gas phases and multi-component absorption compared to the predominantly single-component outer CGM. We find no significant azimuthal dependence for any observed ions, suggesting M31's CGM is shaped by radial processes (e.g., cooling flows, precipitation) rather than disk-aligned outflows. We estimate the total metal mass in M31's cool (\siii, \siiii, \siiv) CGM within \rvir\ to be $ (1.9 \pm 0.3_{\rm stat} \pm 0.7_{\rm sys}) \times 10^7$ M$_\odot$, leading to a cool gas mass of approximately $6 \times 10^9\,(Z/0.3\, Z_\odot)^{-1}$ M$_\odot$. The warmer \ovi\ gas may contain at least 10 times more metal and gas mass. Compared to the COS-Halos $L^*$ galaxies, M31's cool CGM shows lower Si column densities at $R \la 0.4 \,R_{200}$ and lower cool CGM masses, possibly resulting from M31's higher halo mass or different environments.
\end{abstract}


\section{Introduction}\label{s-intro}
The inner circumgalactic medium (CGM) of galaxies represents a critical interface where the complex interplay between galactic disks and their extended halos unfolds. Theoretical models and simulations show that this region profoundly influences disk evolution through processes such as cooling flows of hot gas above the galactic plane and the recycling of material through galactic fountains that may never reach beyond the inner CGM \citep{fraternali17,stern21,hafen22,marasco22}. Fundamentally, this complex interface governs how material flows inward to fuel star formation and outward to enrich the circumgalactic reservoir with energy, mass, and metals. Understanding this interface and the interplay between galactic accretion, recycling, and feedback in the CGM is therefore a key focus for theorists  \citep[e.g.,][]{gurvich23,augustin25,barbani25} and observers \citep[e.g.,][]{howk99,lee01,borthakur15,zheng17,lehner22,nielsen24}. Some recent theoretical studies argue that the classic prediction of a quasi-static hot CGM phase \citep{birnboim03,keres05} is limited to the inner CGM where hot gas cooling times can be short relative to the Hubble time \citep{stern20,stern21,gurvich23,kakoly25}. However, observational studies of this critical transition region have been severely limited by the lack of information at both small {\it and}\ large impact parameters around individual galaxies.

Until a larger space-based telescope (such as the Habitable Worlds Observatory) is approved and operational \citep[e.g.,][]{borthakur25}, M31 is the only massive, $L^*$ galaxy in the universe for which we can simultaneously study the global distribution and properties of the CGM on {\it all}\ scales along many ($>50$) sightlines. This detailed exploration is crucial for connecting the CGM to specific and well-constrained star formation histories, supernova and AGN feedback, metal production budgets, and detailed maps of resolved stellar populations. This unique opportunity includes the ability to probe the critical transition region between the galactic disk and the extended CGM, a boundary region that remains poorly characterized for galaxies beyond the Milky Way (MW). The first sample of our Project AMIGA---the first large survey of the M31 CGM with the Cosmic Origins Spectrograph (COS) onboard of the Hubble Space Telescope (HST)---revealed that M31's CGM extends well beyond $\rvir = 300$ kpc, with near-unity covering factors of cool gas traced by \siiii\ and warm or hot gas traced by \ovi\ to $1.3 \rvir$ and $1.9 \rvir$, respectively (\citealt{lehner20}, hereafter \citetalias{lehner20}; see also \citealt{lehner15}). 

Using 43 sightlines to characterize the CGM around M31 at $R = 25$--569 kpc, Project AMIGA showed that the CGM has a two-component distribution: (1) an outer CGM consisting mainly of highly-ionized gas (\ovi, \civ) and traces of diffuse cool gas (\siiii, \cii) with only a moderate dependence on $R$, and (2) a more ``dynamic" inner CGM ($ 0.08 \la R/\rvir \la 0.25$; $25 \la R \la 75$ kpc) with a strong dependence on $R$, a more complex mixture of multi-phase gas (including low-to-intermediate ions), and more complex kinematics perhaps indicating recent outflows and/or gas recycling. These data also show M31's cool ($T\simeq 10^4$ K) and warm ($T\simeq 10^5$ K) CGM contains more than half of the metal mass of its disk, though with some uncertainty owing to the very few targets probing the inner CGM of M31.

Project AMIGA's characterization of the inner CGM was limited since it was designed to target 7--10 QSOs in each $\Delta R \sim 100$ kpc bin to about \rvir\ \citepalias{lehner20}, leading to only eight sightlines at $R \la 1/3 \rvir$ and one at $R\la 0.08 \rvir$ ($R \simeq 25$ kpc). We were therefore undertook a new survey of CGM absorption toward 13 targets (11 background QSOs and two stars within M31) located within $R \la 1/4 \rvir$ ($R\la 75$ kpc) of M31 to directly probe the thick disk (i.e., corotating gas within a few kpc of M31), investigate the circulation of matter out of and back into M31's disk, characterize the baryonic and metal masses in the inner halo, and bridge the gap between large-scale CGM studies and the region probed by high-velocity cloud (HVC) analogs \citep[e.g.,][]{wakker01,richter17,putman12,lehner22}. In Cycle 29 of HST, we were awarded 137 orbits to obtain G130M and G160M spectra of 11 QSOs sampling the inner region of M31 and two stars in the disk of M31. Combining them with Project AMIGA, our sample increases the number of sightlines by a factor of $\sim$8 at $R \la 0.1 \rvir$ and by a factor of $\sim$2 at $0.1 < R /\rvir \la 0.25$. We refer to the two samples under Project AMIGA umbrella, and when we need to differentiate between them, we will refer to the original sample presented in \citetalias{lehner20} as AMIGA Extended and to our new sample as AMIGA Insider.

In this first paper in a series, we present the analysis of the AMIGA Insider sample (except for the two stars that will be presented in a forthcoming paper). The three science topics we focus on in this paper are: 1) the identification and characterization of the thick disk and its impact on the column density profiles with $R$; 2) the dependence of the column densities and kinematics on the projected radius ($R$) and azimuth ($\Phi$); and 3) the metal and baryon mass estimates of the cool CGM of M31 using the \siii, \siiii, and \siiv\ ions, along with implications from our findings. This paper therefore presents the first comprehensive UV spectroscopic characterization of the transition between M31's thick disk and its extended CGM.

Our paper is organized as follows. In \S\ref{s-data}, we provide more information about the criteria used to assemble our sample of QSOs and explain the various steps to derive the properties (velocities and column densities) of the absorption. Specifically, we analyze the absorption associated with M31 in the AMIGA Insider QSO spectra using two methods, the apparent optical depth (AOD) and Voigt profile fitting (PF) methods in \S\ref{s-aod} and \S\ref{s-pf}, providing a detailed comparison of the component analysis between these two methods in \S\ref{s-comp-aod-pf}. In \S\ref{s-init-results}, we present the column density profiles and kinematics of various ions in the CGM of M31, examining both their radial and azimuthal variations. \S\ref{s-mass-main} is dedicated to estimating the metal and baryon mass of M31's CGM using multiple complementary statistical approaches. In \S\ref{s-disc}, we discuss the implications of our findings, particularly the transition between the thick disk and CGM, the radial and azimuthal structure of the CGM, and comparisons with similar galaxies from the COS-Halos survey. Finally, in \S\ref{s-summary}, we summarize our results and outline future work. The appendices contain additional technical details on the line identification.

To facilitate comparison with other work and simulations, we adopt the following characteristic properties of M31. We refer the reader to \citetalias{lehner20} for more detail. We use the radius $R_{200}$ enclosing a mean overdensity of $200$ times the critical density, and for M31, we adopt $M_{200} = 1.26 \times 10^{12}$~M$_\sun$ (e.g., \citealt{watkins10,vandermarel12}), implying $R_{200} \simeq 230$ kpc. For the virial mass and radius (\Mvir\ and \rvir), we adopt $\Mvir \simeq 1.2 M_{200} \simeq 1.5 \times 10^{12}$ M$_\sun$ and $\rvir \simeq 1.3 R_{200} \simeq 300$ kpc (e.g., \citealt{vandermarel12}). A distance of M31 of $d_{\rm M31} = 752$ kpc based on the measurements of Cepheid variables \citep{riess12} is assumed throughout, and all projected distances are computed using the three-dimensional separation between target coordinates and M31's position and distance. We note that all the distance estimates have varied by a few percent \citep[e.g.,][]{brown04,mcconnachie05,savino22}. The halo mass (and therefore $R_{200}$ and \rvir) is more uncertain; the value adopted here corresponds to about the mean value from various methods \citep[e.g.,][]{bhattacharya23}. For consistency with our previous survey as well as the original design of Project AMIGA, we retained the original values listed above. Finally, except otherwise stated, all the velocities (\vlsr) are expressed in the local standard of rest (LSR) frame.

\input{tab1}

\section{Data and Analysis}\label{s-data}
\subsection{The Sample}\label{s-sample}

The science goals of AMIGA Insider require characterizing the kinematics and metal column densities of M31's inner CGM at $R \la 75$ kpc ($\la 1/4 \rvir$) as a function of azimuthal angle and impact parameter and connect these results to the outer CGM. Our \hst\ Cycle 29 AMIGA Insider program observed 11 new QSOs within $R \la 75$ kpc, complemented by two stars within M31 itself. Combined with the AMIGA Extended sightlines, our sample increases the number of probed sightlines by a factor of $\sim$8 at $R \la 0.1 \rvir$ and by a factor of $\sim$2 at $0.1 < R/\rvir \la 0.25$.\footnote{We note that there are two additional archival QSOs in this region that were observed with COS G130M (PID: 16219). However, these targets were selected to be near detected \hi\ emission between M31 and M33, which may introduce a selection bias, and are therefore not included in the present study.}

The sightlines were selected to systematically probe azimuthal variations across M31's inner CGM. As with Project AMIGA, the sample was partially constrained by the relative scarcity of UV-bright AGNs behind the northern half of M31's CGM due to higher foreground MW dust extinction near the MW disk plane. We optimized QSO target selection from the Million Quasars Catalog (v7.1b, \citealt{flesch23}) for brightness to minimize exposure time while favoring lower redshifts to reduce contamination from unrelated absorption. For targets without existing UV spectra, we required similar GALEX NUV and FUV flux magnitudes to reduce the likelihood of intervening Lyman limit systems (LLS) with optical depth $\tau_{\rm LL}>2$ that could compromise our ability to measure foreground M31 absorption. With this selection strategy, we successfully avoided the inclusion of QSOs with intervening LLS in our sample.

\begin{figure*}[tbp]
\epsscale{1.1}
\plotone{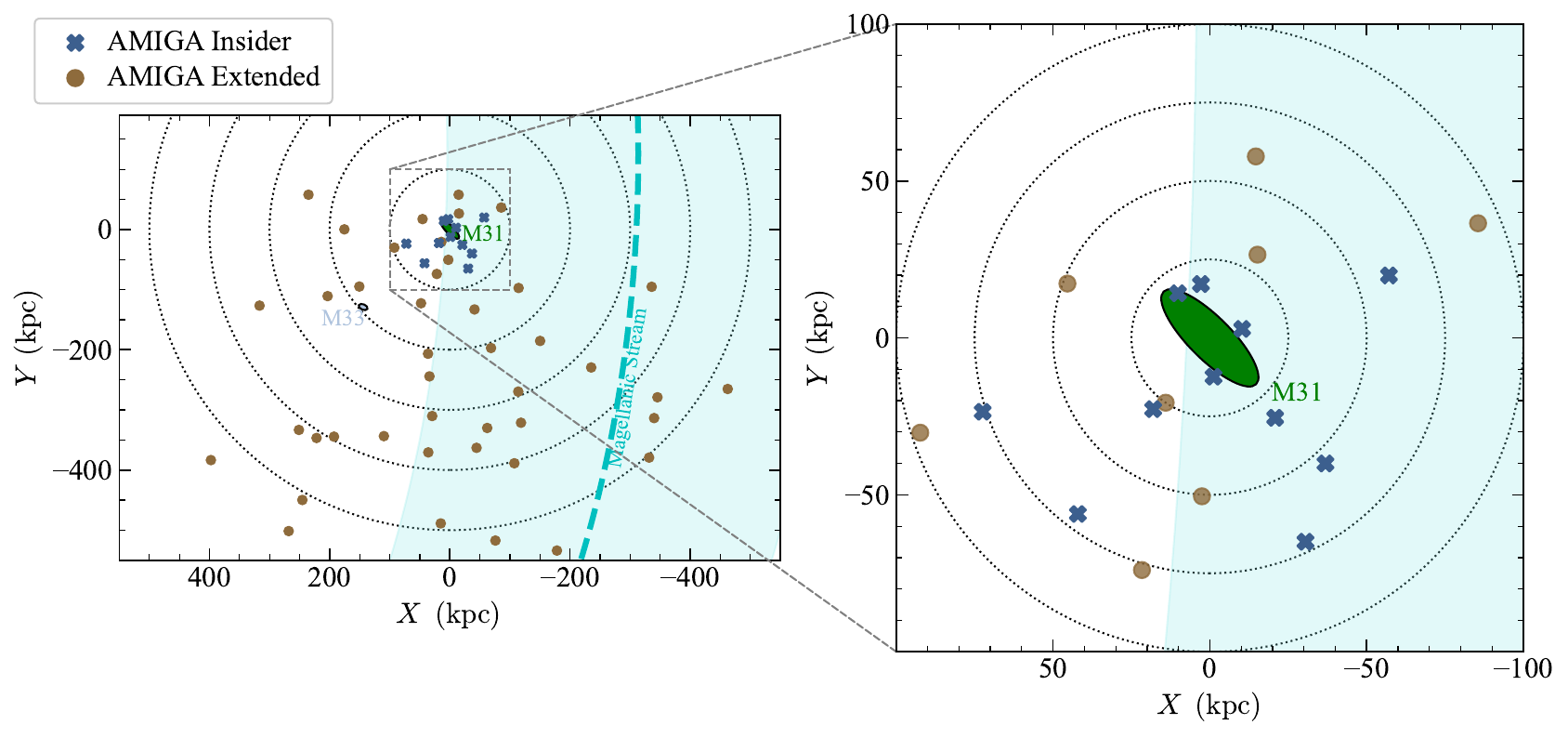}
\caption{Locations of the Project AMIGA pointings relative to the M31--M33 system. The axes show the angular separations converted into physical coordinates relative to the center of M31. North is up and east to the left. The 11 QSO sightlines shown in blue crosses are part of AMIGA Insider and the other 43 shown with brown circles are part of AMIGA Extended. The blue dashed line shows the plane of the MS ($b_{\rm MS} =0\degr$) as defined by \citet{nidever08}. The shaded region within $b_{\rm MS} \pm 25\degr$ of the MS midplane is the approximate region where we identify most of the MS absorption components contaminating the M31 CGM absorption (see \S\ref{s-ms-contamination}).\label{f-map}}
\end{figure*}

In Fig.~\ref{f-map}, we show the locations of QSOs in the M31--M33 system, with filled circles indicating the  AMIGA Extended targets and filled crosses marking the 11 new QSOs. Table~\ref{t-sum} presents properties of the new QSOs, ordered by increasing projected distance from M31, while similar information for the AMIGA Extended targets can be found in Table~1 of \citetalias{lehner20}. For each QSO, we provide: the QSO spectroscopic redshift (\zem), J2000 right ascension (RA) and declination (Dec.), Magellanic Stream (MS) coordinates (\lms, \bms; see \citealt{nidever08} for coordinate system definition), projected distances both radially ($R$) and in cartesian coordinates ($X,Y$), and the signal-to-noise ratio (S/N) per COS resolution element near the \siiii\ transition—a key diagnostic for M31 CGM gas (\citealt{lehner15}; \citetalias{lehner20}).

\subsection{UV Spectroscopic Observations and Line Identification}\label{s-calib}

To search for M31 CGM absorption and to determine the properties of the CGM gas, we use ions and atoms that have their resonance transitions in the UV (see \S\ref{s-prop}). Any transitions with $\lambda>1144$ \AA\ fall in the \hst\ COS bandpass. All the targets in AMIGA Insider were observed with \hst\ using the COS G130M and G160M gratings (that provide a resolution of $\approx 17,000$). For the AMIGA Extended targets, we refer the reader to \citetalias{lehner20}, but most of these have coverage of the entire UV bandpass. 

The data reduction, calibration, and line identification procedures follow the methodology detailed in \citetalias{lehner20} and we refer the reader to that paper. We use pipeline-calibrated HST/COS data products from MAST, with additional processing to address known wavelength calibration issues. Following \cite{wakker15} and \citetalias{lehner20}, absorption lines across different exposures were aligned and the absolute wavelength calibration was determined by comparing UV absorption lines with 21-cm \hi\ emission from our GBT observations (\citealt{howk17} and J.C. Howk et al., 2025, in prep.) or existing \hi\ surveys \citep{kalberla05,kalberla10}. 

Our instrument setup provides near complete wavelength coverage between 1140 and 1730 \AA\ for the 11 targets in our HST program. With the majority of our targets (90\%) having $\zem \la 0.3$, contamination from high-redshift absorbers is minimized and \lya\ remains within the observed wavelength range. 

\subsection{Determination of the M31 CGM Absorption Properties}\label{s-prop}

We search for M31 CGM absorption in the velocity range \dvrange, following the framework established in \citetalias{lehner20}.  This range was chosen based on M31 and MW kinematics. The systemic velocity of M31 in the LSR frame is $\vlsr \simeq -300$ \km. The lower boundary of $-700$ \km\ extends beyond M31's most negative rotation velocities ($\sim -600$ \km, \citealt{chemin09}). The upper limit of $-150$ \km\ marks the transition where the MW contamination becomes too significant, though M31 gas might still be expected  since M31's rotation curve extends to approximately to $-100$ \km\ \citep{chemin09}). \citetalias{lehner20} revealed that most of the M31 CGM components cluster in the velocity range $-350 \la \vlsr\ \la -150$ \km, with the interval $-450 \la \vlsr\ \la -350$ \km\ often showing some contamination from the MS. As shown in \citetalias{lehner20}, this velocity distribution closely matches that of M31's dwarf galaxies, except for a potential gap at $-150 \la \vlsr\ \la -100$ \km\ where some dwarf galaxies reside but CGM detection may be hidden by MW absorption. For the AMIGA Insider sample, we find this velocity framework remains valid, with no CGM absorption detected at $\vlsr\ \la -650$ \km. However, as we will show, our new observations reveal a broader velocity distribution, particularly because several targets lie close enough of M31  where rotational signatures are more prominent.

As in \citetalias{lehner20}, we probe the CGM using a comprehensive set of UV transitions: \oi\ $\lambda$1302, \cii\ $\lambda\lambda$1036, 1334, \civ\ $\lambda\lambda$1548, 1550, \nv\ $\lambda\lambda$1238, 1242, \nni\ $\lambda$1199, \nv\ $\lambda$$\lambda$1238, 1242, \alii\ $\lambda$1670, \siii\ $\lambda\lambda$1190, 1193, 1260, 1304, 1526, \siiii\ $\lambda$1206, \siiv\ $\lambda\lambda$1393, 1402, \sii\ $\lambda\lambda$1250, 1253, 1259, and \feii\ $\lambda\lambda$1144, 1608 (and \ovi\ $\lambda$1031 with \fuse). To determine column densities and velocities, we primarily use the AOD method (see \S\ref{s-aod}), which we complemented with a Voigt PF analysis for comparison and validation of the AOD method. Due to the often complex velocity structure in our search window and potential contamination from the MS (see Fig.~\ref{f-map} and \citetalias{lehner20}), we analyze individual absorption components where COS resolution permits, beginning with a detailed modeling of the QSO continuum.

\subsubsection{Continuum Placement}\label{s-continuum}

We adopted a continuum fitting approach that combined the automated method developed for the COS CGM Compendium (CCC, \citealt{lehner18}) and as described in \citetalias{lehner20}, complemented with manual fitting necessitated by the broader and more complex absorption profiles in the AMIGA Insider sample. For each transition in our pre-defined set, we fitted Legendre polynomials to absorption-free regions near the features of interest, with the continuum defined locally and visually inspected for quality control. The fitting window typically spanned $\pm$1000--2000 \km\ around each absorption transition, though this range was adjusted between $\pm$250--2000 \km\ based on the local continuum complexity. Within these windows, continuum regions were either manually selected or automatically masked following \citet{lehner18}. We tested polynomial orders from 1 to 5, though the relative simplicity of the QSO continua over 500--4000 \km\ regions typically led to best fits with orders 1--3. In cases where the automatic continuum fitting failed due to complex features (e.g., near emission line peaks or in regions with multiple absorption lines or near the Ly$\alpha$ transition for \siiii), we first attempted to adjust the velocity interval to achieve better-constrained fits. If this approach proved insufficient, we resorted to manual selection of continuum regions for fitting.

\subsubsection{Velocity Components and AOD Analysis}\label{s-aod}

\begin{figure*}[tbp]
\epsscale{1}
\plotone{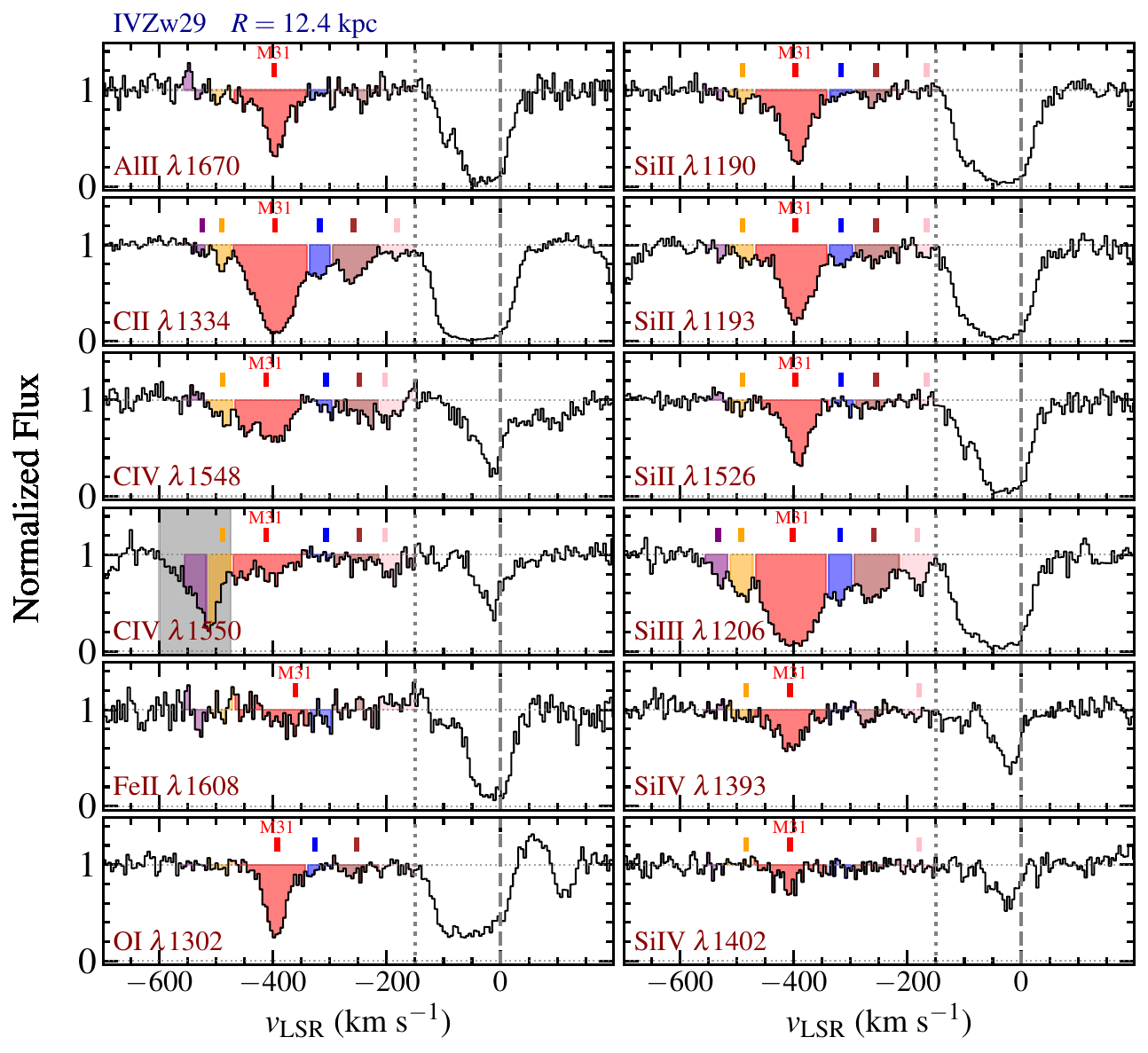}
\caption{Example of normalized absorption lines as a function of the LSR velocity toward IVZw29 showing typical atoms and ions in our survey. Different colors represent distinct velocity components identified at COS G130M-G160M resolution. Tick marks indicate ions with $\ge 2\sigma$ detections (in at least one transition of that ion). Colored regions without tick marks show no detection of that ion in that component. For ions with multiple transitions, the average velocity from detected transition(s) is used for the tick mark position. The thick disk corotating  components are identified in red and labeled ``M31'' (see \S\ref{s-m31-dh}); {\it all the other colored regions in this figure are identified as M31 CGM components}. The MW absorption is observed between  $-150$ \km\ (vertical dotted line) and $+50$ \km\ toward this sightline. At $\vlsr \ga -150$ \km, airglow emission lines can contaminate \oi, and here the MW absorption is contaminated, but not the M31 velocity components. Contaminated regions by other absorption lines or intervening absorbers within the velocity range of interest are grayed out (part of \civ\ $\lambda$1550 in this case). The vertical dashed line marks $\vlsr =0$ \km.
\label{f-example-spectrum}}
\end{figure*}

We first identified velocity components by eye in the \siiii\ $\lambda$1206 absorption profiles within \dvobs, as this ion shows the highest detection rate in our sample (see Fig.~\ref{f-example-spectrum} and supplemental figures in Appendix~\ref{a-supp-fig}). The component structure was initially defined visually using several criteria: (1) distinct absorption features separated by regions where the flux returns to or close to the continuum level, (2) inflection points in blended absorption features, and (3) features with velocity widths exceeding the COS spectral resolution ($\sim$18 \kms). This \siiii-based component structure was then checked against other low ions (e.g., \cii, \siii), and when they are detected they generally match each other. For the high ions like \civ, which may trace a different gas phase, we maintained this component structure for consistency, though it may not necessarily represent its component structure. This approach allows us to systematically examine how different ions follow or deviate from the \siiii-defined velocity structure. A more detailed, ion-by-ion decomposition using Voigt PF is presented in \S\ref{s-pf}, providing complementary insights into the component structure of each ion.

For each identified component, we applied the AOD method \citep{savage91} to determine column densities. The absorption profiles were converted into apparent optical depth per unit velocity, $\tau_a(v) = \ln[F_{\rm c}(v)/F_{\rm obs}(v)]$, where $F_c(v)$ and $F_{\rm obs}(v)$ are the modeled continuum and observed fluxes. The apparent column density per unit velocity follows as $N_a(v) = 3.768 \times 10^{14} \tau_a(v)/(f \lambda(\mbox{\AA}))$ ${\rm cm}^{-2}\,({\rm km\,s^{-1}})^{-1}$, where $f$ is the oscillator strength and $\lambda$ is the wavelength in \AA. We integrated these profiles to obtain total column densities and calculated line centroids using the first moment of the AOD. For non-detections, we quoted 2$\sigma$ upper limits assuming a linear curve of growth.

Following the methodology in \citetalias{lehner20}, we assessed potential contamination through our complete line identification (see Appendix~\ref{a-lineid}). We also checked for unresolved saturation using several criteria. For ions with multiple transitions (\siii, \civ, \siiv, sometimes \sii), we compared column densities from transitions with different $f\lambda$ values. Based on this analysis, we established that absorption features with peak optical depth $\tau_a \ga 0.9$ may show evidence of saturation. For doublets like \civ\ and \siiv, we followed the saturation correction procedure from \citet{lehner18} when needed, though saturation was rare in these ions. For single strong transitions (particularly \siiii\ and \cii), we conservatively adopted lower limits when $\tau_a > 0.9$. All saturated measurements are noted in Table~\ref{t-results}.

\subsubsection{Voigt Profile Analysis}\label{s-pf}

\begin{figure*}[tbp]
\epsscale{1}
\plotone{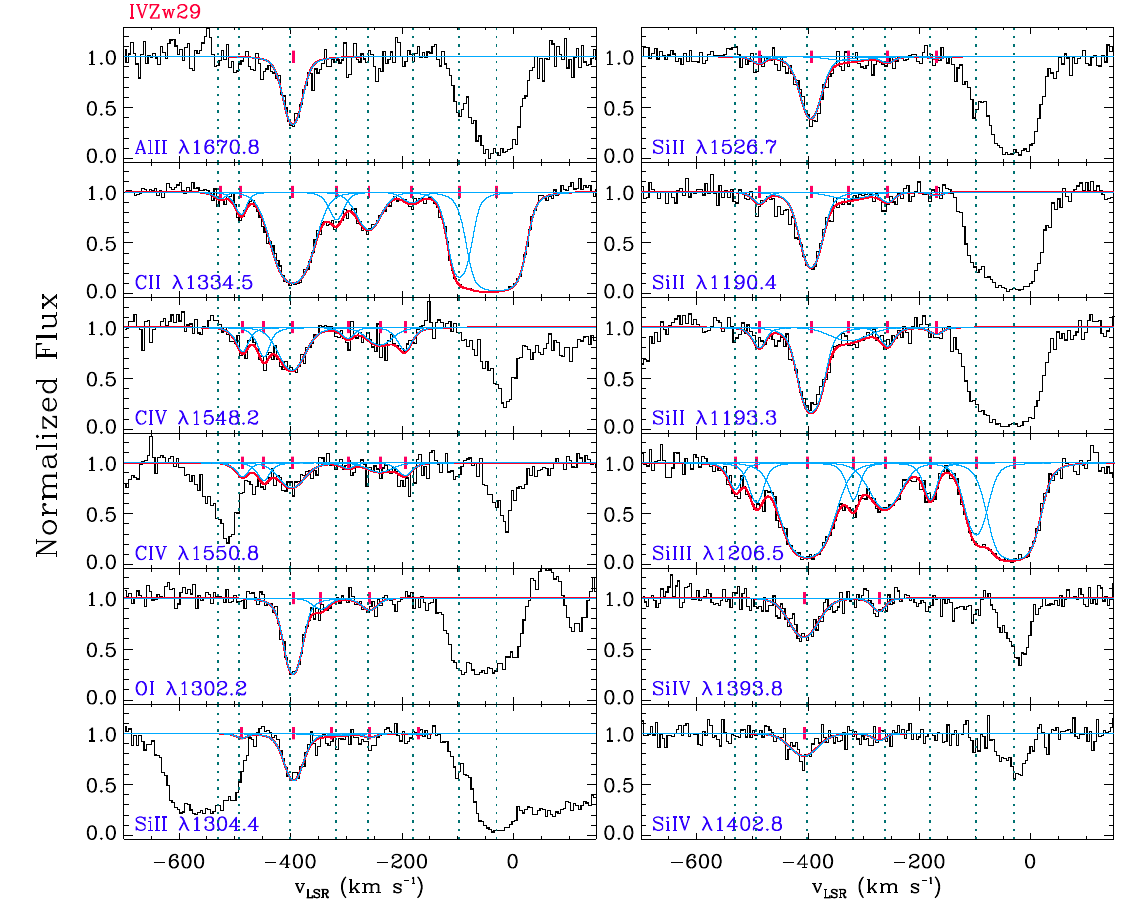}
\caption{Example of normalized absorption lines as a function of the LSR velocity toward IVZw29 with a Voigt PF model to the data. The M31 components are at $\vlsr \la -150$ \km\ and the MW components are at $-150 \la \vlsr \la +50$ \km. Absorption that is not fitted or under-fitted are contaminated regions. The MW components are only fitted if they can affect the M31 profile fit results. The red tick-marks show the velocity centroids for each ion. In each panel, the red line shows the resulting PF while the blue line show the individual components. The green vertical dotted lines show the velocity centroids of \siiii. 
\label{f-example-fit}}
\end{figure*}

We complemented our AOD analysis with Voigt PF using a modified version of the software described in \citet[][see \citealt{lehner11b} for updates]{fitzpatrick97}. Each absorption profile was modeled as a combination of individual Voigt components, with the model profiles convolved with the appropriate COS G130M or G160M instrumental line-spread function (LSF). As the COS LSFs are not purely Gaussian and vary with FUV lifetime positions, we used the tabulated LSFs from the COS instrument handbooks. 

For each component, we fitted three parameters: column density ($N_i$), Doppler parameter ($b_i$), and velocity centroid ($v_i$). Unlike the AOD analysis, we fitted each ion independently without assuming a common component structure across different species or different ions of the same species. The fitting process allowed, however, simultaneous analysis of multiple transitions of the same ion. This method was applied to \siiii\ and, when available, to some of the following ions or atoms: \cii, \civ, \siii, \siiv, \alii, \feii, and \oi. The MW components were only fitted when severely blended with the M31 CGM absorption.

We followed an iterative approach, starting with the minimum number of components needed to reasonably model each profile and also being guided by the results from the AOD analysis. Parameters were generally allowed to vary freely, though in cases of low S/N or complex profiles, we occasionally needed to constrain either the velocity or the Doppler parameter. This constraint was necessary given the COS moderate resolution ($\sim$18 \kms) and the limitations of the PF method in low S/N regions. Any PF results with fixed parameters should be taken with caution. We emphasize that even though the reduced-$\chi^2$ is always $\la 1$ for the results presented here, the moderate COS resolution limits our ability to fully assess saturation because narrower components may still be present, especially in some of the stronger components that may probe the thick disk of M31 (see \S\ref{s-m31-dh}). 

In Fig.~\ref{f-example-fit}, we illustrate our PF results for one of our targets, IVZw29 (the additional targets are all provided in the supplemental material). The figure shows the normalized absorption profiles for multiple ions, with the total Voigt PF shown in red and individual components in blue. The velocity centroids for each component (marked by red ticks) were determined independently for each ion. The green vertical dotted lines, marking the velocity centroids of \siiii, help visualize how the component structure varies across different ions. This figure demonstrates that despite independent fitting, the main components observed in the low ions and \siiii\ align well with each other. However, while the strongest components of \civ\ and \siiv\ align with \siiii, their secondary components show distinct kinematic structure. The complete results of our component fitting analysis are provided in Table~\ref{t-fit}.

\subsubsection{Comparison of PF and AOD Measurements}\label{s-comp-aod-pf}

We performed a systematic comparison between absorption components measured using the PF and AOD methods. Our analysis included seven different ions (\siii, \siiii, \siiv, \cii, \civ, \oi, and \alii) across 10 sightlines where both methods were applied.\footnote{LAMOST005846.82+365514.2 was not included in the PF sample owing to its weak \siiii\ absorption for the given S/N with a detection at $3.1\sigma$. However, that absorber also has a detection of \civ\ aligned with the \siiii\ where the strong and weak transitions are detected at the $5\sigma$ and $3.3\sigma$ levels, respectively.} For these 10 sightlines, we identified 102 PF and 107 AOD (detected) components at $\vlsr \le -150$ \km. (Two PF components for \cii\ are removed as it was necessary to fit them owing to their blending with other components, but they are clearly contaminated by an intervening Ly$\alpha$ absorber in each of these cases.) 

To establish correspondence between components identified by the two methods, we implemented a velocity-based matching algorithm. For each PF component of a given ion, we searched for AOD components within the same sightline with velocities within a threshold of either 15 \km\ (approximately one COS G130M/G160M resolution element) or three times the combined error of the AOD and PF ($ 3 (\sigma^2_{v,{\rm PF}} +\sigma^2_{v,{\rm AOD}})^{0.5}$), whichever was smaller. This approach and the use of a 15 \km\ threshold prevented incorrect associations between components separated by $\Delta v >50$ \km, which could otherwise occur due to large velocity uncertainties in some of the PF results. Through this methodology, we identified 94 matched components across all ions, including 76 regular detections and 18 lower limits in the AOD method; of these components are 14 associated with MS, 80 with M31 CGM or thick disk (see \S\S\ref{s-ms-contamination}, \ref{s-m31-dh}).

\begin{figure*}[tbp]
\epsscale{1}
\plotone{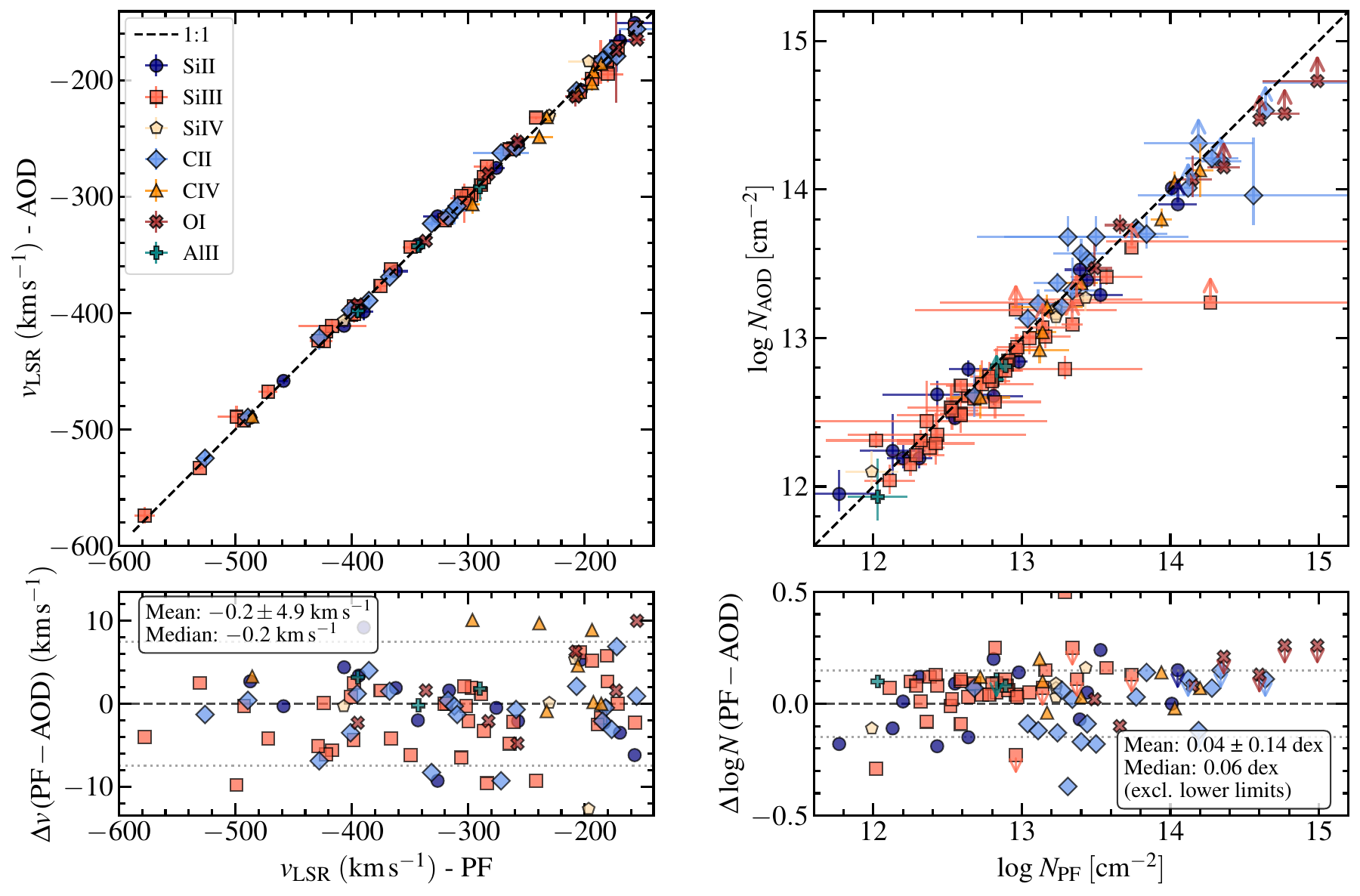}
\caption{Comparison of the velocities ({\it top left}) and column densities ({\it top right}) derived using the AOD and PF methods for the matched components. The bottom panels show the residual plots for the velocities ({\it bottom left}) and column densities ({\it bottom right}). The dotted lines in the bottom panels show the $\pm 7.5$ \km\ and $\pm 0.15 $ dex ranges for the velocity and column density residuals. 
\label{f-comp-fit-aod}}
\end{figure*}

In Fig.~\ref{f-comp-fit-aod}, we show the comparison of column densities for individual components derived from the PF ($x$-axis) and AOD ($y$-axis) methods in the top panels and the respective residuals in bottom panels. For clarity, we ignore the error bars in the residual plots, but note that essentially within $1\sigma$ a large fraction of the data are in agreement. The two methods show a good agreement within $1\sigma$ errors for both velocity and column density measurements for the majority of the components, with mean differences of $\Delta v = \langle v_{\rm PF} -v_{\rm AOD} \rangle= -0.2 \pm 4.9$ \km\ and $\Delta \log N = \langle \log N_{\rm PF} - \log N_{\rm AOD} \rangle = 0.11 \pm 0.37$ dex for all components. Excluding lower limits, the column density agreement improves further to $\Delta \log N = 0.04 \pm 0.14$. Ignoring the errors, 86\% of the velocities are in agreement within $\pm 7.5$ \km\ (about half a COS resolution element) and 82\% of the column densities are in agreement within $\pm 0.15$ dex when lower limits are excluded. We note that we did not remove in this comparison components that are less certain in the PF (e.g., fixed $b$-values, see Table~\ref{t-fit}).  When examining individual ions, we found no significant trends in these differences, except for \civ, which showed a systematic velocity offset of $\Delta v = 4.5 \pm 4.3$ \km. This offset is expected since \civ\ may trace a different gas phase than \siiii\ and other lower ions, and since \siiii\ drove the velocity ranges of each component in the AOD method. 

We observe the same slight systematic effect reported in \citetalias{lehner20}, where PF results yield on average slightly larger column densities than AOD estimates (which can be clearly observed in the residual panel of Fig.~\ref{f-comp-fit-aod}), but this effect is small since $\Delta \log N$ is consistent with zero within $1\sigma$ errors. Some of the profile fit errors are larger than the AOD errors for two main reasons: 1) the line is saturated (a lower limit in the AOD method); 2) low S/N and weak absorption. We emphasize that at COS resolution, all components identified as lower limits in the AOD method remain saturated and very uncertain in the PF method owing to the fact the Doppler broadening remains not well constrained.

We identified 16 AOD components and 11 PF components that had no match. These mismatches can be attributed to three main factors: (1) method-specific biases, where AOD velocity components are driven by \siiii, while PF allows independent parameter fitting for each ion (particularly relevant for \civ, \siiv, and \oi\ that may probe different gas phases); (2) detection threshold differences, where components detected in one method fell below the significance threshold in the other; and (3) blending effects, where heavily blended components were identified differently by the two methods. For example, considering the IVZw29 sightline shown in Fig.~\ref{f-example-spectrum} (AOD) and Fig.~\ref{f-example-fit} (PF), the non-matched components are for \oi, \civ, and \siiv. For \civ, the red component is treated as a single component following the \siiii\ profile decomposition, despite evidence for two components, which is how the PF modeled it. This results in only 1 matched component. For \oi, the mismatch arises from the different treatment of the weak component on the red side of the strong component at $-400$ \km, leading to a 20 \km\ difference, larger than our threshold velocity. Finally, for \siiv, the main detected component at $-400$ \km\ is again a match, but the AOD and PF methods detect different weak components near the 2$\sigma$ detection threshold, leading to 1 unmatched component in each method.

In summary, this comparison demonstrates that both the PF and AOD methods provide consistent measurements for well-detected absorption components, with some small systematic differences in the column densities for the matched components. The two methods also identified a similar number of components. Following the approach in \citetalias{lehner20}, we adopt the AOD measurements for our primary analysis in this paper. This choice is further motivated by the fact that some PF components require fixed velocity or Doppler parameters due to COS resolution limitations and low S/N, making them less reliable. However, the PF results provide valuable additional insights, particularly for understanding kinematic differences between different gas phases (e.g., \siiii\ vs. \civ), which we will investigate in more detail in a future paper.

\subsection{Magellanic Stream Contamination Identification}\label{s-ms-contamination}
\begin{figure}[tbp]
\epsscale{1.2}
\plotone{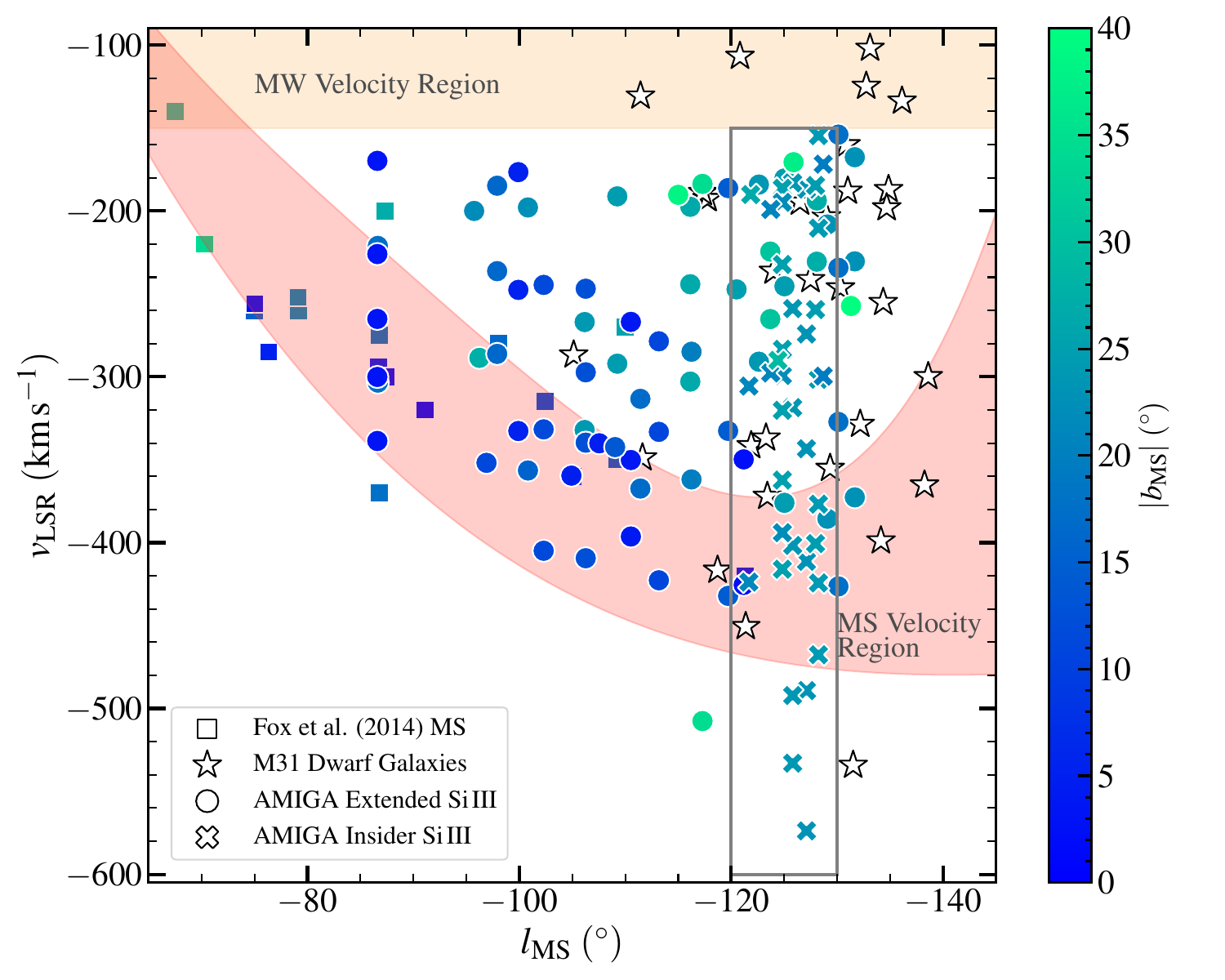}
\caption{The LSR velocity of the \siiii\ components (circles) observed in our sample as a function of the MS longitude \lms, color-coded according to the absolute MS latitude. Shaded regions show the velocities that can be contaminated by the MS and MW (by definition of our search velocity window, any absorption at $\vlsr >-150$ \km\ was excluded from our sample). We also show the data (squares) from the MS survey from \citet{fox14} and the radial velocities of the M31 dwarf galaxies (stars, see \citetalias{lehner20} and references therein). The gray rectangle highlights where all the AMIGA Insider data lie and the majority of those are not contaminated by the MS. 
\label{f-ms-contamination}}
\end{figure}

The region on the sky where M31 and its CGM are found is complex, with potential contamination arising from both the MW and the MS (\citealt{lehner15}; \citetalias{lehner20}; \citealt{kim24}). We have already removed from our analysis any contamination from higher redshift intervening absorbers and from the MW (defined as $-150 \la \vlsr\ \la 100$ \km). To remove the MS contamination, we mainly follow the method described in \citetalias{lehner20}, which uses \hi\ 21-cm emission observations of the MS by \citet{nidever10}. Their \hi\ observations show that the MS extends to about a MS longitude $\lms \simeq -140\degr$. Based on this and previous \hi\ emission surveys \citep{nidever08,nidever10}, \citetalias{lehner20} estimated the upper and lower boundaries of the \hi\ velocity range as a function of \lms, which we reproduce in Fig.~\ref{f-ms-contamination} by the colored area. Components falling in this forbidden band were largely deemed contaminated by the MS (less than about 10\% of the AMIGA Insider components). 

In Fig.~\ref{f-ms-contamination}, we plot the \siiii\ data from AMIGA Extended \citepalias{lehner20}, the MS survey by \citet{fox14}, and the M31 dwarfs from \citet{mcconnachie12}. (\siiii\ is the preferred ion owing to its large detectability frequency.) We also show the AMIGA Insider data that are all in the gray rectangle, which probe a much narrower range of MS longitudes and latitudes with $-129\degr \le \lms \le -121\degr$ and $+20\degr \le \bms \le +25\degr$. At these MS longitudes and latitudes, the MS contamination is expected to be much smaller in the AMIGA Insider sample.

\begin{figure}[tbp]
\epsscale{1.2}
\plotone{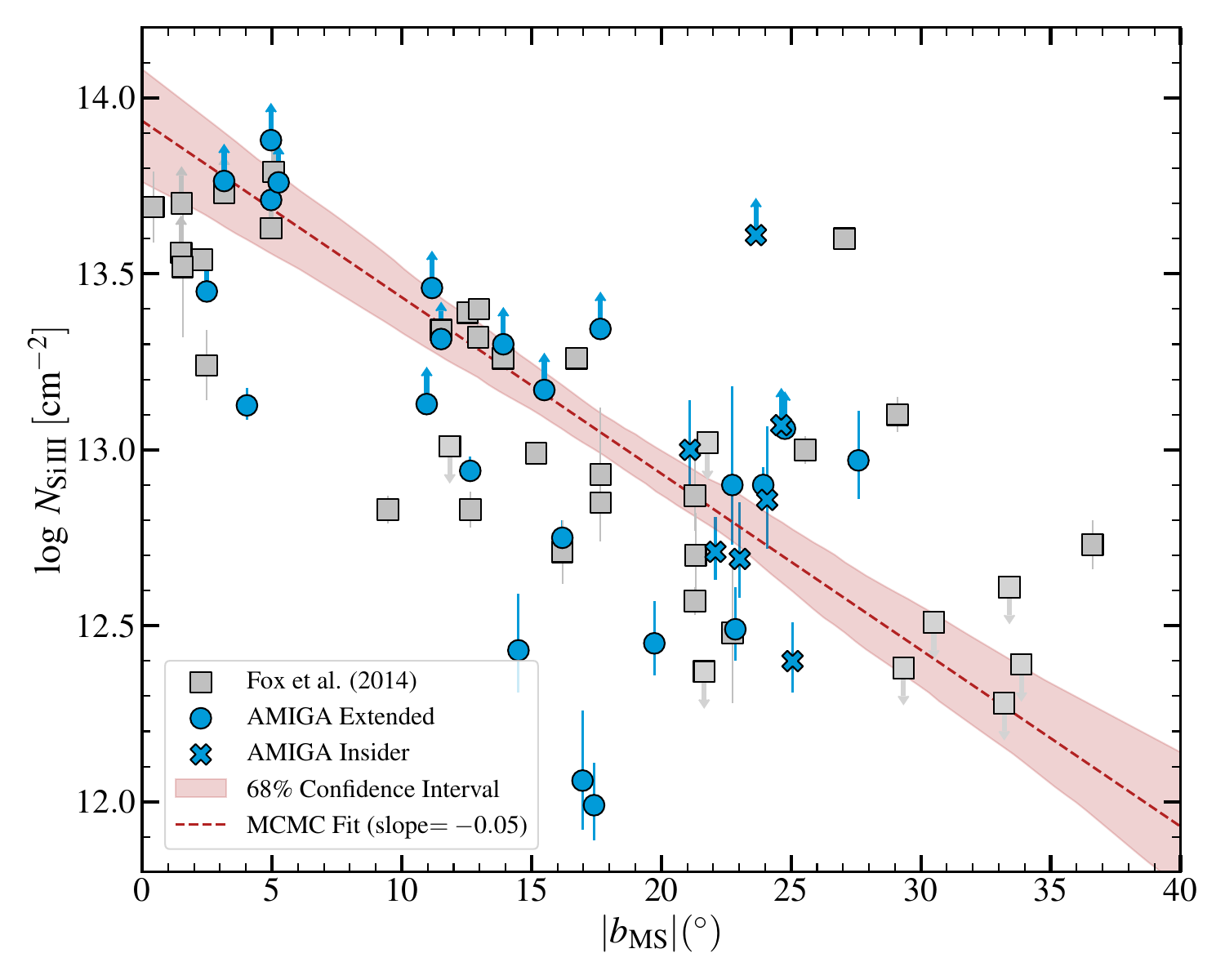}
\caption{The total column density of \siiii\ that is associated with the MS as a function of the absolute MS latitude. We also show the MS survey by \citet{fox14} restricted to data with $-150\degr \le \lms \le -20\degr$. The lighter gray squares with downward arrows are non-detections in the \citeauthor{fox14} sample. The dashed line is the MCMC fit to all the data properly accounting for measurements, upper limits, and lower limits, with the shaded region showing the 68\% confidence interval.
\label{f-col-cont}}
\end{figure}

In the region defined by $-150\degr \le \lms \le -20\degr$, the bulk of the \hi\ 21-cm emission is observed within $|\bms|\la 5\degr$ \citep{nidever10}, and hence the metal ionic column densities are expected to show strong absorption when $|\bms|\la 10\degr$ and weaker absorption as $|\bms|$ increases. The second part of this analysis therefore ensures that the column density of \siiii\ decreases with increasing $|\bms|$. In Fig.~\ref{f-col-cont}, we show the column densities of \siiii\ for the velocity components from AMIGA Insider and AMIGA Extended found within the MS boundary region shown in Fig.~\ref{f-ms-contamination}. We also show in the same figure the results from the \citet{fox14} MS survey. This figure shows the expected trend of decreasing column density with increasing MS latitude with some scatter. Some of the higher latitude absorbers at $|\bms|\ga -25\degr$ with high column densities that depart significantly from the overall trend may not be part of the MS as we argue below. 

\citetalias{lehner20} used a standard least-squares regression to analyze the relationship between column density and MS latitude, which did not handle the asymmetric error bars and censored data properly (see \S\ref{s-mass-model} for more description on the treatment of censored data). Here, we implemented a more rigorous maximum likelihood approach. For each data type, we constructed an appropriate likelihood function that incorporates asymmetric errors and uses the cumulative distribution function for upper limits and the survival function for lower limits. We modeled the relationship as a linear function in log space: $\log N_{\rm Si\,III} = \alpha  |\bms| + \log N_0$, where $\alpha$ is the slope and $\log N_0$ is the intercept.   We include an additional parameter $\sigma$ to account for intrinsic scatter beyond measurement uncertainties (this approach is described more fully \S\ref{s-mass-main}).

\begin{figure}[tbp]
\epsscale{1.2}
\plotone{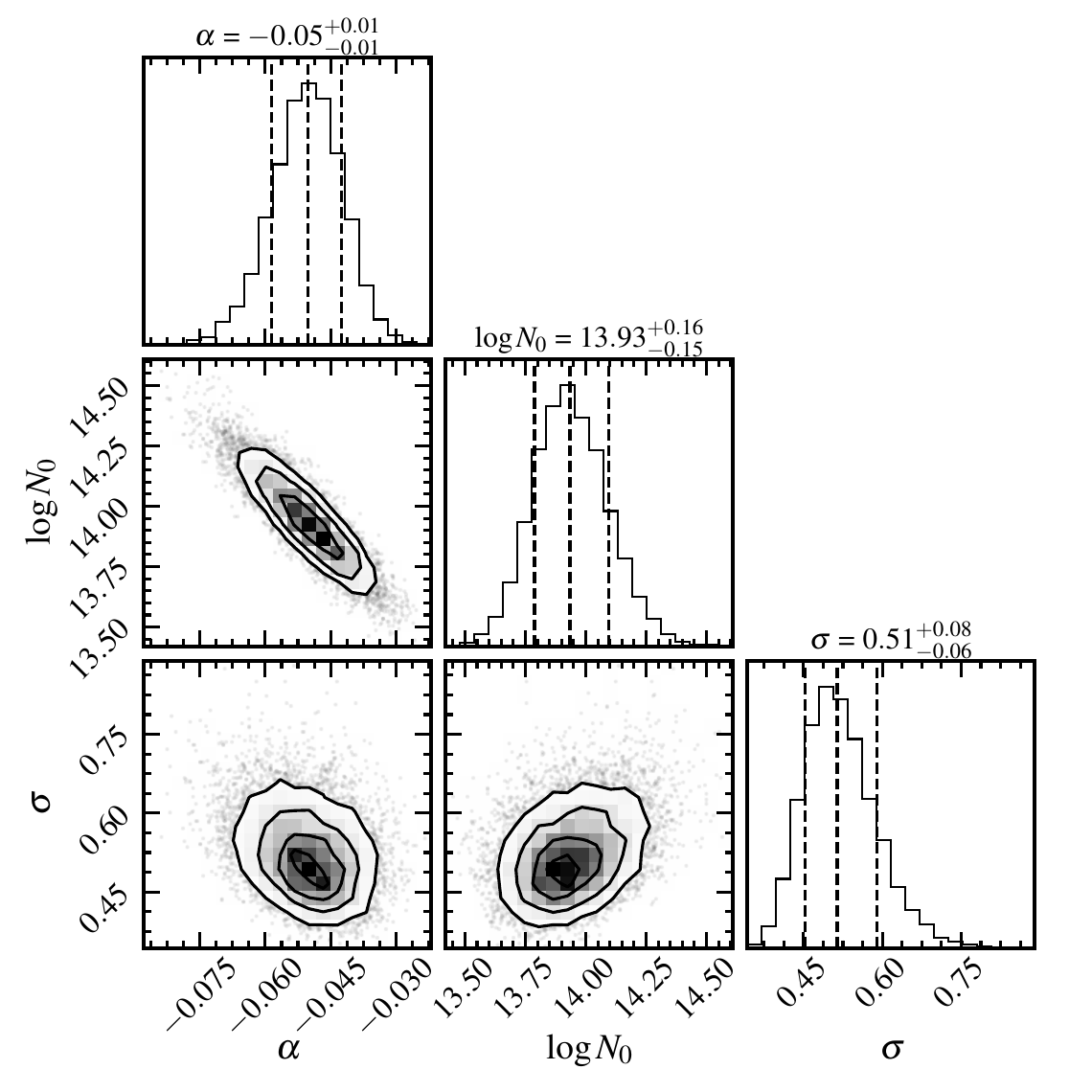}
\caption{Corner plot showing the MCMC posterior distributions for the three parameters in our linear model of $N_{\rm Si\,III}$ as a function of MS latitude. The diagonal panels show marginalized 1D histograms with median and 68\% confidence intervals (dashed lines), while the off-diagonal panels display 2D joint distributions with contours at 68\%, 95\%, and 99.7\% confidence levels.
\label{f-mcmc_corner}}
\end{figure}

We sampled the posterior probability distribution of the model parameters $(\alpha, \log N_0, \sigma)$ using Markov Chain Monte Carlo (MCMC) via the \texttt{emcee} package \citep{foreman-mackey13}. The physical priors constrained the parameter space: $\alpha \in [-2.0, 0.0]$, $\log N_0 \in [11.0, 15.0]$, and $\sigma \in [0.0, 2.0]$. Our MCMC implementation used 32 walkers for 5,100 steps, discarding the first 1,000 steps as burn-in. The confidence bands for the linear fit were generated by drawing random samples from the MCMC posterior distribution, computing model predictions across the $|\bms|$ range, and calculating the 16th and 84th percentiles (68\% confidence interval). The results from the MCMC fit are $\alpha = -0.05 \pm 0.01$, $\log N_0 = 13.93 \pm 0.15$ and $\sigma = 0.51 \pm 0.07$. In Fig.~\ref{f-mcmc_corner}, we show the corner plot for this analysis, with well-defined contours indicating that the MCMC sampling has converged and the parameters are well-constrained by the data. The fit and confidence interval are shown in Fig.~\ref{f-col-cont}. This figure and fit show higher \siiii\ column densities (13.50 to $\,>14.0$ dex) at low $|\bms| < 10\degr$ transitioning to lower column densities ($<12.0$ to 12.5 dex) at high $|\bms| > 30\degr$, where most of the data are upper limits. This gradient spans nearly two orders of magnitude in column density.

In the region $+20\degr \le \bms \le +25\degr$ where the AMIGA Insider data lie, several components with MS-like velocities also have MS-expected column densities. However, we note that one AMIGA Insider data point is 0.8 dex above the cluster at $\bms = 23.6 \degr$ (the lower limit at $\log N_{\rm Si\,III}= 13.6$). Within this \bms\ range, the MCMC fit predicts that $\log N_{\rm Si\,III}= 12.77$ associated with the MS, fully consistent with the mean observed value of $12.79 \pm 0.29$. This data point is several sigma off from these values and also beyond the estimated scatter of $0.51\pm 0.07$ dex. This component is the one observed toward IVZw29 at $-400$ \km\ (see the light blue component in Fig.~\ref{f-example-spectrum}), which is at a velocity consistent with it being associated with the thick disk of M31 based on the predicted rotation velocity of the galaxy at this position (see \ref{s-m31-dh}). Given this statistical outlier status and the kinematic alignment with M31, we a posteriori reassign this component to the thick disk of M31 (including this data point or not would not change quantitatively the linear fit to the data).

We finally note that \citet[][hereafter \citetalias{kim24}]{kim24} have developed a statistical approach to determine whether absorption features should be associated with MS or with M31 CGM and its satellite galaxies using a Wasserstein distance metric. This provides an opportunity to crosscheck our respective identifications of the components with the MS and the CGM of M31.  One of the main differences from \citetalias{kim24}'s association methodology is their requirement that absorbers be gravitationally bound to the galaxy for association with M31's CGM or satellite galaxies. We cannot apply this criterion because the CGM can extend beyond the virial radius (\citetalias{lehner20}; \citealt{wilde21,wilde23}), and we aim to study both inner and outer regions out to about $2\rvir$. They used similar \hi\ data to identify the MS, but what Fig.~\ref{f-ms-contamination} does not reflect well is that the \hi\ gas gets clumpier at $-140\degr \la \lms \la -110 \degr$ (see Fig.~7b in \citealt{nidever10}); this region is treated as more uncertain in \citetalias{kim24}. To compare \citetalias{kim24} identifications with our own, we reclassified any absorber that \citetalias{kim24} associated with M31's satellite galaxies (M33, etc.) as an M31 absorber for consistency.\footnote{We refer the reader to \citetalias{lehner20} for the possible association of some of the M31 dwarf galaxies or M33---the overall conclusion was that the CGM of M33 and the dwarf galaxies do not contribute significantly to the observed absorption associated with the CGM of M31.}

First, when we exclude the uncertain cases in \citetalias{kim24}, we find an agreement rate of \citetalias{kim24}'s classifications with \citetalias{lehner20} of 85.7\% for the M31 CGM identification and 94.3\% for the MS identification.\footnote{We consider only detections and the absorbers flagged as lower limits in \citetalias{lehner20} for that comparison. We also consider all the ions that are detected in that comparison.} The agreement rates of \citetalias{lehner20} classifications with \citetalias{kim24} are 98.8\% for the M31 CGM identification and 70.2\% for the MS identification. The disagreement and the asymmetry come from the fact that \citetalias{kim24} classifies six absorbers (corresponding to 14 ionic components) as M31 CGM that \citetalias{lehner20} classifies as MS, while \citetalias{lehner20} classifies only two absorbers as M31 CGM that \citetalias{kim24} classifies as MS. These six absorbers are all in the forbidden region at $-130\degr \la \lms \la -120 \degr$ where the association with M31 CGM may be more uncertain as discussed above. \citetalias{kim24} classifies them as M31 CGM because they meet the gravitational binding criteria. However, we still favor a MS origin due to some evidence of MS gas in that region. We note for the AMIGA Insider data that all lie in this region, the contamination is minimal because $|\bms|$ is large (see above). For the two other cases, the difference originates from the fact that \citetalias{kim24} used the velocity integration range limits to define the average velocity (i.e., $(v_1+v_2)/2$) rather than the column-density weighted average; in both these cases (that are at the edge of the M31 CGM/MS boundary in Fig.~\ref{f-ms-contamination}), our results would have agreed if the same velocity criteria had been used.

Second, including the uncertain cases in \citetalias{kim24}, the overall agreement between the two methods drops to 61\%. Sixteen absorbers (34 ionic components) are identified as uncertain in \citetalias{kim24} but as M31 CGM in \citetalias{lehner20}. In all but one case, these are not bound to M31 CGM according to their definition, which explains the difference in the identification.\footnote{An exception is one of the absorbers toward KAZ238. Toward that sightline, \citetalias{kim24} and \citetalias{lehner20} associate the most and least negative components to the MS and M31 CGM, respectively (i.e., the two methods are in agreement). The intermediate component is, however, identified as uncertain in \citetalias{kim24}, but as M31 CGM in \citetalias{lehner20}, which is consistent with the classification for the least negative component.} Finally, eight absorbers (24 ionic components) are identified as uncertain in \citetalias{kim24} but as MS in \citetalias{lehner20}. All these targets are far from the MS \hi\ emission but where ionized gas can be plentiful \citep{fox14}, explaining this discrepancy.

This comparison reveals that our method is more conservative regarding M31 identification, in that we favor an MS identification over an M31 CGM identification in regions where there is some evidence of possible MS gas. The key difference between the two methods is that \citetalias{kim24}'s requirement that absorbers be gravitationally bound is too restrictive for our study, which originally aimed to characterize the CGM well beyond the virial radius. Despite these methodological differences, there is a good agreement between approaches in unambiguous regions. The fact that we can explain all discrepancies between the methods validates our methodology as appropriate for studying the full extent of M31's gaseous CGM.

\subsection{Separating the Thick Disk and CGM of M31}\label{s-m31-dh}
A key difference between the AMIGA Insider and AMIGA Extended samples is that several of the Insider sightlines probe the inner regions of M31's CGM at projected distances $R \la 25$ kpc. At these small impact parameters, some of the observed absorption could be associated with the disk or thick disk of M31 rather than its extended CGM. Indeed, in several cases, we also detect this gas in \hi\ 21-cm emission, providing additional evidence for its disk-like nature since it has a small covering factor beyond 25 kpc ($f_c \la 10\%$ at $R\le 50$ kpc; \citealt{howk17}; J.C. Howk et al., 2025, in prep.). In this paper, we define a component as being part of the thick disk if they are within $R\la 30$ kpc and corotate with the M31 gas (according to our model described below). 

Our identification process for components associated with M31's thick disk involves several steps. First, we used the \hi\ velocity field information from \citet[][see their Fig.~8]{chemin09} in conjunction with the position of our targets. For most of the targets at $R\la 30$ kpc, there is evidence of gas with saturated \cii\ and \siiii\ absorption and often detection of \oi\ absorption at velocities consistent with the expected \hi\ gas rotation velocities in these directions. 

For a more systematic and general identification of thick disk components, we developed a model that compares the observed absorption velocity of each component with the predicted rotation velocity of the galaxy (see, e.g., \citealt{prochaska98,kacprzak19a}). We adapted the cylindrical Navarro-Frenk-White (NFW) halo model developed by \citet{french20} to the specific geometry of each M31-QSO system. This model employs an NFW density profile to create a physically motivated, radially declining rotation velocity structure that extends to several virial radii. For each QSO sightline, we generated rotation velocity predictions along the entire line of sight ($D_{\rm los}$) as it passes through M31's CGM.

Our classification process is as follows. For each absorption component, we first determine whether it is consistent with corotation by checking if its velocity falls within the range predicted by the rotation model (with a base tolerance of $\pm$15 km s$^{-1}$) and is in the correct direction relative to M31's systemic velocity based on the sightline position. Components identified as corotating are further evaluated for thick disk association using a distance-dependent velocity tolerance method where we define the velocity scale height (in the $z$ direction) of M31's thick disk as 2 kpc, with a maximum extent of 6 kpc (three scale heights) from the disk plane. For each corotating component, we determine its most likely line-of-sight distance, $D_{\rm los}$, by finding where along the line of sight the modeled velocity most closely matches the observed component velocity. We employ a linearly scaled velocity tolerance that increases with distance from the disk plane from 25 \km\ at the disk plane to 50 \km\ at the maximum thick disk boundary (6 kpc). Components whose velocities fall within this distance-dependent tolerance and whose $D_{\rm los}$ positions are within the thick disk boundaries are classified as thick disk components.

\begin{figure*}[tbp]
\epsscale{1.}
\plotone{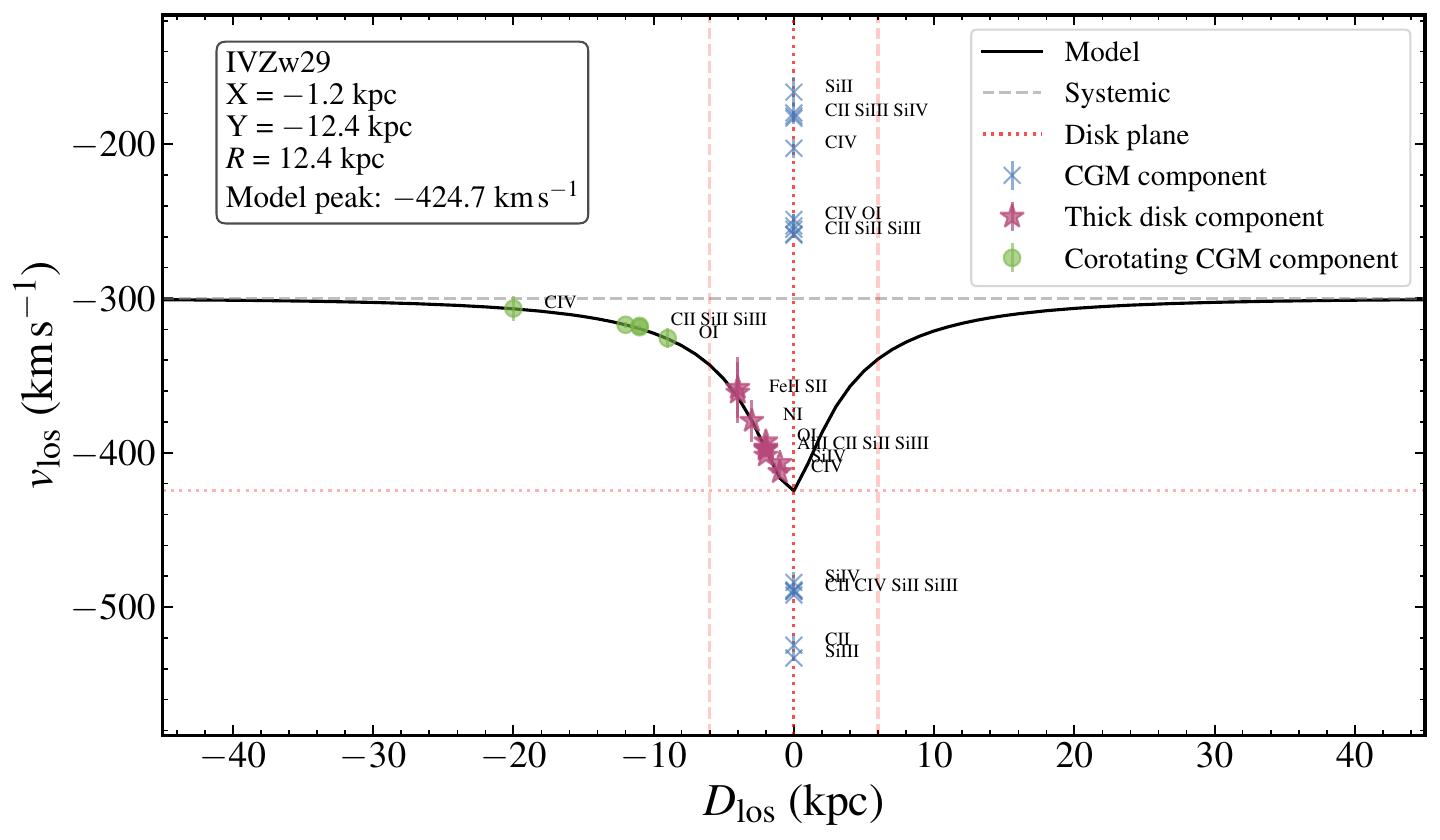}
\caption{Visualization of absorption components along the IVZw29 sightline in relation to M31's rotation model. The black curve shows the predicted line-of-sight velocity ($v_{\rm los}$) as a function of line-of-sight distance ($D_{\rm los}$) from M31's disk plane, with M31's systemic velocity indicated by the gray dashed line at $-300$ \km. The red dotted vertical line marks the disk plane ($D_{\rm los}=0$ kpc). The vertical dashed lines are the 6 kpc thick disk maximum scale height values and the horizontal dotted line is the model peak velocity.  Components are classified and positioned as follows: thick disk components (purple stars) and corotating CGM components (green circles) are placed at $D_{\rm los}$ positions where the model velocity best matches their observed velocities. Due to inherent kinematic degeneracies in the projected rotation signature, positions represent $|D_{\rm los}|$. Components could therefore equally be located at negatively or positive $D_{\rm los}$ values.  We systematically plotted them at negative $D_{\rm los}$ assuming they are on the near side of the galaxy for visualization consistency. Non-corotating CGM components (blue X marks) are arbitrarily placed at the disk plane for visualization purposes only. 
\label{f-IVZw29_components}}
\end{figure*}

In Fig.~\ref{f-IVZw29_components}, we illustrate this classification for IVZw29, showing model rotation curves and the position of observed components. In this figure, we place thick disk and corotating components at their calculated $D_{\rm los}$ positions, while all other components (counter-rotating, high-velocity, or otherwise unassociated) are arbitrarily placed at the disk plane ($D_{\rm los} = 0$ kpc).  This sightline has a small impact parameter ($\rho=12.4$ kpc) and shows multiple absorption components at $\approx [-413, -338]$ \km\ that are classified as thick disk gas based on their velocities and spatial association. Co-rotating components near $-320$ \km\ likely represent gas in the extended rotating CGM beyond the thick disk region. Some of the (non-corotating) CGM components could be the analogs of HVCs observed in the MW.

While our primary classification relies on the $D_{\rm los}$ position and velocity matching, we also apply an impact parameter constraint as a secondary criterion, considering that thick disk material is unlikely to extend far beyond approximately 30 kpc in projected distance. Components at impact parameters exceeding this threshold, even if they otherwise match the velocity criteria, are less likely to represent genuine thick disk material. 

In Table~\ref{t-thick_disk}, we summarize the components consistent with thick disk material, listing the velocity range corresponding to where absorption components are identified as originating from the corotating thick disk. We find thick disk components along six sightlines, primarily from the AMIGA Insider sample, at impact parameters ranging from 10.7 to 29.0 kpc. These components satisfy our kinematic criteria for thick disk association and initial visualization of the profiles: they show corotation with the expected disk motion, have $D_{\rm los}$ positions within 6 kpc of the disk plane, and velocities that match model predictions within our distance-dependent tolerance scale. The velocity ranges of thick disk components vary across sightlines, with more negative velocities generally observed in the western regions (negative $X$ coordinates) compared to the eastern regions. This pattern aligns with M31's rotation structure, where the northeast portion of the disk is receding while the southwest is approaching. In many cases, these same sightlines also have components that do not match the thick disk criteria, representing M31's CGM or HVCs. We will explore this model further and the corotation/non-corotation of the M31 CGM beyond the thick disk region in a future paper.

\section{Column Densities and Kinematics in the CGM of M31}\label{s-init-results}

\subsection{Total Column Densities versus $R$}\label{s-n-vs-r}
\begin{figure*}[tbp]
\epsscale{1.}
\plotone{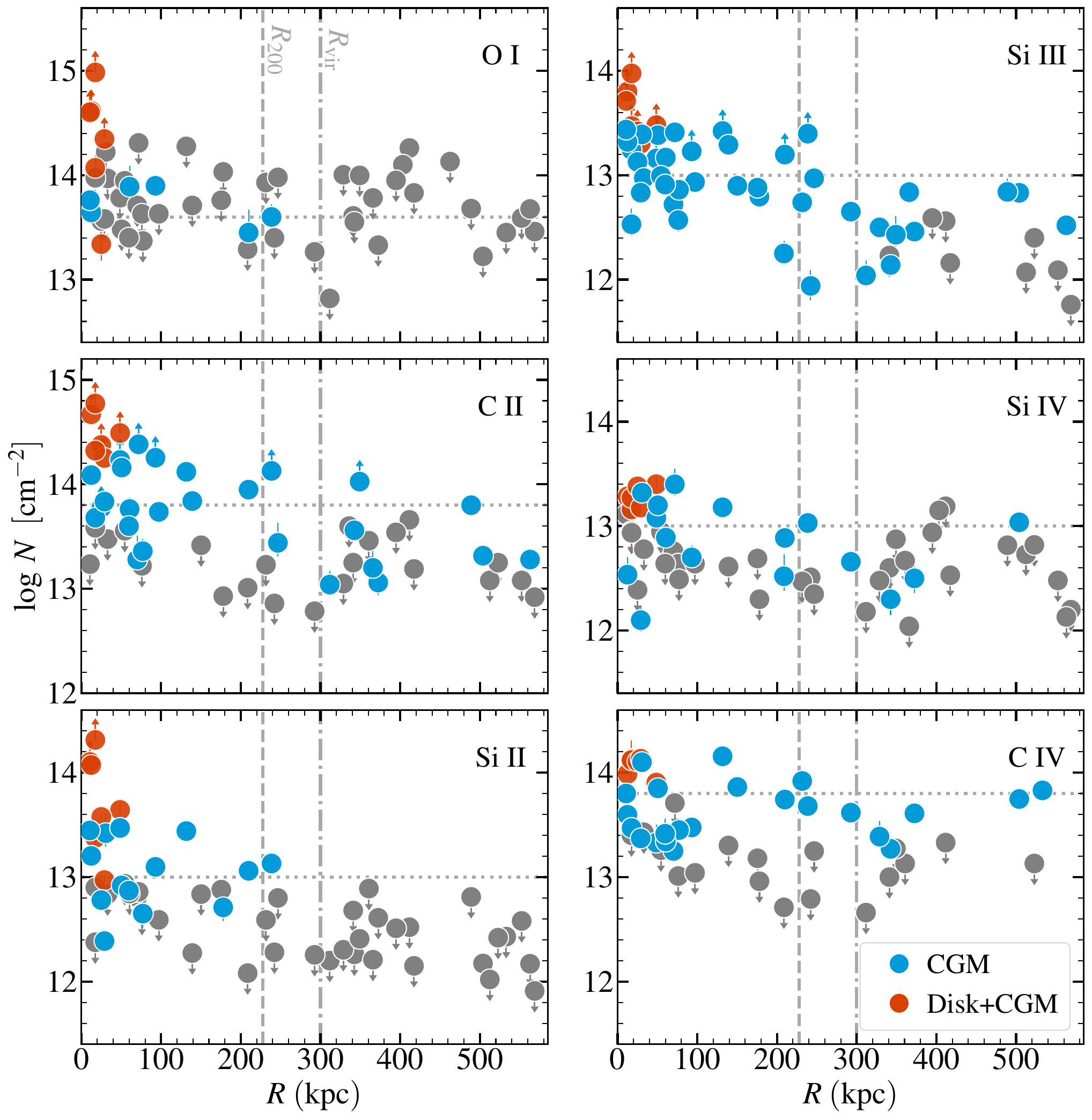}
\caption{Total atomic and ionic column densities of ions as a function of projected distance $R$ from M31. Column densities are shown on a logarithmic scale with a consistent vertical range of 3.2 dex in each panel. Blue circles represent CGM detections, while gray circles with downward arrows indicate non-detections. Red circles show the total column density when  the thick disk components are added to CGM values. Blue or red circles with upward arrows denote saturated absorption resulting in lower limits. The dashed vertical and dot-dashed lines mark $R_{200}$ and \rvir, respectively. The dotted horizontal line in each panel corresponds to the threshold column density ($N_{\rm th}$) separating strong absorption (primarily observed at $R \la R_{200}$) from weak absorption.
\label{f-coltot-vs-rho}}
\end{figure*}

To determine total column densities along each sightline, we combined multiple absorption components using a methodology that accounts for measurement uncertainties and detection limits. For sightlines with a single absorption component, we directly adopted its measured column density, associated errors, and detection classification. For sightlines with multiple components, we implemented the following approach: when all components were firm detections, we summed their column densities in linear space, propagated errors in quadrature, and converted back to logarithmic space. For sightlines with exclusively upper limits, we adopted the most constraining (lowest) upper limit, scaled by the velocity range of the total absorption. When sightlines contained detections and/or lower limits, we employed MCMC survival analysis, sampling from normal distributions with asymmetric errors for detections and truncated distributions for lower limits.\footnote{If one or several components are detected for a given ion, any non-detections in other components of the same ion  have upper limits on the column density that are negligible; therefore the upper limits are ignored in these estimations.} We provide further details in \S\ref{s-mass-estimate} regarding this approach, where we apply it to estimate the total column density of silicon.

In Fig.~\ref{f-coltot-vs-rho}, we display the logarithmic total column densities as a function of $R$ where we observe the total column densities of M31-associated components for \oi, \cii, \siii, \siiii, \siiv, and \civ\ systematically decrease with increasing $R$, with the decline becoming progressively shallower as ionization potential increases. This trend confirms and extends the findings of \citetalias{lehner20} and \citet{lehner15} with our expanded sample. At any $R$, there is large intrinsic scatter (larger than individual uncertainties) for all ions. The AMIGA Insider data clearly demonstrate that thick disk contributions preferentially enhance column densities of neutral and low ions (\oi, \cii, \siii) compared to intermediate (\siiii) and high ions (\civ, \siiv).

To build upon beyond the qualitative analysis in \citetalias{lehner20}, we quantitatively assess the $\log N$ versus $R$ correlations using the non-parametric Kendall's $\tau$ test following the methodology of \citet{isobe86}, which extends standard techniques to handle astronomical data with censoring. We employed the Python code {\tt pymccorrelation} from \citet{privon20}, which estimates correlation coefficient uncertainties using the bootstrap technique described in \citet{curran15}. While the modified Kendall's $\tau$ test is specifically designed for censored data, not all upper limits in our dataset are equally constraining (see Fig.~\ref{f-coltot-vs-rho}), particularly for \oi. We therefore created a filtered sample, excluding upper limits above $\langle \log N_{\rm det}\rangle - 0.5\sigma_{N_{\rm det}}$, where $N_{\rm det}$ represents column densities for absorbers detected at $>2\sigma$ significance. In Table~\ref{t-corr_NvsR}, we summarize the Kendall's $\tau$ coefficients and associated $p$-values for both the CGM-only and combined CGM-thick disk samples.

Nearly all the ions exhibit statistically significant (i.e., $p<0.05$) negative correlations between column density and impact parameter for CGM absorbers alone, and this relationship is universally significant in the combined CGM+thick disk sample. This confirms that column densities systematically decrease with increasing distance from the galaxy. Low ionization species (\oi, \siii, \siiii, \cii) generally show stronger correlation coefficients than higher ionization species (\civ, \siiv), indicating that the radial gradient is more pronounced for low ionization gas---a finding visually apparent in Fig.~\ref{f-coltot-vs-rho}. As discussed and compared with simulations in \citetalias{lehner20}, both trends are also observed in cosmological simulations (see also, e.g., \citealt{oppenheimer08,damle22}). The entire and filtered samples yield consistent results, except for \oi, where the filtered sample shows a stronger and more significant negative correlation (unsurprising given its numerous unconstrained upper limits). Including thick disk components consistently strengthens correlations across all ions, particularly for higher ionization species (\civ, \siiv). This suggests that the entire system (thick disk and CGM) follows a more coherent radial profile than the CGM alone, or alternatively, that the column density profiles for CGM absorbers alone flatten at small impact parameters. This flattening at low $R$ for the CGM-only sample likely may reflect the transition from the hotter to the cooler, denser thick disk environment, revealing how the multiphase CGM of M31 is structurally organized from the inner disk-halo interface to the outer virial radius.

\subsection{Azimuthal Dependence of the Column Densities}\label{s-col-azimuth}
So far, we have ignored the distribution of the targets in azimuthal angle ($\Phi$) relative to the projected minor and major axes of M31, where different physical processes may occur. Observations of the inner CGM ($R\lesssim 50$--100 kpc) of star-forming galaxies have shown that absorption properties may depend on the geometry relative to the sightline (e.g., \citealt{bordoloi11,bordoloi14b,kacprzak12a,bouche12,lan18}). In particular, \citet{lan18} used 200,000 emission line galaxies to show that metal absorption is 5--10 times stronger around star-forming galaxies on small scales ($<100$ kpc) compared to a large sample of luminous red galaxies, and exhibits clear minor-axis enhancement, providing strong statistical evidence for outflow-driven CGM enrichment around star-forming galaxies. Some cosmological simulations also show strong geometrical structure in the CGM shaped by large-scale inflows and outflows, particularly for lower-mass and actively star-forming systems (e.g., \citealt{brook11,stewart11a,nelson15,peroux20}). However, theoretical predictions are not universal and depend strongly on galaxy mass and star formation activity. For more massive galaxies that have settled into a quiescent mode of star formation, FIRE simulations predict little azimuthal dependence in CGM properties \citep{muratov15,stern21}, as the cold streams and strong outflows that drive azimuthal asymmetries in lower-mass systems are not expected to be prominent in M31-like halos. We now examine whether M31's CGM exhibits the minimal azimuthal dependence predicted for massive, quiescent systems.

\begin{figure*}[tbp]
\epsscale{1.2}
\plotone{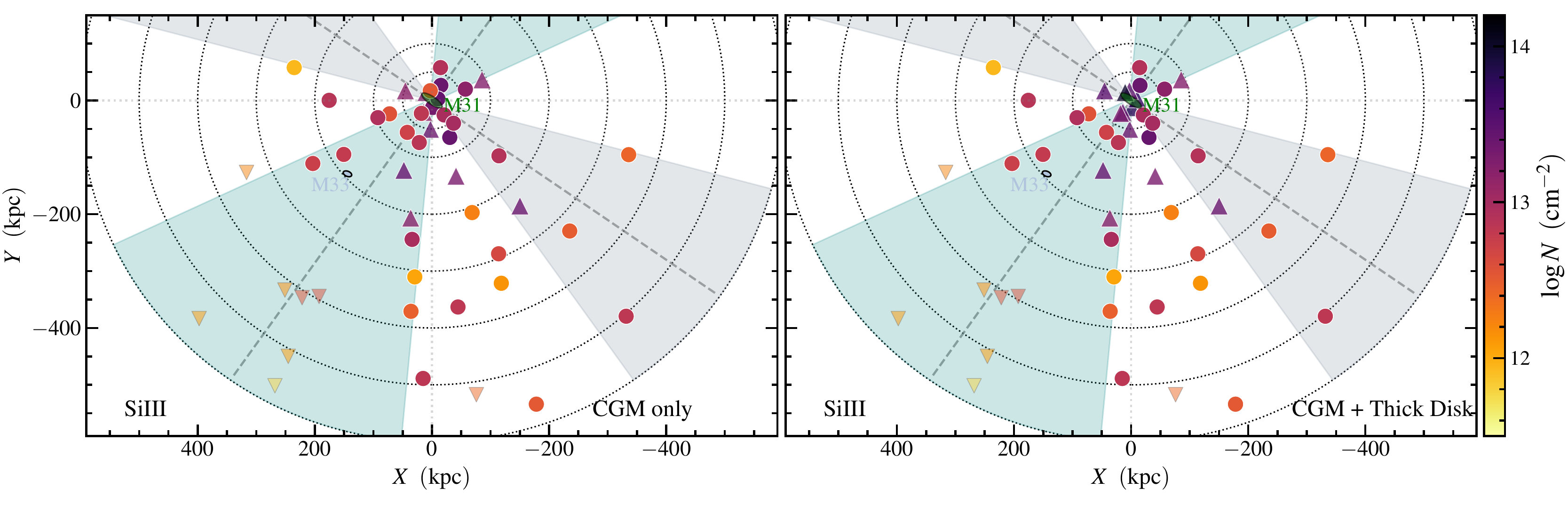}
\caption{Positions of the Project AMIGA targets relative to M31, where the axes show the physical impact parameter from the center of M31. The data points are color-coded according to the vertical color bar with the total column densities of \siiii, showing the CGM-only components on the {\it left} and the CGM plus thick disk components on the {\it right}. Circles represent detections, upward triangles are lower limits on the column densities, and downward triangles are non-detections. The colored cones indicate regions $\pm 30\degr$ along the projected minor axis and $\pm 20\degr$ along the projected major axis.
\label{f-colmap}}
\end{figure*}

In Fig.~\ref{f-colmap}, we display the spatial distribution of \siiii\ total column densities in the $X$--$Y$ plane around M31. The left panel shows the CGM-only components, while the right panel includes the identified thick disk components added to the CGM components. As expected, the right panel exhibits higher column densities in the inner CGM regions. The most prominent trend is the radial dependence characterized in the previous section: column densities generally decrease with increasing impact parameter $R$. As previously noted in \citetalias{lehner20}, there is no strong large-scale azimuthal dependence, except beyond \rvir, where we find only non-detections along the projected minor axis, while most detections occur in the white region between the minor and major axis cones.

\begin{figure*}[tbp]
\epsscale{1.2}
\plotone{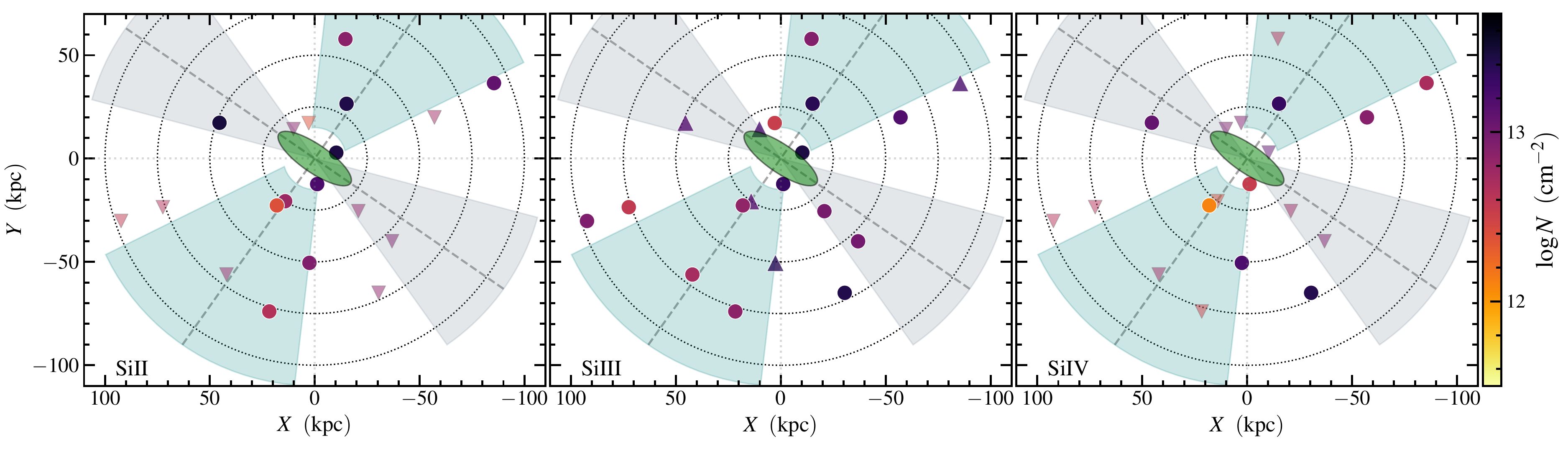}
\caption{Similar to Fig.~\ref{f-colmap}, but for the CGM components only of \siii, \siiii, and \siiv, and within only the inner regions of the M31 CGM.  
\label{f-colmapinner}}
\end{figure*}

In Fig.~\ref{f-colmapinner}, we examine the inner regions where azimuthal ($\Phi$) dependence might be more pronounced, displaying the spatial distributions of total CGM column densities for \siii, \siiii, and \siiv\ (left, middle, and right panels, respectively). At the smallest impact parameters ($R \sim 55$ kpc, $0.18 \rvir$), our sample includes seven sightlines within the projected minor axis cone and four along the major axis. Visual inspection suggests marginally higher column densities along the minor axis compared to the major axis. To quantitatively assess this potential difference, we applied a parametric survival analysis to properly account for both detections and lower limits (e.g., \citealt{feigelson85}; implemented via the {\tt survreg} function in R, \citealt{therneau21}). The resulting column density statistics for each silicon ion are presented in Table~\ref{t-colaxis}. The mean column densities for \siiii\ and \siiv\ are higher by approximately 0.2 dex along the minor axis, though with no corresponding difference in median values and substantial standard deviations. Log-rank tests comparing the distributions along minor and major axes yield $p > 0.05$ for all three ions, indicating no statistically significant differences. Our sample has limited size at $R\la 55$ kpc, but we find no compelling evidence for significant azimuthal variation at any radius in M31's CGM for any of the studied ions in our sample. This suggests that M31's CGM is not currently structured by large-scale bipolar outflows, consistent with its relatively quiescent star formation history and supporting theoretical predictions for massive, settled galaxies.

\subsection{Azimuthal and Radial Dependence of the Kinematics}\label{s-kin}
\begin{figure}[tbp]
\epsscale{1.25}
\plotone{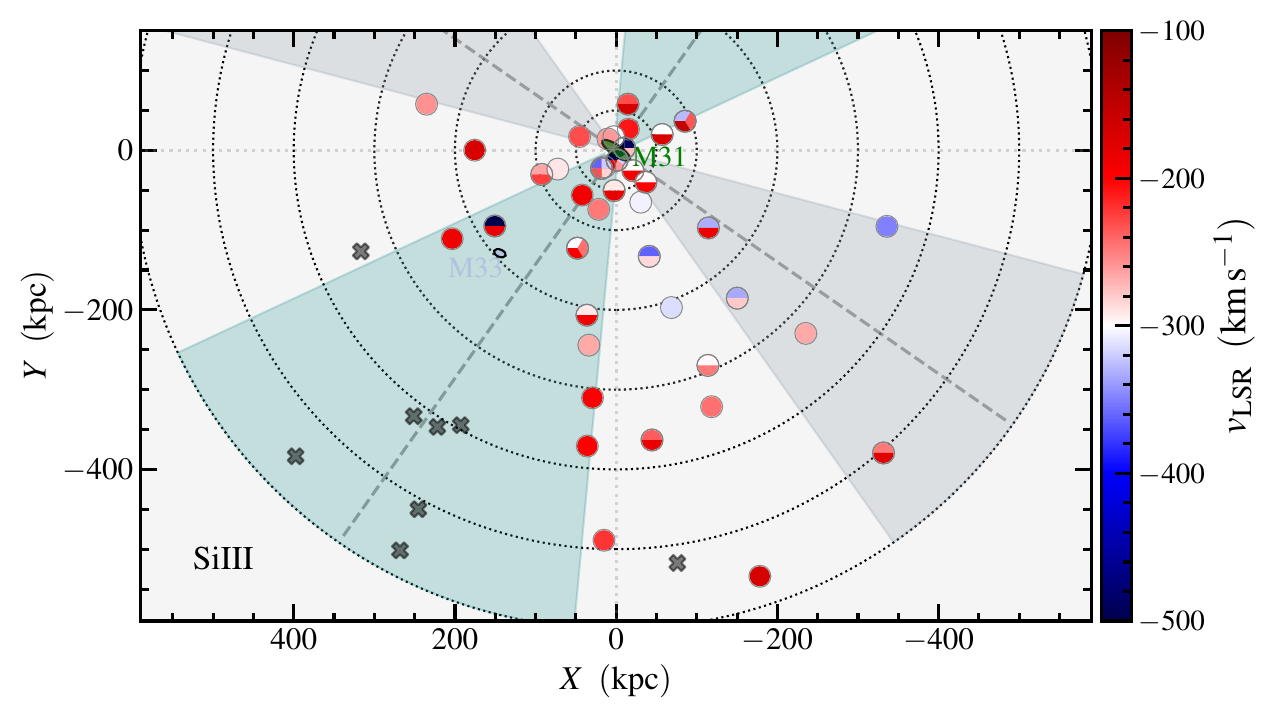}
\caption{Positions of the Project AMIGA targets relative to M31, where the axes show the physical impact parameter from the center of M31. The LSR velocities of each velocity component are color-coded according to the vertical color bar. Circles represent detections while crosses indicate non-detections. Circles with multiple colors indicate that the observed absorption along those sightlines has more than one component. Both CGM and thick disk components are included in this visualization. The systemic velocity of M31 is $\vlsr = -300$ \km.
\label{f-vel-dist}}
\end{figure}

\begin{figure}[tbp]
\epsscale{1.2}
\plotone{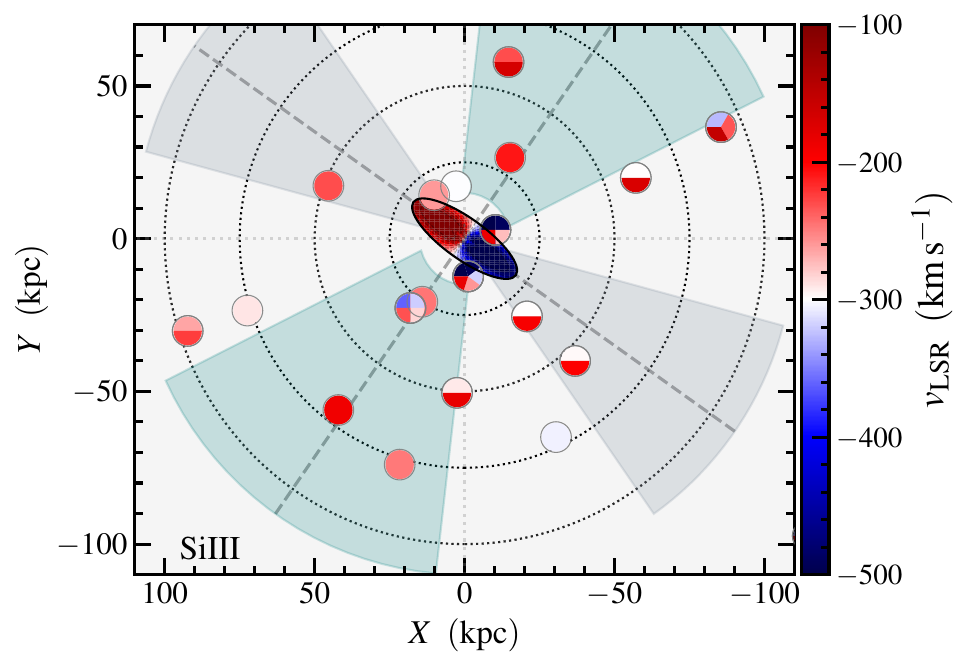}
\caption{Same as Fig.~\ref{f-vel-dist}, but focusing on the inner region of the CGM of M31. The velocity field of the M31 disk is shown based on results from \citet{chemin09}.
\label{f-vel-dist-inner}}
\end{figure}

To further explore the spatial and radial structure of the M31 CGM and thick disk, we analyze the distribution of \siiii\ absorption component velocities in Fig.~\ref{f-vel-dist}, which shows the spatial distribution of \siiii\ LSR velocity components associated with M31. Importantly, this analysis includes both CGM and thick disk components to provide a comprehensive view of M31's gaseous environment. We focus again on \siiii\ as it provides the most detections compared to other ions. In Fig.~\ref{f-vel-dist-inner}, we provide a closer view of the inner region of M31, displaying the velocity field based on results from \citet{chemin09}. In both figures, circles with multiple colors indicate that the observed absorption appears in more than one component.

The first striking feature from Figs.~\ref{f-vel-dist} and \ref{f-vel-dist-inner} is the predominance of components with $\vlsr \ge -300$ \km. While some components associated with the MS could potentially blend with M31's CGM, at any MS longitudes there is no evidence of any absorption at $\vlsr \la -420$ \km\ (see Fig.~\ref{f-ms-contamination}), except for a few near M31 itself (some corotating components, but not all of them, see Fig.~\ref{f-IVZw29_components}). This confirms that much of the M31 CGM kinematics lie at $\vlsr \ge -300$ \km, meaning that in the lower right quadrant most of the CGM gas is counter-rotating. Using semi-analytic parametric models, \citet{afruni22} demonstrated that the \siiii\ kinematics (as well as the column densities and covering factors) are well reproduced by an inflow model where gas infalling from the virial radius has low radial and tangential velocities and falls toward the center with spiraling motion (see also \S\ref{s-disc-radial} for further discussion of these and other models).

The second remarkable feature is the radial dependence in kinematic complexity. When considering both thick disk and CGM components together, we find a statistically significant decline in the number of absorption components with increasing distance from the galaxy. This also implies that the full-width of the absorption profiles is larger in the inner than outer CGM regions.  A Spearman's rank correlation analysis of component count as a function of $R$ yields $\rho = -0.54$ with $p = 0.0001$, confirming a moderately strong negative correlation between component count and impact parameter. A Kruskal-Wallis test ($p = 0.011$) further confirms significant differences in component distributions across different radial bins of 50 kpc. These statistical indicators show that the combined gas structure becomes progressively less complex at larger galactocentric distances, transitioning from a multi-component absorption in the inner regions ($R \lesssim 100$ kpc) to predominantly single-component absorption at larger radii.

When we restrict our analysis to only CGM components (excluding thick disk components), the statistical significance diminishes, with the Spearman's rank correlation analysis yielding $\rho = -0.27$ with $p = 0.086$. Nevertheless, even within the CGM alone, we observe a trend of a decreasing of the number of components with increasing $R$: on average, the number of components at $R\la 150$ kpc is about 2--3, while it decreases to approximately one at $R>150$ kpc. This suggests that the inner CGM maintains greater kinematic complexity than the outer regions, though the transition is more pronounced when thick disk components are included.

While kinematic complexity clearly decreases with increasing radius, it shows no systematic variation with azimuthal angle. We find no statistically significant difference between the number of components along the minor and major projected axes within the cones defined in Figs.~\ref{f-vel-dist} and \ref{f-vel-dist-inner}. The bottom right quadrant does show a hint of a larger number of components with $\vlsr \lesssim -300$ \km\ (blue components); however, our statistical analysis of M31's kinematics reveals no major azimuthal dependencies, regardless of whether we include thick disk components or restrict our analysis to CGM components only.

These kinematic properties demonstrate that the gas structure varies systematically with distance from the host galaxy, with the thick disk having a significant impact on the observed complexity in the inner regions. Together, the spatial distribution, component structure, and velocity patterns reveal a transition from the thick disk to the CGM---from a complex, multiphase medium strongly influenced by galactic processes in the inner regions to a simpler, more diffuse structure at larger distances where the influence of the host galaxy  may diminish. The absence of strong azimuthal patterns in the velocity structure suggests that both the CGM and thick disk kinematics may be dominated by radial effects rather than orientation-dependent processes, at least at the current sensitivity and spatial sampling of our observations.

\section{Metal and Baryon CGM Mass of M31}\label{s-mass-main}

\subsection{Silicon Column Density versus $R$}\label{s-si-column-vs-R}

To estimate the total metal mass in the cool CGM of M31, we need to first characterize the column density profile $N_{\rm Si}$ as function of $R$ over the full volume since the cool CGM metal mass is $M^{\rm cool}_{\rm Z} \propto  \int R\, N_{\rm Si}(R) \, dR$. This requires a robust characterization of how silicon column density varies with projected distance from M31. Using \siii, \siiii, and \siiv, we can estimate the total column density of Si within the ionization energy range 8--45 eV without any ionization modeling. This ionizing range directly probes the bulk of the cool photoionized CGM of M31, but does not provide any information on the warmer gas with $T\ga 10^5$ K.  

For each target, we identified absorption components of the silicon ions associated with the CGM of M31 (see \S\ref{s-data}). We exclude the identified M31 thick disk components (see \S~\ref{s-m31-dh}) from our primary analysis and separately calculate their contribution to determine their impact on the total CGM metal mass. To determine the total silicon column density along each sightline, we therefore sum $N_{\rm Si} = N_{\rm Si\,II} +  N_{\rm Si\,III} + N_{\rm Si\,IV}$, where for each ion $N=\sum_i N_i$ with $N_i$ being the column density of each component. For each sightline, we  identified all available Si ion measurements associated with the CGM of M31 and categorize them according to detection status: positive detections, lower limits (from saturation), or upper limits (from non-detections). 

The summation is straightforward when there are no censored data (i.e., no saturated components, corresponding to lower limits, and no non-detections of one or more ions, corresponding to upper limits). In that case, we used direct error propagation to derive $N_{\rm Si}(R)$. In \citetalias{lehner20},  a conservative approach was adopted to estimate the cases with only upper limits or with upper limits and detections with a simple of sum of these. However, with some information about the ionic ratios of $N_{\rm Si\,II}/N_{\rm Si\,III}$ and $N_{\rm Si\,IV}/N_{\rm Si\,III}$, we can use a statistical Monte Carlo (MC) approach to get a distribution of possible total column densities using information from detected and non-saturated ions.

In \citetalias{lehner20}, using a survival analysis (the Kaplan–Meier, KM, estimator) to account for censored measurements (upper limits on these ratios), they estimated: $\langle \log N_{\rm Si\,II}/N_{\rm Si\,III}\rangle = (-0.50 \pm 0.04) \pm 0.23$ (mean, error on the mean from the KM estimator, and standard deviation) and $\langle \log N_{\rm Si\,IV}/N_{\rm Si\,III}\rangle = (-0.49 \pm 0.07) \pm 0.20$. With enhanced data coverage in the inner regions, we can now statistically assess radial trends for these ratios by employing a log-rank test between different radial bins (here, inner: $R < 50$ kpc; intermediate: $50 \le R < 150$ kpc; outer: $R > 150$ kpc). The log-rank test is a non-parametric test specifically designed to compare survival distributions while accounting for censoring. For the \siii/\siiii\ ratio, we find a marginally significant difference ($p = 0.066$) between the inner and intermediate regions, and a highly significant difference ($p = 0.002$) between the inner and outer regions, while no significant difference exists between the intermediate and outer regions. For the \siiv/\siiii\ ratio, none of these comparisons show statistically significant differences ($p \gg 0.05$), suggesting no clear radial trend for this ratio. Following this result and using a modified KM estimator from the Python {\tt lifelines} package \citep{davidson-pilon20}, we estimate these ionic ratios for data separated at the boundary of 50 kpc: 
\begin{itemize}[wide, labelwidth=!, labelindent=0pt]
    \item $R < 50$ kpc: $\langle \log N_{\rm Si\,II}/N_{\rm Si\,III}\rangle = (-0.18 \pm 0.05) \pm 0.16$ and $\langle \log N_{\rm Si\,IV}/N_{\rm Si\,III}\rangle = (-0.52 \pm 0.06) \pm 0.12$;
    \item $R \ge 50$ kpc: $\langle \log N_{\rm Si\,II}/N_{\rm Si\,III}\rangle = (-0.56 \pm 0.06) \pm 0.12$ and $\langle \log N_{\rm Si\,IV}/N_{\rm Si\,III}\rangle = (-0.60 \pm 0.08) \pm 0.24$.
\end{itemize}
These results confirm the log-rank test findings, with a significantly higher $N_{\rm Si\,II}/N_{\rm Si\,III}$ ratio (by a factor of $\sim$2.4) at $R < 50$ kpc compared to $R \ge 50$ kpc. We also examined the \siii/\siiv\ ionic ratio, which shows an even more dramatic radial contrast, finding at $R<50$ kpc and $R\ge 50$ kpc, $\langle \log N_{\rm Si\,II}/N_{\rm Si\,IV}\rangle = (+0.10 \pm 0.09) \pm 0.35$ and $ (-0.49 \pm 0.13) \pm 0.33$, respectively. While \siiii\ is the dominant silicon ion at all radii, the contribution from  \siii\ becomes notably more important in the inner regions, consistent with the inference in \citetalias{lehner20} but now demonstrated with a larger sample at $R \la 75$ kpc. These empirically derived ratios are used throughout our analysis where appropriate.

For all scenarios with censored data, we implemented an MC approach with 10,000 realizations to propagate uncertainties and handle limits. We used a statistical approach based on truncated normal distributions to sample from censored data, which ensures proper statistical treatment of both upper and lower limits (see \S\ref{s-mass-model}). 

When only \siiii\ was detected and not saturated with \siii\ and \siiv\ as upper limits, we sampled the \siiii\ column density within its measurement errors. For the undetected ions, we used the radius-dependent empirical ionic ratios relative to \siiii. These ratios were sampled from normal distributions with the means and standard deviations provided above, ensuring that any sampled column densities for undetected ions did not exceed their respective upper limits. We computed the total silicon column density
as the sum of these three ion contributions. From our MC samples, we determined the median, 16th, and 84th percentiles to represent the total Si column density and its uncertainties. 

In cases where \siiii\ was detected along with either \siii\ or \siiv, we used the direct measurements for the detected ions while applying the same MC approach for any undetected ion. This involved sampling from the radius-dependent ionic ratios while handling appropriately any upper limits, providing a more constrained estimate of the total Si column density.

For sightlines where \siiii\ was saturated but with detections of other Si ions, we used these other detections to estimate the \siiii\ column density. For each MC realization, we sampled the detected ions within their measurement errors and then used the inverse of our empirical ionic ratios to estimate the likely \siiii\ column density, respecting the \siiii\ lower limit from the AOD measurement. We calculated the total Si column density summing all three ions, and we treated it as a lower limit since \siiii\ was saturated.

When \siiii\ was saturated with no other detections, we simply used the \siiii\ lower limit as our estimate for the total Si column density. This estimate was marked as a lower limit in our results.

For sightlines with only upper limits for all Si ions, we used statistical sampling from the truncated portions of normal distributions. Starting with the \siiii\ upper limit, we sampled potential column densities below this limit. Then, using our empirical ionic ratios, we estimated contributions from \siii\ and \siiv, always ensuring these stayed below their respective upper limits. The resulting total Si column density was treated as an upper limit.

\begin{figure*}[tbp]
\epsscale{1}
\plotone{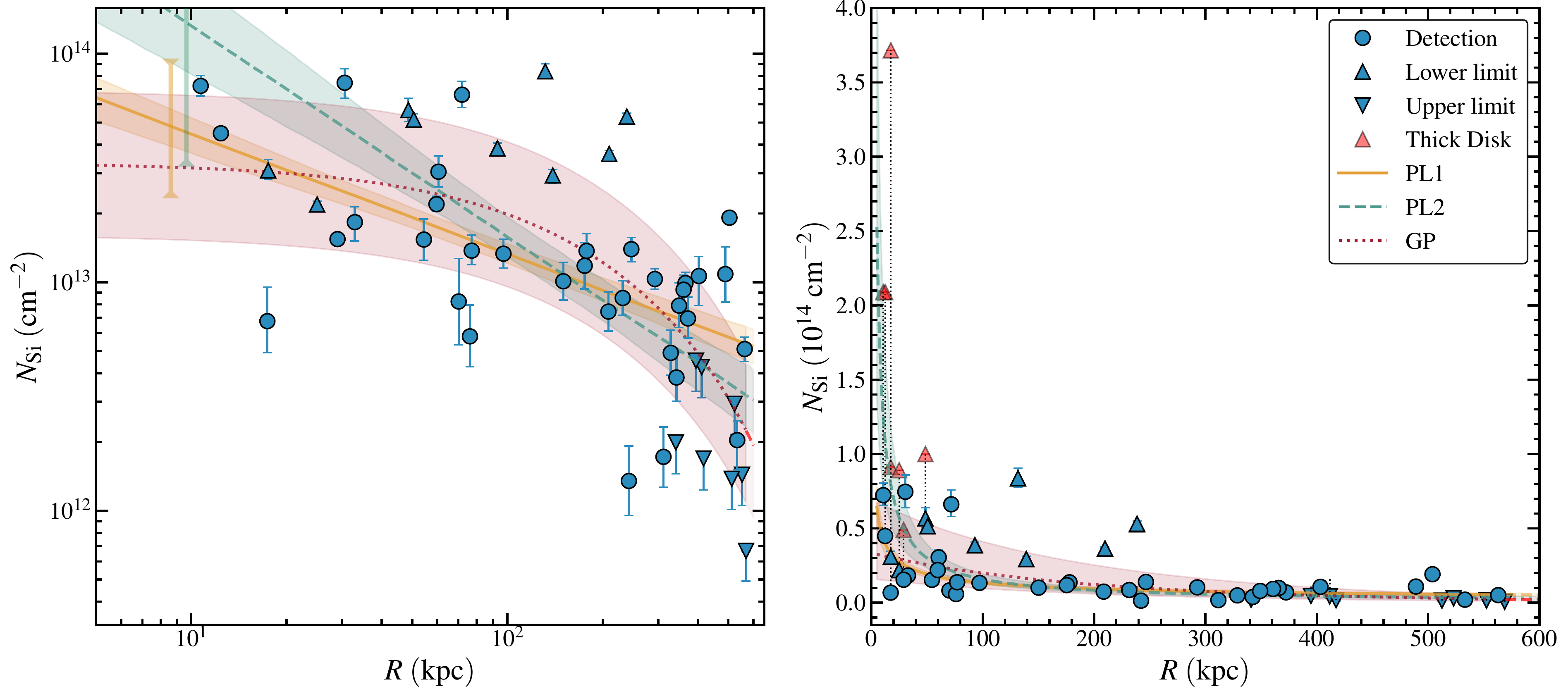}
\caption{Total column densities of Si (i.e., $N_{\rm Si} = N_{\rm Si\,II} + N_{\rm Si\,III} + N_{\rm Si\,IV}$) as a function of projected distance from M31. The {\it left}\ panel shows the data in logarithmic scale, while the {\it right}\ panel presents the same data in linear scale. The yellow-solid (PL1), green-dashed (PL2), and red-dotted (GP) curves represent the penalty-based power law, survival analysis power law, and Gaussian Process models, respectively. Shaded areas show 68\% confidence intervals: the GP band includes intrinsic scatter, while the PL1 and PL2 bands show only parameter uncertainty for clarity. The vertical yellow and green bars illustrate the  68\% confidence intervals that include the intrinsic scatter for PL1 and PL2, respectively. Blue circles, upward triangles, and downward triangles indicate detections, lower limits, and upper limits, respectively. Red triangles show measurements that include thick-disk components ({\it right} panel), with vertical dotted lines connecting these points to their CGM-only counterparts, illustrating the substantial contribution of the thick disk to the total Si column density in the inner region.
\label{f-coltotsi-vs-rho}}
\end{figure*}

We find that the total silicon column density exhibits a clear decline with increasing projected distance $R$ from M31 with considerable scatter. For each sightline, we list this total silicon column density as a function of $R$ in Table~\ref{t-tot-si} and display $N_{\rm Si}$ vs. $R$ in Fig.~\ref{f-coltotsi-vs-rho}. In the next section, we explore three models to describe this column density trend.

\begin{figure*}[tbp]
\epsscale{1}
\plottwo{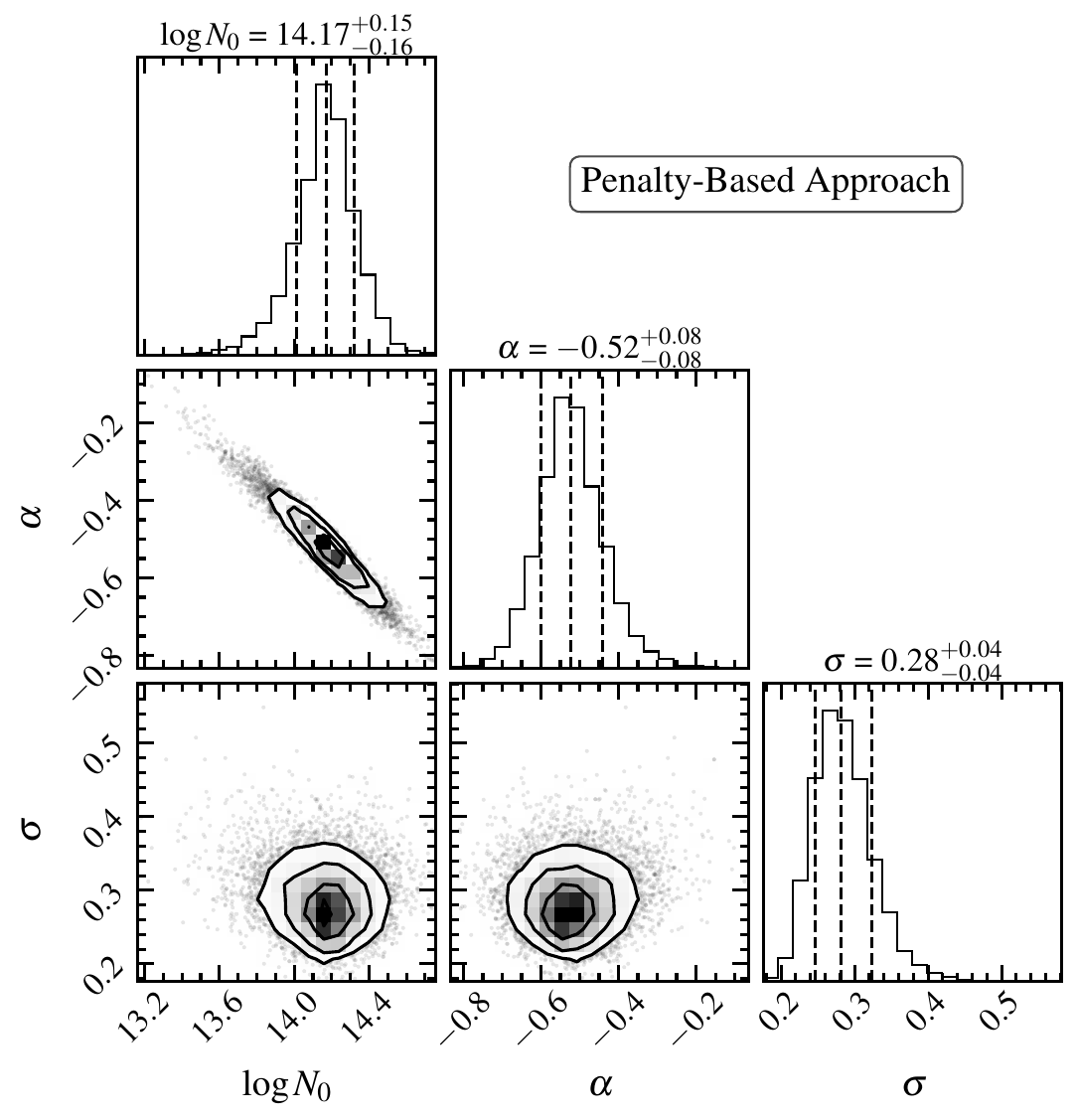}{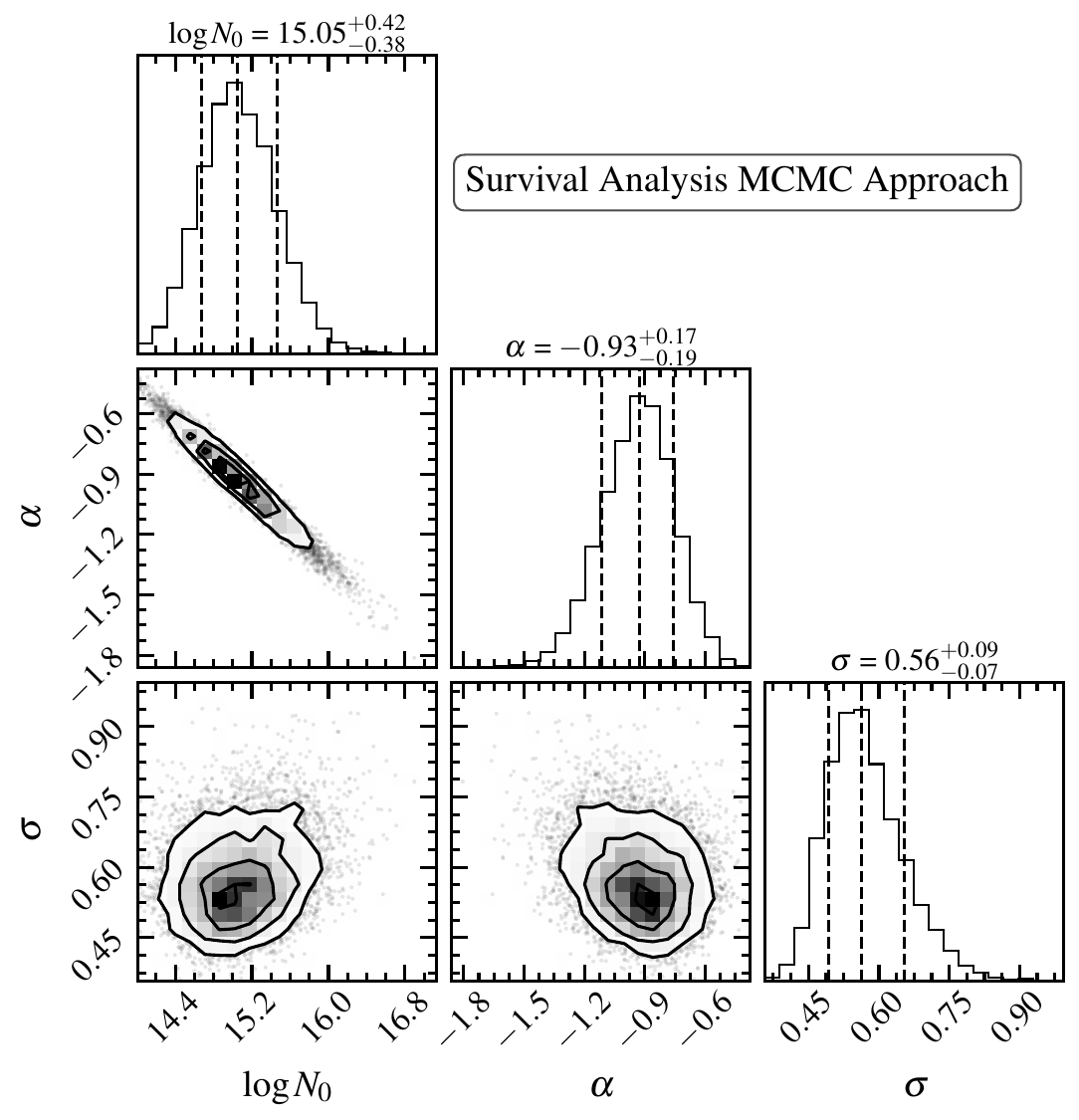}
\caption{Corner plots showing posterior distributions and parameter correlations for our radial power law profile fits to the total Si column densities. \textit{Left:} results from the penalty-based likelihood approach (PL1 model); \textit{right:} results from the survival analysis MCMC approach (PL2 model).
\label{f-corner-powerlaw}}
\end{figure*}

\subsection{Modeling of the  $N_{\rm Si}(R)$ Profile}\label{s-mass-model}

Our dataset presents two significant complexities: the presence of numerous upper and lower limits (approximately one-third of all measurements), requiring robust treatment of censored data, and a substantial scatter in $N_{\rm Si}$ that complicates the functional modeling with respect to $R$. To address these, we implement three complementary approaches: (1) a penalty-based power law fit (PL1), (2) a MCMC power law fit with statistical treatment of censored data (PL2), and (3) a non-parametric Gaussian Process (GP) model that makes no a priori assumptions about functional form and treats the censored data as in the PL1 model;. Each method offers distinct advantages for our dataset---the parametric power law models provide physically intuitive characterizations, while the flexible GP model can capture potential features that simple power laws might miss and allow for an intrinsic scatter. This methodological diversity allows us to assess systematic uncertainties arising from modeling choices, while the different statistical treatments of censored data help quantify the impact of upper and lower limits on our results. By comparing these approaches, we can derive more robust estimates of the total metal mass in M31's cool CGM and quantify the systematical uncertainties in these measurements.

\subsubsection{Power Law Models}\label{s-powelaw}
Power law profiles are commonly used to characterize column density distributions in the CGM of galaxies (e.g., \citealt{chen10a,werk14}) and provide a physically intuitive parameterization model. We therefore model the silicon column density as a function of projected distance using the functional form:
\begin{equation}\label{e-powerlaw}
\log N({\rm Si}) = \log N_0 + \alpha \log(R),
\end{equation}
where $N_0$ represents the normalization (column density at unit radius) and $\alpha$ the power law slope. We include an additional parameter $\sigma_{\rm int}$ to account for intrinsic scatter beyond measurement uncertainties.

In the first variation (PL1), we use a penalty-based likelihood approach as detailed in Appendix~\ref{a-likelihood}. Using the MCMC package \texttt{emcee}, we sample the posterior distributions and find best-fit values of $\log N_0 = 14.17^{+0.15}_{-0.16}$, $\alpha = -0.52 \pm 0.08$, and $\sigma_{\rm int} = 0.28 \pm 0.04$ dex. In the second variation (PL2), we employ the survival analysis MCMC approach described in Appendix~\ref{a-likelihood}, which ensures proper statistical weighting of censored data without relying on ad-hoc penalties. The resulting best-fit parameters are $\log N_0 = 15.05^{+0.42}_{-0.38}$, $\alpha = -0.93^{+0.17}_{-0.19}$, and $\sigma_{\rm int} = 0.56^{+0.09}_{-0.07}$ dex, indicating a steeper decline with radius, larger intrinsic scatter, and much larger $N_0$ than the penalty-based approach.

\subsubsection{Gaussian Process Model}\label{s-gp}
To complement our parametric power law models, we implement a non-parametric GP approach that uses a custom likelihood function to properly handle censored data that follows that of the PL1 model (see Appendix~\ref{a-likelihood}). Unlike parametric models that impose a specific functional form, the GP approach offers flexibility to capture variations in the column density profile without assuming a particular structure. This helps us assess whether the CGM profile exhibits features that might not be well-described by simple power laws. Our GP model uses a radial basis function (RBF) kernel to characterize spatial correlations in the silicon column density:
\begin{equation}
k(r, r') = \exp\left(-\frac{(r - r')^2}{2l^2}\right)\,,
\end{equation}
where $r$ and $r'$ are projected distances from M31 for any two data points, and $l$ is the ``characteristic" length scale.   The GP model's RBF kernel uses this characteristic length scale parameter that is optimized during model fitting, starting from an initial value of 200 kpc and allowing variation within bounds of 50--400 kpc. This optimized length scale determines the spatial correlation scale in the GP model and represents the radial distance over which column densities are significantly correlated. We tested the sensitivity of our GP model to both the initial length scale and its allowed range during optimization. Varying the initial length scale from 50 to 200 kpc changed the mass estimate by approximately 1.6\%, while reducing the allowed optimization range had a smaller effect of less than 1\%. In all cases, the changes remained well within the statistical uncertainties. 

Our GP implementation follows a two-stage approach. First, we fit a standard GP all the data assuming they all detections to establish the initial hyperparameters. Second, we optimize the custom log-marginal likelihood function that properly incorporates all data types. We use the L-BFGS-B optimization\footnote{The L-BFGS-B algorithm is a bounded optimization method that efficiently finds parameter values that maximize the likelihood of the observed data, with bounds ensuring physically reasonable hyperparameter values.} algorithm with multiple restarts (25 attempts) to avoid local optima and find the hyperparameters that maximize this composite likelihood.

Unlike our power law approaches where we sample intrinsic scatter as a free parameter during MCMC, the GP model treats intrinsic scatter ($\sigma_{\rm int}$) as a hyperparameter optimized during model fitting alongside the kernel parameters. The optimization yields $\sigma_{\rm int} = 0.32$ dex, representing the magnitude of stochastic variations in the CGM that cannot be explained by measurement uncertainties or the spatial correlations captured by the RBF kernel. This intrinsic scatter is incorporated into all GP predictions by adding $\sigma_{\rm int}^2$ to the diagonal of the covariance matrix, ensuring that the GP uncertainty bands reflect both parameter uncertainty and intrinsic scatter.

The resulting GP profile can be approximated analytically as:
\begin{equation}
\log N({\rm Si}) = 13.50 - \log\left[1 + \left(\frac{R}{139\,{\rm kpc}}\right)^{0.34}\right]\, \left(\frac{R}{139\,{\rm kpc}}\right)^{0.69}.
\end{equation}
We chose the broken power law functional form for the analytical approximation because it can capture the non-monotonic radial behavior revealed by the GP model—specifically the inner flattening followed by steepening at larger radii. The 139 kpc characteristic radius in this formula represents a transition scale in the fitted approximation and is distinct from the GP's intrinsic correlation length scale. This fitted parameter emerges from the overall shape of the GP-predicted profile rather than being directly related to the RBF kernel's length scale parameter. We emphasize, however, our mass calculations use the full non-parametric GP predictions rather than this analytical approximation.

Comparing the three model profiles in Fig.~\ref{f-coltotsi-vs-rho}, we observe notable differences in their functional forms. The most striking feature is the behavior at small radii ($R < 30$ kpc), where the GP model exhibits a significant flattening compared to both power law models: the GP curve maintains a more plateau-like profile in the inner CGM before transitioning to a steeper decline at intermediate radii ($>100$ kpc). In contrast, by design, both power law models monotonically decline from the smallest radii, with PL2 (dashed green line) predicting substantially higher column densities than PL1 (solid orange line) in this inner region due to its steeper slope ($\alpha = -0.91 \pm 0.18$ compared to $\alpha = -0.52 \pm 0.08$).  At large radii ($R \ga 350$ kpc), the GP model shows evidence of a steeper decline compared to the power laws, which maintain their constant slopes throughout the entire radial range. 

In \S\ref{s-mass-thickdisk}, we explore the impact of thick disk components on the mass estimates. This comparison also serves as an important diagnostic for potential outlier influence on our GP model---a concern for flexible non-parametric approaches. The GP profile consistently exhibits an inner flattening at $R \la 50$ kpc regardless of whether thick disk components are included, showing only the expected overall normalization shift.  However, at $R\ga 50$ kpc, the model  that includes the thick disk components has a steeper negative slope, following more the PL2 model. 

The distinct flattening of the GP profile at small radii is significant as it suggests a possible transition region between the thick disk and CGM, where the CGM column density gradient becomes shallower when the thick disk component takes over. This behavior cannot be captured by simple power law models and may reflect the physical regime change discussed in the previous section. Despite these structural differences, all three models yield integrated masses within a factor 1.2--1.8 of each other (see Section~\ref{s-mass-results}), indicating that the total metal content of M31's CGM is robustly constrained even though local features in the distribution may vary between models. 

\subsection{The Cool CGM Mass of M31}\label{s-mass-estimate}

\subsubsection{Mass Integration}\label{s-mass-integration}

Using our three complementary approaches to modeling the silicon column density profile, we can estimate the total silicon mass within M31's CGM. For each method, we integrate the model profile over a fiducial radius of 390 kpc or about $1.3 \rvir$. This fiducial radius is selected because the covering factor of \siiii\ is nearly 100\% within $1.3 \rvir$ (\citetalias{lehner20} and see Fig.~\ref{f-colmap}). For a given column density profile $N_{\rm Si}(R)$ as a function of projected radius, the total metal mass of the cool CGM can be calculated via: 

\begin{equation}
M^{\rm cool}_{\rm Z} = 2\pi \mu_{\rm Si} m_{\rm Si} \int_{R_{\rm min}}^{R_{\rm max}} N_{\rm Si}(R)\, R  \, dR,
\end{equation}
where $\mu_{\rm Si} = 1/0.064$ represents the inverse solar mass fraction of silicon \citep[e.g.,][]{asplund09}, $m_{\rm Si} = 28 \times 1.67 \times 10^{-24}$ g is the mass of a silicon atom, and we integrate from $R_{\rm min} = 5$ kpc to $R_{\rm max} = 390$ kpc. While $R_{\rm max}$ is our fiducial radius for the CGM of M31, we also provide estimates over different maximum radii (including $R_{200}$, \rvir, and $1/2\rvir$) to 390 kpc. (Although \siiii\ is detected beyond 390 kpc, the detections are mostly localized in one quadrant of the CGM, i.e., we cannot assume a spherical distribution beyond this radius; we therefore do not estimate beyond 390 kpc for the Si ions.)

\subsubsection{Statistical and Systematic Uncertainties}\label{s-mass-uncertainties}
Our multi-method/modeling approach allows us to quantify both statistical (random) and systematic uncertainties in our mass estimate. For each method, we derive statistical uncertainties through different approaches that properly account for the large intrinsic scatter ($\sigma_{\rm int} \approx 0.3$--0.6 dex) revealed by our analyses. For the power law models (PL1 and PL2), we use scatter-corrected mass calculations that account for the log-normal distribution of column densities around the mean profile. We use the MCMC posterior samples directly (10,000 samples after burn-in and thinning) to propagate parameter uncertainties and their covariances through the mass integration. 

Crucially, at each radius we sample multiple realizations from the log-normal scatter distribution and average the resulting linear column densities before integrating. This scatter correction is essential because averaging $10^{\log N}$ values with large scatter around the mean profile gives different results than simply using $10^{\langle\log N\rangle}$, leading to higher mass estimates when properly accounted for. For the GP model, we use posterior sampling that preserves spatial correlations inherent in the GP—unlike independent sampling methods. These correlated uncertainties partially cancel during integration, explaining why GP mass uncertainties are smaller than those from parametric methods. We use 10,000 posterior samples for all methods.

We identify three main sources of systematic uncertainty. First, variation between our three modeling approaches represents a systematic uncertainty of $\sim$32.2\%, which dominates the error budget. This quantifies how our mass estimate depends on the mathematical form (power law vs. Gaussian process) and statistical treatment of the errors and limits (penalty-based vs. survival analysis). Second, we assess dependence on the maximum radius used for GP profile fitting of $N_{\rm Si}(R)$ by testing different radii from \rvir\ to 569 kpc (corresponding to our survey limit). This contributes $\sim$2.1\% systematic uncertainty. Third, we tested different approaches to handling censored data within our GP model, varying constraint strengths by ±50\%, yielding $\sim$2.3\% systematic uncertainty. Adding these components in quadrature yields a total systematic uncertainty of $\sim$32.4\%.

\subsubsection{Metal and Baryon CGM Mass Estimates for the Cool Gas}\label{s-mass-results}

From each of our methods, we derive the following metal masses integrating the column density profiles from 5 to 390 kpc:
\begin{itemize}[wide, labelwidth=!, labelindent=0pt]
    \item Penalty-based power law (PL1): $M^{\rm cool}_{\rm Z} = (1.83^{+0.27}_{-0.21}) \times 10^7 \;{\rm M}_\odot$;
    \item Survival analysis MCMC (PL2): $M^{\rm cool}_{\rm Z} = (3.27^{+1.36}_{-0.79}) \times 10^7 \;{\rm M}_\odot$;
    \item Gaussian Process model (GP): $M^{\rm cool}_{\rm Z} = (2.25^{+0.09}_{-0.08}) \times 10^7 \;{\rm M}_\odot$.
\end{itemize}
The larger errors on the PL2 model is partly due to the larger scatter term, $\sigma_{\rm int}$ (0.6 dex compared to 0.3 dex compared to the other two models).  Rather than simply averaging the three estimates, we sample from their respective probability distributions 100,000 times using a MC approach that accounts for the different error ranges in the positive and negative direction. Our combined estimate from these three methods from 5 to 390 kpc is:
\begin{equation}\label{e-mass-estimate}
M^{\rm cool}_{\rm Z} = (2.56 \pm 0.43_{\rm stat} \pm 0.83_{\rm sys}) \times 10^7 \;{\rm M}_\odot.
\end{equation}
In Table~\ref{t-mass}, we summarize the average estimated metal mass over different radii from these three models of $N_{\rm Si}(R)$ (in particular for $R_{200}$ and \rvir). Our revised estimates are consistent with the GP analysis in \citetalias{lehner20}, but with now better constraints in the inner regions and more robust confidence intervals. 

\begin{figure}[tbp]
\epsscale{1.2}
\plotone{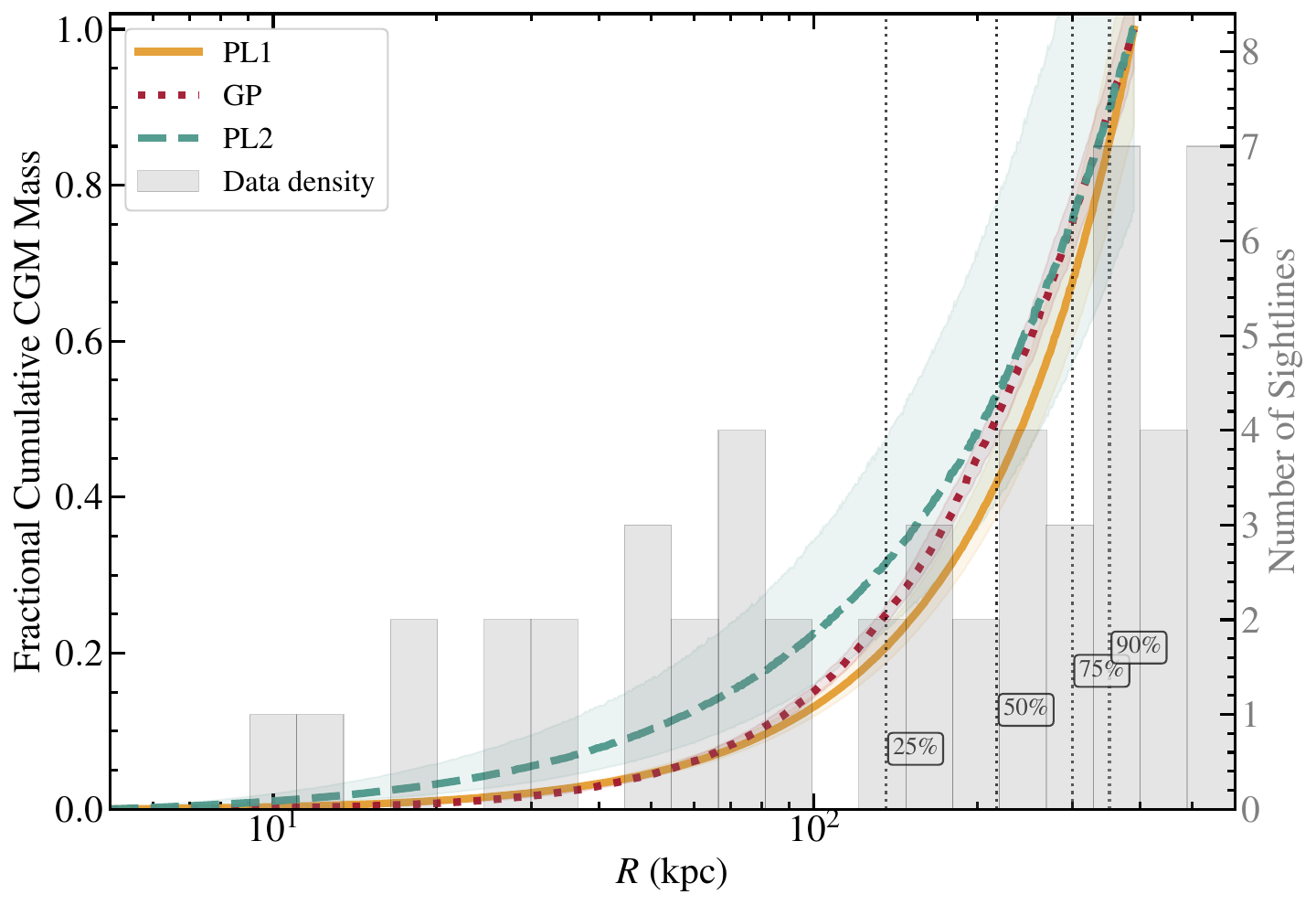}
\caption{Fractional cumulative CGM metal mass as a function of impact parameter for the three modeling approaches (using the fiducial integration from 5 to 390 kpc). Solid, dashed, and dotted lines show results for the PL1, PL2, and GP approaches, respectively. Shaded regions indicate 1$\sigma$ confidence intervals. Vertical dashed lines mark radii containing 25\%, 50\%, 75\%, and 90\% of the total CGM mass (based on the GP model). Gray histogram shows the distribution of observational sightlines, illustrating data coverage as a function of impact parameter.
\label{f-cdf-mass}}
\end{figure}

In Fig.~\ref{f-cdf-mass}, we show the fractional cumulative CGM metal mass within $0.016 \le  R/\rvir \le 1.3$ (5 to 390 kpc) as a function of the impact parameter for the three modeling approaches. Despite the factor of $\sim$1.8 difference in total mass between the PL1 and PL2 models, the fractional cumulative profiles show remarkably similar shapes, with systematic but modest offsets in the characteristic radii. All methods indicate that 25\% of the CGM metal mass is contained within $\sim$110--155 kpc, 50\% within $\sim$205--245 kpc, 75\% within $\sim$300--325 kpc, and 90\% within $\sim$350--365 kpc. The PL2 method consistently predicts the most centrally concentrated profile, while PL1 shows the most extended distribution, with GP results following PL1-like behavior at small radii  ($R \la 100$ kpc) and transitioning to PL2-like behavior at large radii  ($R \ga 250$ kpc). We emphasize that these results reflect integration to 390 kpc ($1.3\rvir$); the relative behavior between methods can vary somewhat with different integration limits.

Following \citetalias{lehner20}, the total mass of the cool gas can be obtained by converting the total observed column density of Si  to total hydrogen column density via $N_{\rm H}  = N_{\rm H\,I} + N_{\rm H\,II} =   N_{\rm Si}\, ({\rm Si}/{\rm H})_\sun^{-1}\, ({\rm Z/Z}_\sun)^{-1}$, and therefore,
$$
M^{\rm cool}_{\rm gas} =  2\pi\, m_{\rm H}\, \mu\,  \Big(\frac{\rm Si}{\rm H}\Big)_\sun^{-1}  \Big(\frac{Z}{Z_\sun}\Big)^{-1}\, \int R\, N_{\rm Si}(R)\, dR\,,
$$
where  $\mu \simeq 1.4$ (to correct for the presence of He), $m_{\rm H} = 1.67\times 10^{-24}$ g is the hydrogen mass, and $\log ({\rm Si/H})_\sun = -4.49$ is the solar abundance of Si \citep{asplund09}. Inserting the values for each parameter, $M^{\rm cool}_{\rm gas}$ can be simply written in terms of $M^{\rm cool}_{\rm Z}$: $M^{\rm cool}_{\rm gas} \simeq 10^2 (Z/Z_\sun)^{-1} M^{\rm cool}_{\rm Z}$. Using the best value above, we derive $M^{\rm cool}_{\rm gas} \simeq 1.5 \times 10^9 (Z/Z_\sun)^{-1}\;{\rm M}_\sun$ within $1.3 \rvir$. The values listed in Table~\ref{t-mass} can be multiplied by a factor 100 to obtain the baryon mass of the cool CGM for M31 to different radii.

\subsubsection{Mass Contribution from Thick Disk Components}\label{s-mass-thickdisk}
In our treatment above, we explicitly removed the thick components. Here we review their possible impact on the CGM mass by including these components in the total estimate of the Si column densities. In Fig.~\ref{f-coltotsi-vs-rho}, we show with red triangles (right panel) the $N_{\rm Si}$ values when including components identified as part of M31's thick disk. While most CGM components have column densities below $10^{14}$ cm$^{-2}$, thick disk components can reach substantially higher values. The vertical dotted lines connect these points to their CGM-only counterparts, illustrating the significant contribution of thick disk gas in the inner region.

To assess the impact of these components on the total mass estimate, we repeated our analysis including thick disk components and find:
\begin{itemize}[wide, labelwidth=!, labelindent=0pt]
    \item PL1: $M^{\rm cool}_{\rm Z} = (1.95^{+0.33}_{-0.25}) \times 10^7 \;{\rm M}_\odot$;
    \item PL2: $M^{\rm cool}_{\rm Z} = (5.42^{+3.58}_{-1.63}) \times 10^7 \;{\rm M}_\odot$;
    \item GP: $M^{\rm cool}_{\rm Z} = (2.38 \pm 0.10) \times 10^7 \;{\rm M}_\odot$.
\end{itemize}
Not surprisingly, the penalty-based power law is the least affected while the survival analysis approach shows the largest difference, reflecting their respective treatments of lower limits. Our combined estimate from these three methods (from 5 to 390 kpc) is $M^{\rm cool}_{\rm Z} = (3.59 \pm 1.10_{\rm stat} \pm 1.59_{\rm sys}) \times 10^7 \;{\rm M}_\odot$. Therefore, considering the three models, the five thick disk components highlighted in Fig.~\ref{f-coltotsi-vs-rho} may have an impact on the mass estimates, increasing the total silicon mass by a factor of $\sim$1.4 compared to the fiducial CGM-only estimate (Eqn.~\ref{e-mass-estimate}). However, it is important to note that within the 68\% confidence interval, the estimated values with or without the thick disk component overlap. If only the PL1 and GP models were considered, the difference would be much smaller, with a factor of 1.05 increase. 

\subsection{Metal and Baryon CGM Mass of the \ovi-Bearing Gas}\label{s-ovi-mass}
In \citetalias{lehner20}, metal and baryon masses were estimated for the \ovi-bearing gas (\ovi\ was observed toward 11 QSOs with the Far Ultraviolet Spectroscopic Explorer). However, upon revisiting those calculations, we identified several issues with the \ovi\ measurements that warranted reanalysis: 1) MW H$_2$ contamination was not accounted for in five of the 11 sightlines;\footnote{RX\_J0048.3+3941, IRAS\_F00040+4325, MRK335, UGC12163, and MRK1502.} 2) for one additional sightline (NGC7469), an H$_2$ contamination correction was applied, but one component was erroneously omitted from the total column density summation. We note that some of these \ovi\ measurements were initially and correctly reported in \citet{lehner15}, where the H$_2$ contamination was properly treated. Additionally, lower and upper limits were not optimally treated with survival analysis techniques in the summation of the components (though a posteriori this is a small effect at the 0.01--0.06 dex level). While these issues have minimal impact on the previously derived masses and overall conclusions as we show below, addressing them provides an opportunity to apply our improved methodological framework to ensure consistency across our analyses. To correct for H$_2$ contamination, we followed the methodology described in \citet{lehner15}, finding consistent results across independent analyses. This correction decreased the affected column densities by 0.14--0.20 dex, with uncertainties of $\pm$0.1--0.2 dex depending on contamination level \citep{wakker03}. These uncertainties dominate the error budget for $N_{\rm O\,VI}$ measurements where H$_2$ contamination is present.

\begin{figure*}[tbp]
\epsscale{1}
\plotone{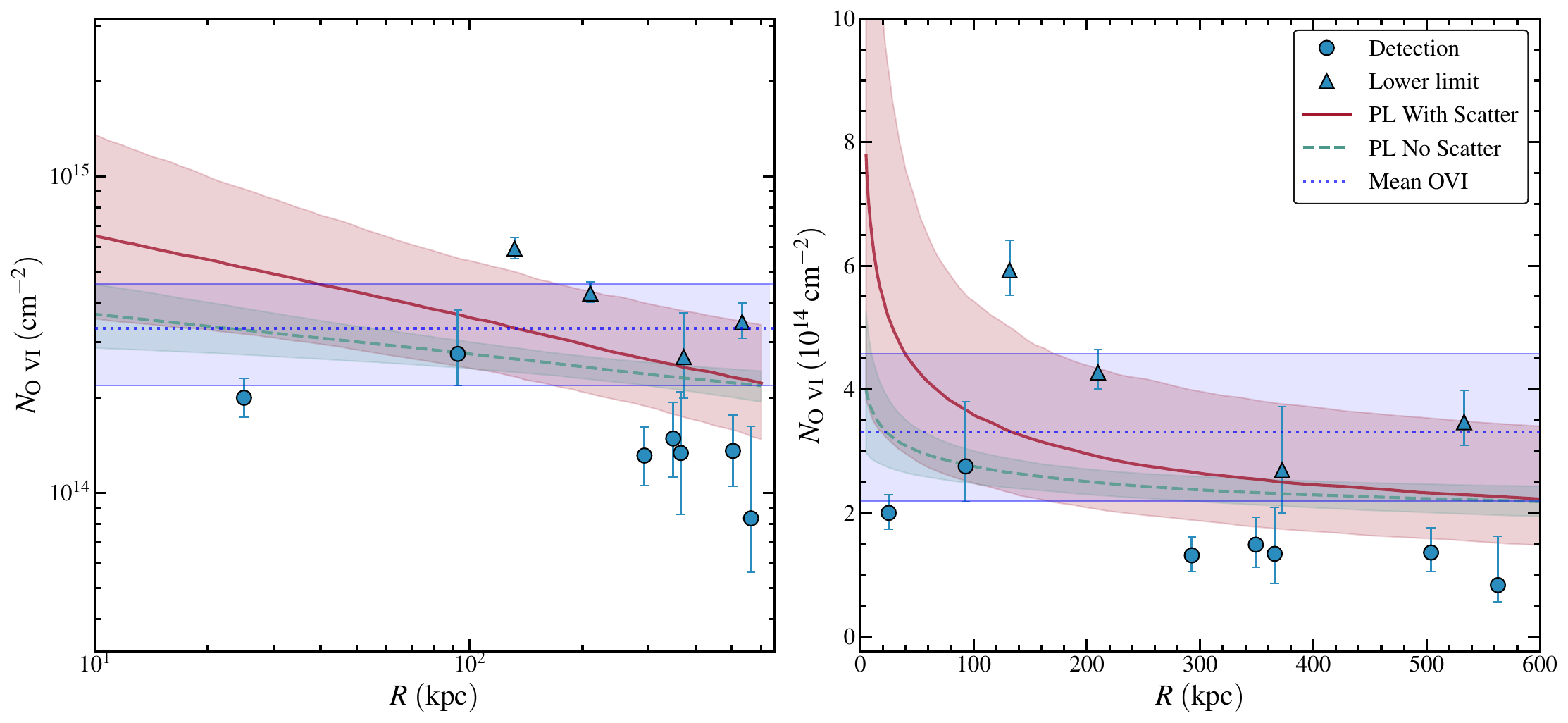}
\caption{Total column densities of \ovi\ as a function of projected distance from M31. The {\it left}\ panel shows the data in logarithmic scale, while the {\it right}\ panel presents the same data in linear scale. The red-solid and green-dash curves represent survival analysis power law without and with a scatter term, respectively, with shaded areas showing the 68\% confidence intervals. The blue dotted line and shaded area show the mean and 68\% confidence interval of the \ovi\ column density (estimated from a survival analysis). 
\label{f-coltotovi-vs-rho}}
\end{figure*}

Using the same approach as described in \S\ref{s-si-column-vs-R}, we then sum the \ovi\ components associated with the CGM of M31. The new estimates are summarized in Table~\ref{t-novi}, with the last column providing the differences compared to the \citetalias{lehner20} results. With changes of values being both lower and higher, the actual mean of the logarithmic \ovi\ column density is $\mlnovi = 14.46 \pm 0.20$ (error on the mean from the survival analysis), the same value as in \citetalias{lehner20}.  Therefore, the results regarding the \ovi\ in \citetalias{lehner20} still hold. If we estimate the mean in linear space using the survival analysis, we find $\mlnovi = 14.52\,^{+0.14}_{-0.21}$, which we adopt for the remainder of our analysis. In Fig.~\ref{f-coltotovi-vs-rho} we show the total column densities of \ovi\ as a function of projected distance from M31. We also plot the mean and error on the mean from the survival analysis.

\begin{figure}[tbp]
\epsscale{1.2}
\plotone{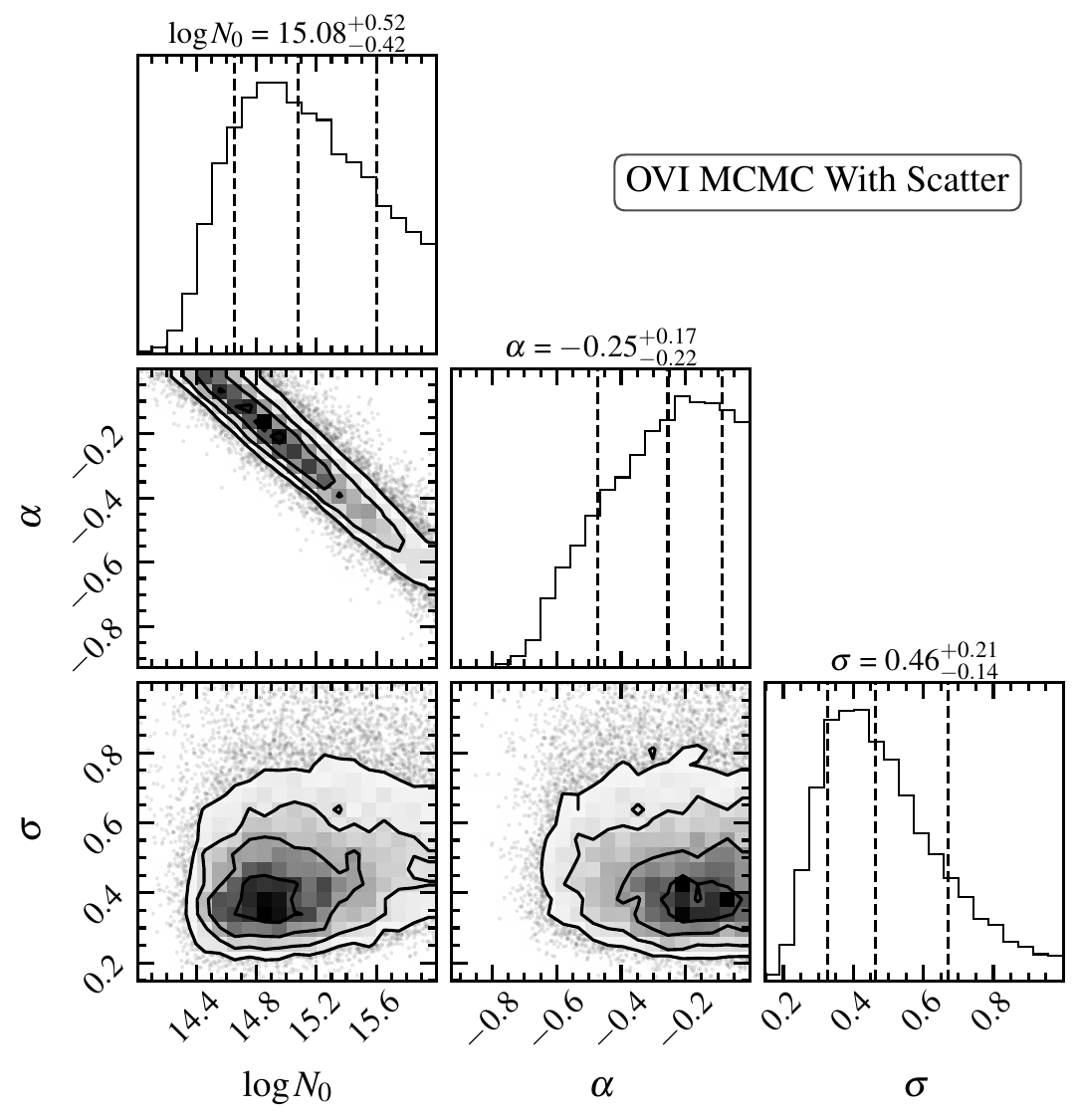}
\caption{Corner plots showing posterior distributions and parameter correlations for the power law profile fits to the \ovi\ column density profile with a scatter term.
\label{f-corner-powerlaw-oviw}}
\end{figure}

\begin{figure}[tbp]
\epsscale{1}
\plotone{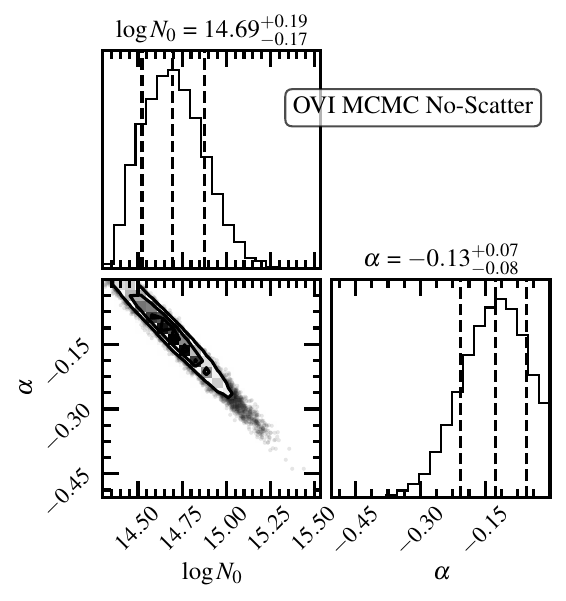}
\caption{Corner plots showing posterior distributions and parameter correlations for the power law profile fits to the \ovi\ column density profile without a scatter term.
\label{f-corner-powerlaw-oviwo}}
\end{figure}

From the estimates of the \ovi\ column densities, the \ovi\ mass of the CGM can be directly calculated via:
\begin{equation}\label{e-movi}
M_{\rm O\,VI} = 2\pi  m_{\rm O} \int_{R_{\rm min}}^{R_{\rm max}} \, N_{\rm O\,VI}(R) \, R \, dR \,, 
\end{equation}
where $m_{\rm O} = 16 \,m_p$ is the mass of an oxygen atom. The total oxygen mass is then $M_{\rm O} =M_{\rm O\,VI}/f_{\rm O\,VI}$, where $f_{\rm O\,VI}$ is the ionization fraction of oxygen in the \ovi\ state (which is always $\la 0.2$ for any ionization model, \citealt{gnat07,lehner11b,oppenheimer13a}), and the metal mass is $M^{\rm warm}_{\rm Z} = \frac{M_{\rm O}}{\mu_{\rm O}}$, where $\mu_{\rm O} = 0.55$ is the oxygen mass fraction in metals. The warm CGM gas mass in terms of the oxygen mass is then:
\begin{equation}
 \frac{M^{\rm warm}_{\rm gas}}{M^{\rm warm}_{\text{Z}}} = \frac{m_{\text{H}}}{m_{\text{O}}} \,\mu \, \left(\frac{\text{O}}{\text{H}}\right)_\odot^{-1} \,\left(\frac{\text{Z}}{\text{Z}_\odot}\right)^{-1} \, \mu_{\rm O} \,.
\end{equation}
Thus, $M^{\rm warm}_{\rm gas} \approx 100 M^{\rm warm}_{\rm Z}\, (Z/Z_\odot)^{-1}$ (where we used  ${\rm (O/H)}_\odot = 4.90 \times10^{-4}$ from \citealt{asplund09}). 

The \ovi\ mass depends solely on the column density of \ovi\ and its integration over $R$, while all the other masses derived from $M_{\rm O\,VI}$ are lower limits because $f_{\rm O\,VI}\la 0.2$ and $Z/Z_\odot$ is likely subsolar in the CGM of M31 (e.g., \citealt{prochaska17,bregman18,berg23}). To model $N_{\rm O\,VI}(R)$, we use three methods, but because of the limited data points (11 total, five lower limits), we use a constant model adopting the survival mean $\mlnovi = 14.52\,^{+0.14}_{-0.21}$ shown in Fig.~\ref{f-coltotovi-vs-rho} and two MCMC power-law models, one with a scatter term (similar to that for Si) and one without. In Figs.~\ref{f-corner-powerlaw-oviw} and \ref{f-corner-powerlaw-oviwo}, we show the resulting corner plots from the MCMC analysis. Both fits show a much shallower slope than for the cooler gas. The Bayesian Information Criterion (BIC) difference between the models with and without scatter is $\Delta {\rm BIC} = 29.8$, strongly indicating that the model with scatter is statistically preferred, despite having one more parameter. Integrating Eqn.~\ref{e-movi} from 5 to 300 kpc (\rvir), we find: 
\begin{itemize}[wide, labelwidth=!, labelindent=0pt]
    \item Power law with scatter: $M_{\rm O\,VI} = (1.10^{+0.57}_{-0.27})\times 10^7 \; {\rm M}_\odot$;
    \item Power law without scatter: $M_{\rm O\,VI} = (0.92^{+0.07}_{-0.08}) \times 10^7 \;{\rm M}_\odot$;
    \item Constant model: $M_{\rm O\,VI} = (1.19 \pm 0.48) \times 10^7 \;{\rm M}_\odot$.
\end{itemize}
We find excellent agreement within 68\% confidence interval between all three modeling approaches for gas within \rvir. However, the impact of model choice becomes increasingly significant when extending to larger radii. At 390 kpc (1.3\rvir) and 569 kpc (1.9\rvir), the power law model with scatter yields $(1.7^{+0.8}_{-0.5}) \times 10^7 \;{\rm M}_{\odot}$ and $(2.2^{+1.0}_{-0.7}) \times 10^7 \;{\rm M}_\odot$ respectively, compared to $(2.0 \pm 0.7) \times 10^7 \;{\rm M}_\odot$ and $(4.3 \pm 1.6) \times 10^7 \;{\rm M}_\odot$ for the constant column model. Based on theoretical expectations for column density profiles \citepalias{lehner20}, the power law model likely provides more reliable mass estimates at large radii.\footnote{We note that the most inner \ovi\ absorber toward RX\_J0048.3+3941 has velocities consistent with a thick disk component. Given the small sample of \ovi\ absorbers, we included it in our calculation. However, if this absorber is removed from the sample, the mass derived from the power-law models would increase by a factor of $\sim$1.2, yielding values similar to those from the constant column density model within \rvir\ due to the steeper slopes that result from excluding this inner data point that anchors the models at small $R$.}

Using the values in Table~\ref{t-mass}, we find $M_{\rm O\,VI} \simeq 0.6\times M^{\rm cool}_Z$ within \rvir. This suggests that a substantial fraction of M31's CGM metals reside in the warmer gas phase traced by \ovi. The metal and baryon masses scale from $M_{\rm O\,VI}$ as $M^{\rm warm}_Z \ga 10\, M_{\rm O\,VI} \,(0.2/f_{\rm O\,VI})$ and $M^{\rm warm}_{\rm gas} \approx 10^3\, M_{\rm O\,VI}\,  (Z/Z_\odot)^{-1} \, (0.2/f_{\rm O\,VI})$, highlighting the potential for the warm phase to dominate the total CGM metal and baryon budgets between the cool ($\sim 10^4$ K) and warm ($\sim 10^5$ K) gas-phases. We, however, emphasize that the sample of \ovi\ absorbers is small. In particular, the targets are not distributed uniformly in the CGM of M31, but mostly in one of the quadrants of the CGM (see Fig.~13 in \citealt{lehner20}). 

We note that theoretical models and simulations predict that M31-like halos should be dominated by an even hotter phase ($T \sim 10^6$ K) that is not traced by the ions we observe \citep{wijers24,sultan25}. Indeed, extrapolating the X-ray results to \rvir, \citet{bregman18} find for the hot X-ray gas for similar stellar masses as M31 in the range $M^{\rm hot}_{\rm gas}\simeq1$--$10\times 10^{11}$ M$_\sun$, a factor of a few larger than the warm gas (however, this depends on the metallicity adopted for the \ovi-bearing gas and $f_{\rm O\,VI}$).

\section{Discussion}\label{s-disc}
\subsection{The Transition between Thick Disk and CGM of M31}\label{s-disc-thick}

Project AMIGA provides the first UV absorption characterization of M31's gaseous thick disk and its transition to the extended CGM. The thick disk components, detected in sightlines at $R \la 30$ kpc, exhibit distinct and unique properties that definitively separate them from the general CGM population. These components show significantly higher column densities in low and intermediate ions (\oi, \cii, \siii, \siiii), with $\log N$ values typically 0.5--1.0 dex higher than CGM components at similar impact parameters. This enhancement is particularly pronounced in the neutral and singly ionized species (\oi, \siii) and the detection of weaker ions such as \sii, \alii, and \feii\ (see Table~\ref{t-results}), consistent with a higher density, lower ionization state medium that is kinematically coupled to the galactic disk. The combination of the strength of the absorption and the fact that these components have velocities aligned with corotation of M31's disk strongly supports their association with the gaseous thick disk rather than the CGM. Our rotation model implies by design that these components trace material within approximately 6 kpc (three velocity scale heights) of the disk plane. 

An additional distinguishing feature of the thick disk components is that they are the only absorption systems in our survey that exhibit strong \oi\ absorption. In three cases, we find further evidence of \hi\ 21-cm emission spatially and kinematically aligned with the \oi\ absorption features, which will be discussed in more detail in a future paper. This \hi-\oi\ correspondence is not observed in any CGM components outside the thick disk region, and in fact \hi\ 21-cm emission is rarely detected beyond 25 kpc (\citealt{howk17}; J.C. Howk et al., 2025, in prep.). The presence of significant \oi\ absorption, which closely traces neutral hydrogen due to charge exchange reactions, provides further confirmation that these components represent a distinct component with more neutral gas compared to the more ionized CGM components.

While the detection of both \hi\ emission and \oi\ absorption is significant, the $9\arcmin$ beam size of the GBT observations (approximately 2 kpc at the distance of M31) prevents us from directly deriving accurate metallicities of the thick disk material. For example, toward IVZw29, we estimate $\log N_{\rm H\,I}=19.5$ in the component at $\vlsr \simeq -400$ \km\ associated with the \oi\ absorption ($\log N_{\rm O\,I}\ge 14.51$), which would produce an unlikely very low metallicity of $\ga -1.7$ dex solar if the \hi\ and \oi\ traced exactly the same gas (it is a lower limit because \oi\ is saturated, but it is unlikely to be saturated by much more than 0.1 dex based on the comparison with the \siii\ transitions). If we assume a metallicity of 0.5 solar (or $-0.3$ dex relative to solar), the expected \hi\ column density associated with the \oi\ absorption would need to be $\log N_{\rm H\,I} \simeq 18.1$, suggesting substantial beam dilution in the GBT measurements, especially for a target so close to M31 ($R\simeq 12.4$ kpc). While material in the thick disk might be expected to have  metallicities closer to solar, the models by \citet{afruni22} indicate that the CGM metallicities could be quite low, of the order of $\log Z/Z_\odot \sim -1.5$. Galactic fountain material originally can also trigger the condensation of the CGM that can be low-metallicity gas even relatively close to the disk \citep{fraternali15}. Constraining the metallicity of the observed gas around M31 is beyond the scope of this paper, but it will be one of our goals  in a future paper. 

The non-parametric GP model of the silicon column density profile (see \S\ref{s-si-column-vs-R}) provides additional evidence for a transition between the CGM and the gaseous thick disk of M31, revealing a flattening of the radial profile at $R \la 50$ kpc that simple power law models cannot capture (see Fig.~\ref{f-coltotovi-vs-rho}). This flattening may reflect the physical regime where the thick disk begins dominating the CGM, creating a distinct spatial domain with different physical and kinematic properties. Such disk-CGM transitions have been observed in other systems. \citet{nielsen24} recently used Keck KCWI observations to investigate this transition in emission around a nearby galaxy and found two distinct surface brightness radial profiles for gas within the disk versus the CGM, providing further evidence that these regions represent separate physical domains with unique properties.

In addition to this corotating thick disk component, we also detect CGM gas at these small impact parameters that is not corotating and has typically weaker absorption (see, e.g., Fig.~\ref{f-IVZw29_components}). These are likely the ionized boundaries around the \hi\ clouds detected with the GBT within 30--50 kpc of M31 \citep{thilker04} and may represent analogs of HVCs observed in the MW. In a future paper, we will provide  a more systematic characterization of this cloud population near M31. 

The key boundary between disk and halo plays a crucial role in gas circulation between the star-forming disk and extended CGM. Theoretical models and simulations of galaxy evolution consistently identify this interface as a key region where feedback processes propel material from the disk into the halo, and where condensing CGM gas can re-accrete onto the disk with relatively low metallicity (e.g., \citealt{fraternali15,marinacci11}). In a future paper with the combination of the observations of the two stars within M31, we will characterize the properties of this transition region, in particular, its relative abundances (including an estimate of the dust depletion) and kinematics. From the present analysis, the multi-phase nature of the thick disk is evident from the simultaneous detection of neutral oxygen (\oi), low ions (\cii, \siii, \sii, \alii, \feii), and intermediate ions (\siiii), and high ions (\siiv, \civ) within the same components (see Figs.\ \ref{f-example-spectrum}, \ref{f-example-fit}). This co-existence of multiple ionization states indicates a complex medium where different gas phases are either spatially mixed or in close proximity, supporting scenarios where cooler gas condenses within a warmer environment through thermal instabilities or through the mixing of outflows with ambient halo gas \citep[e.g.,][]{afruni22,voit18}. Additionally, recent simulations of M31-like halos suggest that large-scale cooling flows in the hot CGM can also produce multi-phase structure, including cooling as hot inflow circularizes into a disk \citep{stern24,sultan25}.

\subsection{Radial and Azimuthal Structure of the CGM and Thick Disk of M31}\label{s-disc-radial}
Our expanded sightline coverage reveals the complex structure of M31's gaseous environment, from its thick disk through its extended CGM, as a function of distance. A radial trend emerges in the kinematic complexity of both the thick disk and CGM, with the number of distinct absorption components decreasing significantly with increasing impact parameter, and, therefore, the total velocity of the absorption profiles is also decreasing significantly with increasing $R$. When considering both thick disk and CGM components together, this trend is statistically significant (see \S\ref{s-kin}). The region where the thick disk dominates ($R \la 25$ kpc) shows the most complex absorption with typically 3--5 components per sightline, the inner CGM ($25 \la R \la 100$ kpc) typically shows 2--3 absorption components, while the outer CGM ($R \ga 150$ kpc) predominantly exhibits single-component absorption. This transition from complex to simpler kinematic structure likely reflects the progression from the dynamically active thick disk, through the inner CGM influenced by galactic processes such as outflows and fountain flows, to the more quiescent outer CGM.

The column density profiles for all ions show systematic decreases with impact parameter, but with varying slopes that depend on ionization potential as already observed in \citetalias{lehner20}. The thick disk components at $R \la 25$ kpc show the highest column densities, particularly in low ions, appearing as significant enhancements above the general CGM trend. Beyond this radius ($R > 25$ kpc), low-ionization species (\oi, \cii, \siii) exhibit steeper radial gradients compared to higher ionization species (\civ, \siiv), indicating a shift in the ionization balance with distance from the host galaxy as first noted by \citetalias{lehner20}. The \ovi-bearing phase, which likely traces a more extended, diffuse component of the CGM, shows the shallowest radial profile, extending significantly beyond the virial radius. A similar trend was also observed recently in the Large Magellanic Cloud \citep{krishnarao22,mishra24}. These differentiated radial profiles are systematically observed in cosmological simulations showing similar trends (see discussion in \citetalias{lehner20} and, e.g., \citealt{peeples19,oppenheimer18a,ji20}). 

While a simple power law provides a reasonable first-order approximation to the overall trend of the Si column density profiles, the data show increased scatter at intermediate radii ($50 \la R \la 150$ kpc) and a possible steepening at large radii ($R \ga 200$--300 kpc). This non-uniform behavior demonstrates that different physical processes dominate at different radial ranges. In the inner regions, the elevated column densities reflect contributions from both thick disk components and nearby CGM gas, potentially including disk outflows and fountain flows. At intermediate radii, the scatter may result from a heterogeneous distribution of CGM clouds. The possible steepening at large radii could indicate a transition to the intergalactic medium or the edge of M31's gravitational influence within the Local Group environment. These distinct radial zones reveal the complex, multi-phase nature of gas surrounding M31 and demonstrate the limitations of simple, monotonic models in describing the full range of CGM properties.

Despite the more detailed spatial sampling enabled by our multiple sightlines, we find no strong azimuthal dependence in either the kinematics or column densities of both thick disk and CGM components. Our statistical analysis in \S\ref{s-col-azimuth} reveals no significant difference in column densities along the minor versus major projected axes, though our sample size at $R \la 55$ kpc (particularly along the major axis) remains very limited. Similarly, we find no statistically significant difference in the number or velocity distribution of absorption components between these axes.  Based on these statistical analyses, we find that both M31's thick disk and CGM are shaped more by radial effects than by specific disk-aligned structures such as bipolar outflows or accretion preferentially along the disk plane. 

The absence of strong azimuthal patterns indicates that in M31, whose star formation activity is currently moderate compared to more actively star-forming galaxies, the structural and kinematic organization of both the thick disk and CGM is dominated by processes other than ongoing strong outflows.  Our observations are consistent with theoretical expectations for quiescent, massive galaxies like M31, where outflows are weak \citep{muratov15,muratov17, faucher-giguere18,stern21}, so the structural and kinematic organization of both the thick disk and CGM is dominated by large-scale cooling flows in the hot gas \citep{stern19,stern24,hafen22, sultan25}.

Another possible mechanism could be accretion of low-metallicity intergalactic medium as modeled by \citet{afruni22}, who used semi-analytic models to investigate the origin of M31's cool CGM and found that most cool gas clouds follow infall trajectories rather than organized disk-aligned flows. The lack of strong azimuthal dependence we observe, combined with the radial dependence, aligns well with this theoretical framework. While this model provides a compelling explanation for the overall CGM structure (including the observed distribution of the kinematics around M31), \citet{afruni22} acknowledged that their accretion-only models underestimated the observed number of components per sightline, especially in the inner CGM at $R \la 150$ kpc. They proposed that additional physical processes, such as condensation of hot coronal gas through thermal instabilities (e.g., the precipitation model by \citealt{voit17,voit18}), may be necessary to account for the observed added kinematic complexity. 

Our detection of thick disk components with high column densities and corotation signatures provides another piece of this puzzle, suggesting that while the extended CGM may indeed be dominated by rotating cooling flow, accreted material, or precipitation from the hot gas, the most inner regions ($R \la 30$ kpc) show  evidence of disk-related processes. These thick disk components potentially represent the interface where outflows from the disk interact with the accreting CGM, adding to the overall kinematic complexity observed in the inner regions.

\subsection{Comparison with the COS-Halos Survey}\label{s-cos-halos}

Traditional CGM surveys typically sample one sightline per galaxy (see also, e.g., \citealt{nielsen13,liang14,borthakur16,bordoloi14,turner15,heckman17,chen20,berg23}), with some nearby cases utilizing up to 3--4 sightlines (e.g., \citealt{bowen16,keeney17,sameer22}). These studies assemble sizable samples of absorbers associated with particular galaxy populations to assess how column densities vary with radius and derive average surface densities and mass budgets. In contrast, Project AMIGA has now compiled more sightlines for a single galaxy, M31, than most entire surveys obtain for dozens of galaxies.

The COS-Halos survey represents one of the most comprehensive studies of CGM properties around $L^*$ galaxies at $z\sim 0.2$ with a clean and straightforward selection process \citep{tumlinson11a,tumlinson13,werk13,werk14,prochaska17}. Targeting 44 galaxies with $0.3 < L/L^* < 2$, this survey has provided crucial insights into CGM properties as a function of galaxy mass, star formation activity, and impact parameter. Most star-forming galaxies in the sample have $11.5 \la \log M_{200} \la 12.5$, comparable to M31's mass (though about 60\% have $\log M_{200} <12.1$), while passive galaxies have $13.0 \la \log M_{200} \la 13.7$. With our expanded dataset, we can now make a detailed comparison between M31's CGM and that of typical ``$L^*$" galaxies at $z\sim 0.2$.

\begin{figure}[tbp]
\epsscale{1.2}
\plotone{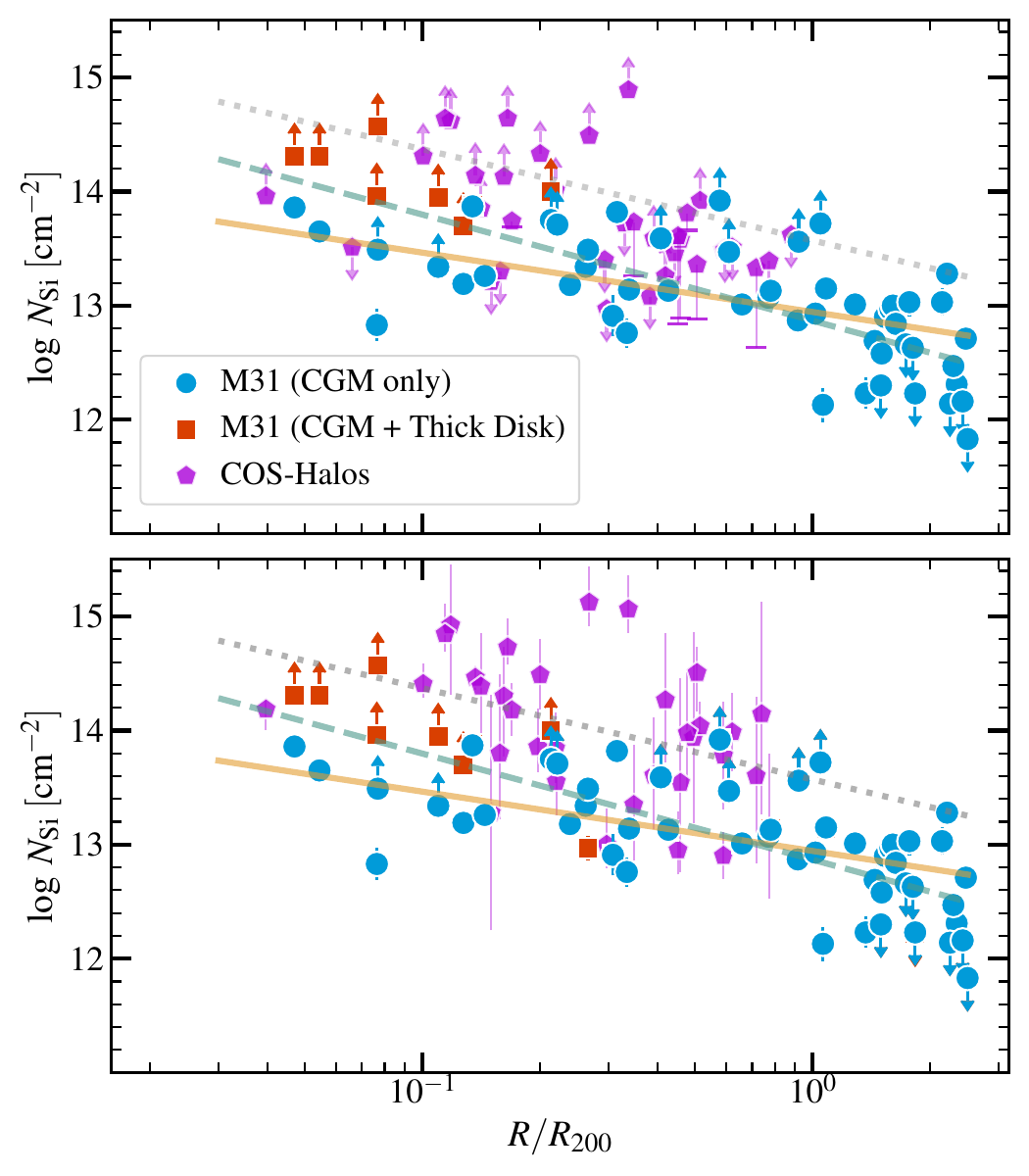}
\caption{Comparison of the total column densities of Si from M31 and the COS-Halos galaxies as a function of $R/R_{200}$.  {\it Top panel}: $N_{\rm Si}$ are directly constrained by the estimates on $N_{\rm Si\,II}$, $N_{\rm Si\,III}$, and $N_{\rm Si\,IV}$ from the observations for both Project AMIGA and COS-Halos (using the results in \citealt{werk13}, see \citetalias{lehner20}). {\it Bottom panel:} same as top panel but $N_{\rm Si}$ for the COS-Halos data is derived from Cloudy photoionization models from \citet{werk14}. The dotted line represents the best fit between $N_{\rm Si}$ and $R/R_{200}$ using the COS-Halos modeled data \citep{werk14,peeples14}). The solid and dashed lines are the PL1 and PL2 models to the M31 data (see \S\ref{s-mass-model}).
\label{f-cos-halos}}
\end{figure}

In Fig.~\ref{f-cos-halos}, we present the total Si column densities as a function of $R/R_{200}$ for both the COS-Halos galaxies and M31 that was first shown in \citetalias{lehner20}, but now including our AMIGA Insider data and identified thick disk components. Following \citetalias{lehner20}, we present two cases for the COS-Halos sample: (1) total Si column densities directly constrained by observations of \siii, \siiii, and \siiv\ (top panel, see \citealt{lehner20}); and (2) Si column densities derived from photoionization models by \citet[][bottom panel]{werk14}. The latter estimates were used to determine metal and baryonic masses in the CGM of COS-Halos galaxies by \citet{peeples14} and \citet{werk14}. The observed sample (case 1) includes 35 sightlines, while the modeled sample (case 2) contains 33, with partial overlap between them.

The most striking difference, previously noted in \citetalias{lehner20} but further reinforced with our expanded dataset, is the prevalence of very high Si column densities in COS-Halos at $R/R_{200} \la 0.4$ that are largely absent in Project AMIGA data in the CGM components. The main reason for this difference is that unlike in M31 the COS-Halos \siii\ transitions are saturated, resulting in higher lower limits than in M31, where saturation occurs primarily in \siiii\ for the CGM components. These high Si column densities in COS-Halos correspond to strong \hi\ absorbers ($\log N_{\rm H\,I} \ga 18$) that are not observed in M31's CGM (see Fig.~5 and Section~4.1 in \citealt{howk17} and J.C. Howk et al., 2025, in prep.); they are, however, observed in the thick disk of M31, but at smaller impact parameters.  At larger impact parameters ($R/R_{200} \ga 0.4$), both surveys show similar scatter in the observationally-constrained Si column densities, where the gas is more highly ionized (top panel). The photoionization-modeled COS-Halos Si column densities (bottom panel) also exhibit systematically higher values, extending this difference out to $R/R_{200} \la 0.6$. 

These differences in column density profiles translate directly into significant differences in the inferred CGM masses. The best-fit model to the COS-Halos observations (gray dotted line in Fig.~\ref{f-cos-halos}) yields column densities approximately 5--10 times higher than our PL1 (solid line) and PL2 (dashed line) models for M31 at any $R\la R_{200}$. This difference is consistent with the low-ionization CGM metal mass of $2.3 \times 10^7$ ${\rm M}_\odot$ within 150 kpc derived for COS-Halos galaxies \citep{peeples14} compared to $0.8\times 10^7$ ${\rm M}_\odot$ for M31 over the same radius (see Table~\ref{t-mass}), a factor $\sim$3 times larger for COS-Halos. Similarly, the most recent cool gas mass estimate for COS-Halos galaxies is $M^{\rm cool}_{\rm CGM} = (9.2 \pm 4.3) \times 10^{10}$ ${\rm M}_\odot$ within $R_{\rm vir}$ \citep{prochaska17}, consistent with the lower limit in \citet{werk14}. Adopting the median CGM metallicity of $0.3\,Z_\odot$ from \citet{prochaska17}, the COS-Halos cool CGM mass exceeds that of M31 by a factor of $\sim$15 within $R_{\rm vir}$.

Several scenarios might explain these differences in CGM properties between M31 and the COS-Halos galaxies. First, the higher column density absorbers in COS-Halos could be fully or partly associated with smaller galaxies closer to the sightline than the targeted COS-Halos galaxy. Such smaller galaxies at smaller impact parameters would naturally explain some of the strong \siii\ absorption observed at relatively large projected distances from the primary COS-Halos targets. \citet{bregman18} noted, in comparison with other low-redshift galaxy/absorber surveys \citep{stocke13,bowen02}, a higher frequency of high \hi\ column density absorbers in the COS-Halos survey. Supporting this interpretation, \citet{tumlinson13} found that six COS-Halos galaxies are in groups. Notably, these group galaxies include two sub-DLA systems and three with strong, multi-component \hi\ absorption—systems that have high (lower limit on) silicon column densities. 

At the same time, M31's CGM may genuinely contain less $\sim10^4\,{\rm K}$ gas at these radii than typical $L^*$ galaxies at $z \sim 0.2$, reflecting intrinsic differences in galaxy evolution, properties, and environment. M31's relatively quiescent recent star formation history \citep{williams17} likely produces less energetic outflows compared to the more actively star-forming COS-Halos galaxies. The Local Group environment, where M31 interacts with the MW and other galaxies, may also create different CGM conditions compared to more isolated systems \citep[e.g.,][]{nuza14}. Interestingly, \citet{bish21} found that \civ\ in the MW's inner CGM is also markedly lower than in other $L^*$ galaxies, suggesting that both Local Group spirals may share similar CGM characteristics distinct from typical blue COS-Halos galaxies.  

This distinction is also potentially a manifestation of the classic ``hot halo formation" process predicted since the late 1970s \citep{white78,white91,birnboim03,keres05},  which in FIRE simulations occurs at inner CGM radii when the halo mass exceeds a threshold of $\approx10^{12}\,{\rm M}_\odot$ \citep{stern21,gurvich23,kakoly25}. About $2/3$ of star-forming COS-Halos galaxies have $M \la 10^{12}$ M$_\sun$ \citep[e.g.,][]{mcquinn18} in which the inner CGM in FIRE is predominantly cool and turbulent, whereas FIRE simulations of $\sim 10^{12.1}$ M$_\sun$ halos like the Milky-Way and M31 have inner CGM dominated by quasi-static gas at $T\sim10^6\,{\rm K}$, potentially explaining the observed differences in the cooler gas content.

\clearpage
\section{Summary and Future Work}\label{s-summary}

We present the first results from AMIGA Insider, a survey of the inner circumgalactic medium of M31 using 11 additional QSO sightlines, including five within 11--18 kpc. All of these QSOs were observed with HST COS G130M and G160M and the data were uniformly reduced and analyzed. All of the absorption features observed in these spectra were identified, and we provide the entire line identification in  Appendix~\ref{a-lineid}. Our analysis employs both the AOD and Voigt PF methods, finding excellent agreement between these complementary approaches for the component-by-component analysis (\S\ref{s-comp-aod-pf}). We compare our identification of M31 CGM and MS components in these directions with the independent analysis by \citetalias{kim24}, finding good agreement in unambiguous regions despite methodological differences and understanding any differences in more ambiguous regions. Combined with our earlier AMIGA Extended survey, we have information on M31's CGM from its disk to $\sim 2\rvir$ along an unprecedented 54 QSO sightlines. Our key findings are:

\begin{enumerate}[wide, labelwidth=!, labelindent=0pt]
\item We have detected the thick disk of M31 in absorption, finding components within projected distances of $\sim$10--30 kpc that show distinct kinematics consistent with corotation with the galactic disk. These thick disk components exhibit higher column densities in low and intermediate ions compared to CGM components at similar impact parameters, while high ions show similar column densities in both thick disk and CGM components. Some of the thick-disk components have strong \oi\ absorption and \hi\ 21-cm emission observed at the same velocities, a combination not observed in any of the CGM components in our sample. 

\item Column densities of all ions decrease with increasing impact parameter, with slopes that become progressively shallower as ionization potential increases: low ions (\oi, \cii, \siii) show steeper gradients than high ions (\civ, \siiv), reflecting changing ionization conditions with distance from M31.  

\item We also detect significant radial variations in gas ionization state using survival analysis with log-rank tests. Both the \siii/\siiii\ and \siii/\siiv\ column density ratios show strong radial gradients, being $\sim$2 times and $\sim$4 times higher, respectively, at $R\la 50$ kpc than at larger radii. On the other hand, the \siiv/\siiii\ ratio shows no significant trend with $R$. This implies that the inner CGM has a more complex physical conditions than the outer regions, with singly ionized species even dominating the inner CGM and indicating cooler and denser gas conditions closer to the galaxy. The coexistence of low and high ionization species (\oi\ through \ovi) in the inner regions, therefore, reveals a more complex multiphase structure than the outer CGM (mostly \siiii, \civ, \ovi).

\item The kinematic complexity (number of velocity components per sightline) of the CGM decreases significantly with increasing radius, transitioning from multi-component absorption near the thick disk to $\sim$2 components in the inner regions ($R\la 100$ kpc) to predominantly single-component absorption at larger radii. This also implies that the breadths of the absorbers are larger at small than large impact parameters. 

\item We find no statistically significant azimuthal dependence in column densities or kinematics on the large scale. There is a hint that column densities of \siiii\ and \siiv\ are somewhat higher along the projected minor axis than along the major projected axis at $R\la 55$ kpc, but our sample size at these small impact parameters is too small for statistically robust results. 

\item Based on the current data, the spatial distribution, component structure, and velocity patterns together reveal a transition from the thick disk to the CGM---from a complex, multiphase medium that is consistent with being more strongly influenced by galactic processes in the inner regions to a simpler, more diffuse structure at larger distances where the host galaxy's influence most likely  diminishes. These patterns show no strong evidence that M31's CGM structure is organized by disk-aligned features (such as a strong bipolar outflow). Instead, the lack of azimuthal dependence combined with clear radial structure appears broadly consistent with a cooling flow in M31's halo, radial accretion of low-metallicity intergalactic medium, or thermal precipitation from the hot corona. 
 
\item Using non-parametric and parametric modeling approaches, we estimate the total metal mass in the cool CGM (traced by \siii, \siiii, and \siiv) within 1.3\rvir\ to be $M^{\rm cool}_{\rm Z} = (2.56 \pm 0.43_{\rm stat} \pm 0.83_{\rm sys}) \times 10^7$ ${\rm M}_\odot$, leading to a cool gas mass of approximately $8 \times 10^9\,(Z/0.3\, Z_\odot)^{-1}$ M$_\odot$ over the same radius. The \ovi-bearing warm phase, when corrected for ionization fraction ($f_{\rm  O\,VI} \le 0.2$), potentially contains $\sim$10 times more metal mass than the cool phase (assuming a similar metallicity in both gas phases), suggesting that the warmer CGM, and potentially even hotter phases not traced by our observations, may dominate M31's total metal and baryon budgets. However, the \ovi\ estimate should be taken with caution owing to the small and spatially non-uniform sample of the \ovi\ absorbers (a total of 11 absorbers and only five within \rvir).

\item Our increased sample at small impact parameters reinforces previous findings that M31's cool CGM has overall lower Si column densities compared to the COS-Halos sample of $L^*$ galaxies at  $R/R_{200} \la 0.4$, but better matched Si column densities at larger $R/R_{200}$. This results in a cool CGM metal mass that is a factor of  three times smaller over a radius of 150 kpc and substantially smaller cool gas mass within \rvir\ compared to the COS-Halos galaxies. This difference may reflect either environmental effects in the COS-Halos sample or fundamental halo mass differences: many COS-Halos star-forming galaxies have $M_{200} \la 10^{12.1}$ M$_\sun$ where cooler gas ($T \la 3\times 10^4$ K) may dominate the CGM mass, while M31's higher mass ($M \sim 10^{12.1}$ M$_\sun$) places it in a regime where hotter gas ($T\ga 5\times 10^5$ K) likely dominates the CGM mass budget.

\end{enumerate}

Future work will explore the physical origins of the observed CGM properties using detailed ionization modeling, characterize the relative abundances and possibly abundance of the M31 CGM and thick disk, investigate the multiphase structure and corotation of the CGM beyond the thick disk region, and characterize the role of M31's past star formation history in shaping its current CGM properties. Our continuing analysis will also incorporate data from the two hot stars within M31's disk to probe the innermost regions of M31's CGM and the disk-halo interface, thereby possibly constraining the gas flow in and out of M31.

\section*{Acknowledgments}
Support for this research was provided by NASA through grant HST-GO-16730 from the Space Telescope Science Institute, which is operated by the Association of Universities for Research in Astronomy, Incorporated, under NASA contract NAS5-26555. CAFG was supported by NSF through grants AST-2108230 and AST-2307327; by NASA through grants 21-ATP21-0036 and 23-ATP23-0008; and by STScI through grant JWST-AR-03252.001-A. JS was supported by the Israel Science Foundation (grant No. 2584/21).  RA acknowledges funding by the European Research Council through ERC-AdG SPECMAP-CGM, GA 101020943.  DJP greatly acknowledges support from the South African Research Chairs Initiative of the Department of Science and Technology and National Research Foundation. 

\software{astropy \citep{price-whelan18,price-whelan22}, emcee \citep{foreman-mackey13}, matplotlib \citep{hunter07}, lifelines \citep{davidson-pilon20}, pymccorrelation \citep{privon20}, scikit-learn \citep{pedregosa11}, survreg \citep{therneau21}}
\facilities{HST(COS); FUSE, Green Bank Telescope}
\bibliographystyle{aasjournal}

\input{bib}
\input{tab2}
\input{tab3}
\input{tab4}
\input{tab5}

\input{tab6}
\input{tab7}
\input{tab8}
\input{tab9}
\clearpage

\appendix

\makeatletter
\renewcommand{\thefigure}{A\@arabic\c@figure}
\setcounter{figure}{0}
\renewcommand{\thetable}{A\@arabic\c@table}
\setcounter{table}{0}

\section{AMIGA Insider Spectra and Profile Fits \label{a-supp-fig}}

\begin{figure*}[tbp]
\epsscale{1}
\plotone{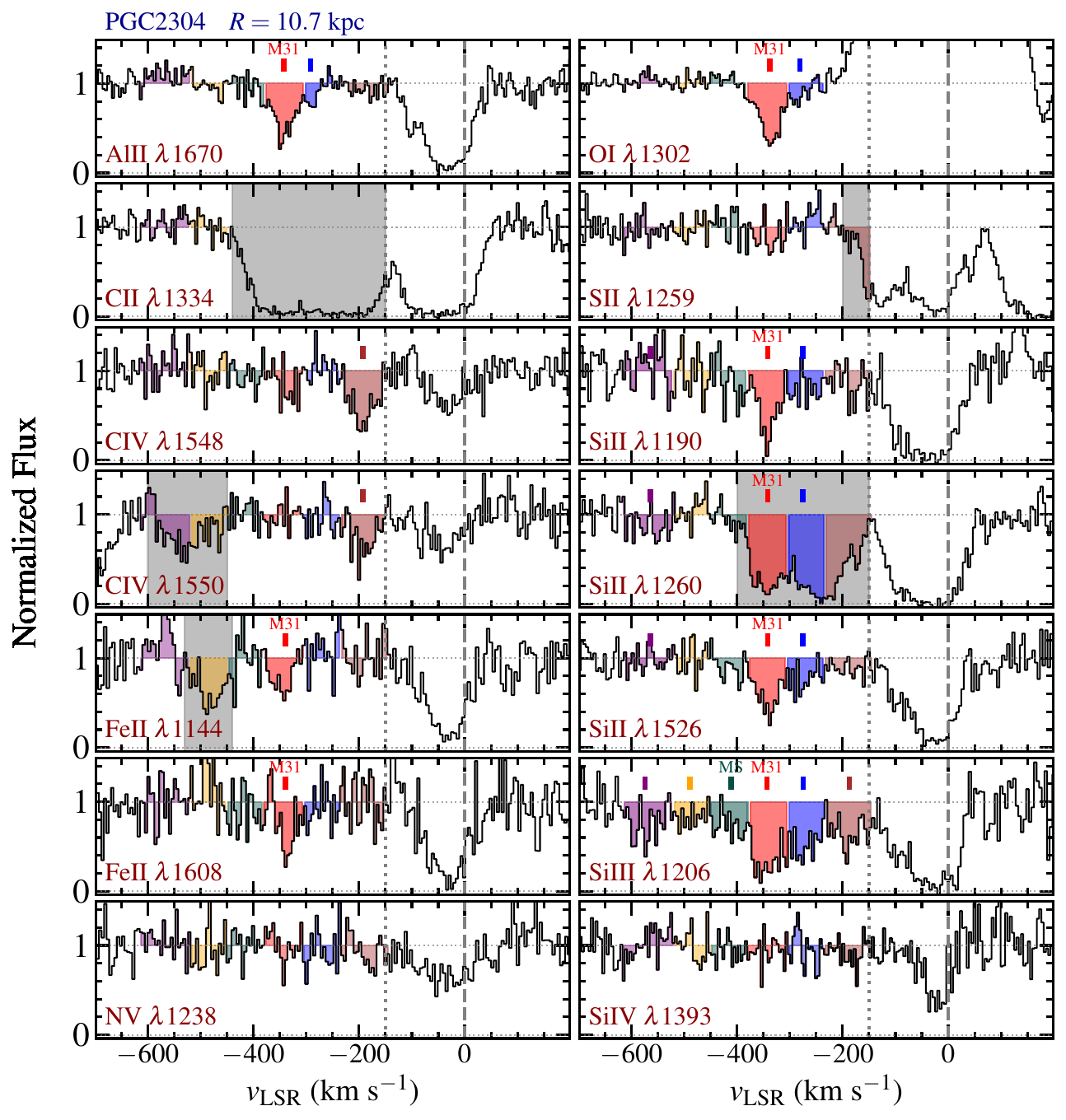}
\caption{Normalized absorption lines as a function of the LSR velocity toward PGC2304. Different colors represent distinct velocity components identified at COS G130M-G160M resolution. Tick marks indicate ions with $\ge 2\sigma$ detections (in at least one transition of that ion). If present,``MS'' components are identified in green and labeled ``MS'' are components associated with the MS.  If present, thick disk corotating components are identified in red and labeled ``M31''. {\it All the other colored components are associated with M31 CGM}. Colored regions without tick marks show no detection of that ion in that component. For ions with multiple transitions, the average velocity from detected transition(s) is used for the tick mark position. The MW absorption is observed between  $-150$ \km\ (vertical dotted line) and $+50$ \km\ toward this sightline. Contaminated regions within the velocity range of interest are grayed out. The vertical dashed line marks $\vlsr =0$ \km.
\label{f-appendix-spectrum}}
\end{figure*}

As part of the supplemental figures associated with this work, we provide for each absorber from the AMIGA Insider sample a figure  as shown in Figs.~\ref{f-example-spectrum} and \ref{f-appendix-spectrum} where we plot the normalized profiles of metal-line transitions for which we estimated the column densities. Each color corresponds to a component and the  velocity range of the absorption over which the velocity profile is integrated to derive the column densities and kinematics. Components associated with the thick disk of M31 and MS are labeled as red ``M31" and green ``MS", respectively. 

\begin{figure}[tbp]
\epsscale{0.5}
\plotone{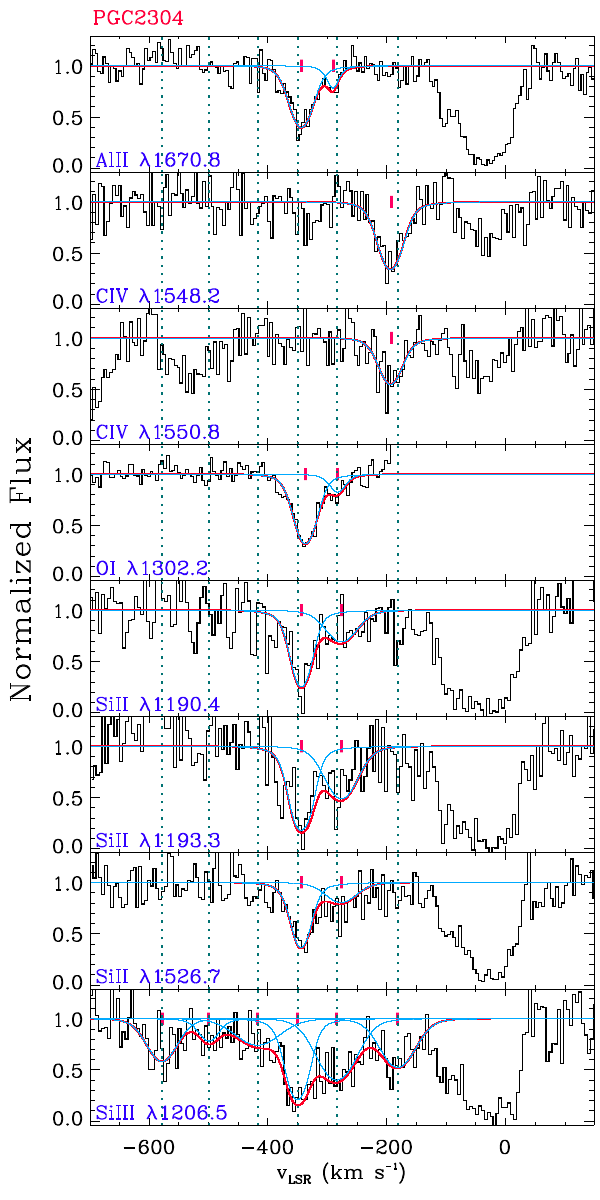}
\caption{Normalized absorption lines as a function of the LSR velocity toward PGC2304 with a Voigt PF model to the data. The red tick-marks show the velocity centroids for each ion. In each panel, the red line shows the resulting PF while the blue line show the individual components. The green vertical dotted lines show the velocity centroids of \siiii. 
\label{f-appendix-fit}}
\end{figure}

In a separate file, we also provide a figure as shown in Figs.~\ref{f-example-fit} and \ref{f-appendix-fit}, which shows the normalized profiles and the Voigt profile fitting. Note that the component $-408$ \km\ for \cii\ toward RX\_J0049.8+3931 is 
is entirely contaminated by a Ly$\alpha$ absorber, but because of blending it is treated as \cii\ in the profile fit and shown as such in the figure. Toward the same sightline, RX\_J0049.8+3931, the \oi\ absorption is severely blended with the airglow emission and the results should be treated with caution. 

Similar figures can be found in \citetalias{lehner20} for the absorbers in the AMIGA Extended sample. 

\section{Line Identification}\label{a-lineid}
In Table~\ref{t-linelist}, we provide the line identification for each absorption feature detected at about the 2$\sigma$ level (the line list is complete at this level, but can include also less significant absorption). The table is ordered by increasing impact parameters as for Table~\ref{t-sum} and for each QSO in order of increasing observed wavelength (second column). In this table, we define the various types of absorption features as follows (third column): ``INTERSTELLAR" is any absorption from the Local group environment (mostly from the MW, MS, and M31 CGM); ``IGMABS" is any intervening intergalactic medium (IGM) absorber separated from the QSO redshift by $\Delta v > 3000 $ \km; ``PROXIMATE" is a  proximate/associated absorber within $500< \Delta v < 3000 $ \km\ from the QSO redshift; ``INTRINSIC" is an absorber  within $\Delta v < 500 $ \km\ from the QSO redshift. Any ``UNIDENTIFIED" feature at the $>2\sigma $ level is marked with that denomination. Finally, ``OTHER" includes known fixed-pattern noise feature (``FPN"). FPN appears in the fourth column, which is otherwise used to list the atom or ion detected. The fifth column gives the rest wavelength of the atom/ion. The sixth column provides the information regarding frame into which the velocity (sixth column) and redshift (seventh column) are defined (``L": LSR frame--- any absorption at $|\vlsr|\le 700$ \km, and otherwise ``H": heliocentric frame). Note that the velocities are independently measured, but the redshift is set by the velocity of the first line (in case of an absorber that is not interstellar, that would be Ly$\alpha$). Please note that the velocity centroids for the M31 and MS components listed in Table~\ref{t-linelist} may be somewhat different from those listed Table~\ref{t-results} owing to some refinements in the velocity structure as part  of the AOD measurements. Finally the last two columns give the approximate equivalent widths ($W_{\lambda}$) and errors that are only provided as guidelines, i.e., these should not be used for quantitative scientific purposes since the continuum placement is only approximate. 

\section{Treatment of Errors and Censored Data in the Modeling of $N_{\rm S\MakeLowercase{i}}(R)$}\label{a-likelihood}

For the penalty-based likelihood approach (PL1, GP), our implementation uses a modified likelihood function that applies asymmetric treatment reflecting the different physical nature of the constraints: strong penalties for violating lower limits (which represent firm physical constraints from saturated absorption) and standard likelihood treatment for upper limits (which represent observational sensitivity limits from non-detections). The penalty-based log-likelihood function has the following components:
\begin{itemize}[wide, labelwidth=!, labelindent=0pt]
    \item For detections (data points with measured values and uncertainties):
    $$
    \ln \mathcal{L}_{\text{det}} = -\frac{1}{2} \sum_i \left[ \frac{(y_i - \mu_i)^2}{\sigma_{\text{tot},i}^2} + \ln(2\pi\sigma_{\text{tot},i}^2) \right]
    $$
    where $y_i = \log N_i$, $\mu_i = \log N_{\text{model}}(R_i)$, and $\sigma_{\text{tot},i}^2 = \sigma_i^2 + \sigma_{\text{int}}^2$.
    
    \item For lower limits (at least one of the ions has its absorption saturated):
    $$
    \ln \mathcal{L}_{\text{lower}} = \sum_i
    \begin{cases}
    0, & \\ \text{if } \mu_i \geq y_{\text{limit},i} \\
    -\frac{3}{2} \left[ \frac{(y_{\text{limit},i} - \mu_i)^2}{\sigma_{\text{tot},i}^2} + \ln(2\pi\sigma_{\text{tot},i}^2) \right], & \\ \text{otherwise}
    \end{cases}
    $$
    where the factor 3 increases the penalty weight for violating lower limits, making model predictions below these limits $3\times$ more costly in the likelihood function. While the factor of 3 is somewhat arbitrary, it balances fitting detections against respecting physical constraints. We tested penalty factors from 2--10 and found that our mass estimates are robust to this choice.
    
    \item For upper limits (from non-detections):
    \begin{equation}
    \ln \mathcal{L}_{\text{upper}} = \sum_i
    \begin{cases}
    0, & \\ \text{if } \mu_i \leq y_{\text{limit},i} \\
    -\frac{1}{2} \left[ \frac{(y_{\text{limit},i} - \mu_i)^2}{\sigma_{\text{tot},i}^2} + \ln(2\pi\sigma_{\text{tot},i}^2) \right]\,. & \\\text{otherwise}
    \end{cases}
    \end{equation}
\end{itemize}
No additional penalty is applied to upper limit violations since these represent observational sensitivity limits rather than firm physical constraints.

For the survival analysis MCMC approach (PL2), we employ a survival analysis MCMC statistically rigorous treatment of censored data using cumulative distribution functions. For detections, we use standard Gaussian likelihoods as in the penalty-based approach, but the lower and upper limits are now treated as follows:
\begin{itemize}[wide, labelwidth=!, labelindent=0pt]
    \item For upper limits, we incorporate the probability that the true value lies below the limit using the normal cumulative distribution function (CDF):
    $$
    \ln \mathcal{L}_{\text{upper}} = \sum_i \ln \Phi\left(\frac{y_{\text{limit},i} - \mu_i}{\sqrt{\sigma_i^2 + \sigma_{\text{int}}^2}}\right)
    $$
    where $\Phi$ is the standard normal CDF.
    
    \item For lower limits, we use the survival function (1-CDF) to represent the probability that the true value exceeds the limit:
    $$
    \ln \mathcal{L}_{\text{lower}} = \sum_i \ln \left[1-\Phi\left(\frac{y_{\text{limit},i} - \mu_i}{\sqrt{\sigma_i^2 + \sigma_{\text{int}}^2}}\right)\right].
    $$
\end{itemize}

The total log-likelihood is the sum of these three components. This approach provides computational efficiency while incorporating the information from all available constraints.

\clearpage
\input{tabA1}


\end{document}

%% file: tab1.tex
\begin{deluxetable*}{lccccccrrc}
\tabletypesize{\scriptsize}
\tablecaption{Sample Summary \label{t-sum}}
\tablehead{\colhead{Target} & \colhead{$z_{\rm em}$} & \colhead{RA} & \colhead{Dec} & \colhead{$l_{\rm MS}$} & \colhead{$b_{\rm MS}$} & \colhead{$R$} & \colhead{$X$} & \colhead{$Y$} & \colhead{S/N}\\ 
\colhead{} & \colhead{} & \colhead{($\mathrm{{}^{\circ}}$)} & \colhead{($\mathrm{{}^{\circ}}$)} & \colhead{($\mathrm{{}^{\circ}}$)} & \colhead{($\mathrm{{}^{\circ}}$)} & \colhead{ ($\mathrm{kpc}$)} & \colhead{($\mathrm{kpc}$)} & \colhead{($\mathrm{kpc}$)} & \colhead{ }}
\startdata
PGC2304 & $0.072$ & $9.638$ & $41.481$ & $-127.1$ & $23.0$ & $10.7$ & $-10.3$ & $ 2.8$  & 9.3 \\
IVZw29 & $0.103$ & $10.567$ & $40.327$ & $-125.8$ & $23.6$ & $12.4$ & $-1.2$ & $-12.4$  & 30.1 \\
QSO0043+4234 & $0.191$ & $10.979$ & $42.575$ & $-128.2$ & $24.1$ & $17.4$ & $ 2.8$ & $17.2$  & 11.0 \\
RX\_J0046.9+4220 & $0.306$ & $11.731$ & $42.347$ & $-128.0$ & $24.6$ & $17.5$ & $10.2$ & $14.2$  & 7.1 \\
RX\_J0049.8+3931 & $0.142$ & $12.470$ & $39.525$ & $-124.8$ & $25.1$ & $29.0$ & $18.1$ & $-22.7$  & 22.1 \\
LAMOST003432.52+391836.1 & $0.138$ & $8.635$ & $39.310$ & $-124.8$ & $22.1$ & $32.9$ & $-20.8$ & $-25.5$  & 13.2 \\
WISEA002827.05+380942.0 & $0.448$ & $7.113$ & $38.162$ & $-123.7$ & $20.8$ & $54.4$ & $-36.9$ & $-40.0$  & 10.0 \\
UVQSJ001903.85+423809.0 & $0.112$ & $4.766$ & $42.636$ & $-128.7$ & $19.5$ & $60.5$ & $-57.1$ & $19.9$  & 10.5 \\
LAMOST005846.82+365514.2 & $0.283$ & $14.695$ & $36.921$ & $-121.8$ & $26.7$ & $70.1$ & $42.1$ & $-56.1$  & 11.2 \\
PGC2077956 & $0.145$ & $7.805$ & $36.280$ & $-121.7$ & $21.1$ & $71.8$ & $-30.5$ & $-65.0$  & 7.6 \\
MS0108.4+3859 & $0.323$ & $17.819$ & $39.251$ & $-124.4$ & $29.2$ & $76.1$ & $72.4$ & $-23.5$  & 10.9 \\
\enddata
\tablecomments{The 11 sightlines are all from our large HST Cycle 29 program PID 16730. All the projected distances are computed using the three dimensional separation (coordinates of the target and distance to M31 assumed to be 752 kpc). The coordinates $l_{\rm MS}$ and $b_{\rm MS}$ are the MS longitudes and latitudes as defined by \citet{nidever08}. S/N is given per COS resolution element (assuming $R\sim 17,000$) and measured in the continuum near  \siiii\ $\lambda$1206.
}
\end{deluxetable*}

%% file: tab2.tex
\startlongtable
\begin{deluxetable*}{llcccccccrrr}
\tabletypesize{\scriptsize}
\tablecaption{Summary of the AOD Results\label{t-results}}
\tablehead{\colhead{Target} & \colhead{Ion} & \colhead{$v_1$} & \colhead{$v_2$} & \colhead{$v$} & \colhead{$\sigma_v$} & \colhead{$\log N$} & \colhead{$\sigma_{ \log N}^1$} & \colhead{$\sigma_{ \log N}^2$} & \colhead{$f_{\rm N}$} & \colhead{$f_{\rm MS}$} & \colhead{$f_{\rm M31}$}\\ 
\colhead{} & \colhead{} & \colhead{(\km)} & \colhead{(\km)} & \colhead{(\km)} & \colhead{(\km)} & \colhead{[cm$^{-2}$]} & \colhead{} & \colhead{} & \colhead{}& \colhead{}& \colhead{}}
\startdata
PGC2304 & \ion{Al}{2} & $-615.0$ & $-520.0$ & $999.0$ & 999.0 & 11.92 & 0.18 & 0.30 & $-1$ & $ 0$ & $ 0$ \\
PGC2304 & \ion{C}{2} & $-615.0$ & $-520.0$ & $999.0$ & 999.0 & 13.12 & 0.18 & 0.30 & $-1$ & $ 0$ & $ 0$ \\
PGC2304 & \ion{C}{4} & $-615.0$ & $-520.0$ & $999.0$ & 999.0 & 13.15 & 0.18 & 0.30 & $-1$ & $ 0$ & $ 0$ \\
PGC2304 & \ion{Fe}{2} & $-615.0$ & $-520.0$ & $999.0$ & 999.0 & 13.75 & 0.18 & 0.30 & $-1$ & $ 0$ & $ 0$ \\
PGC2304 & \ion{N}{5} & $-615.0$ & $-520.0$ & $999.0$ & 999.0 & 13.32 & 0.18 & 0.30 & $-1$ & $ 0$ & $ 0$ \\
PGC2304 & \ion{O}{1} & $-615.0$ & $-520.0$ & $999.0$ & 999.0 & 13.27 & 0.18 & 0.30 & $-1$ & $ 0$ & $ 0$ \\
PGC2304 & \ion{S}{2} & $-615.0$ & $-520.0$ & $999.0$ & 999.0 & 14.19 & 0.18 & 0.30 & $-1$ & $ 0$ & $ 0$ \\
PGC2304 & \ion{Si}{2} & $-615.0$ & $-520.0$ & $-564.2$ & 10.3 & 12.53 & 0.14 & 0.21 & $ 0$ & $ 0$ & $ 0$ \\
\enddata
\tablecomments{The velocities $v_1$ and $v_2$ correspond to the integration of the absorption component. The $1\sigma$ errors $\sigma_{ \log N}^1$ and $\sigma_{ \log N}^2$ are the positive and negative errors, respectively, on the logarithm of the column density. The flag $f_{\rm N}$ has the following definition: $0=$ detection (not saturated or contaminated); $-1 =$ upper limit (an entry with 999.0 in the velocity and velocity error columns indicates no data because the absorption is not detected); $-2 =$ lower limit (due to saturation of the line). The flag $f_{\rm MS}$ has the following definition: $0=$ not contaminated by the MS; $-1 =$ contaminated by the MS (see \S\ref{s-ms-contamination}). The flag $f_{\rm M31}$ has the following definition: $0=$ associated with the CGM of M31 (if $f_{\rm MS} = 0$); $-1 =$ associated with the thick disk of M31 (if $f_{\rm MS} = 0$; see \S\ref{s-m31-dh}).
}
\end{deluxetable*}

%% file: tab3.tex
\startlongtable
\begin{deluxetable*}{llcccccccrr}
\tablecaption{Summary of the PF Results\label{t-fit}}
\tablehead{\colhead{Target} & \colhead{Ion} & \colhead{Comp.} & \colhead{$v$} & \colhead{$\sigma_v$} & \colhead{$b$} & \colhead{$\sigma_b$} & \colhead{$\log N$} & \colhead{$\sigma_{\log N}$}& \colhead{$f_{\rm MS}$} & \colhead{$f_{\rm M31}$}\\
\colhead{} & \colhead{} & \colhead{} & \colhead{(\km)} & \colhead{(\km)} & \colhead{(\km)} & \colhead{(\km)} & \colhead{[cm$^{-2}$]} & \colhead{}& \colhead{}& \colhead{}}
\startdata
PGC2304 & \ion{C}{4} & $ 1$ & $-192.5$ & $ 3.0$ & $20.1$ & $ 4.9$ & 13.94 & 0.07 & $ 0$ & $ 0$ \\
PGC2304 & \ion{Si}{2} & $ 1$ & $-343.2$ & $ 2.4$ & $14.4$ & $ 3.3$ & 14.05 & 0.13 & $ 0$ & $ 0$ \\
PGC2304 & \ion{Si}{2} & $ 2$ & $-275.9$ & $ 7.4$ & $29.7$ & $12.8$ & 13.44 & 0.10 & $ 0$ & $ 0$ \\
PGC2304 & \ion{Si}{3} & $ 1$ & $-577.8$ & $ 8.6$ & $28.0$ & $20.2$ & 12.83 & 0.17 & $ 0$ & $ 0$ \\
PGC2304 & \ion{Si}{3} & $ 2$ & $-499.0$ & $16.5$ & $16.9$ & \nodata & 12.36 & 0.81 & $ 0$ & $ 0$ \\
PGC2304 & \ion{Si}{3} & $ 3$ & $-416.9$ & $29.2$ & $45.6$ & \nodata & 12.73 & 0.35 & $-1$ & $ -1$ \\
PGC2304 & \ion{Si}{3} & $ 4$ & $-349.6$ & $ 6.6$ & $16.6$ & \nodata & $(>)\,13.37 :$ & 0.44 & $ 0$ & $ -1$ \\
PGC2304 & \ion{Si}{3} & $ 5$ & $-283.8$ & $11.3$ & $36.8$ & $25.0$ & 13.16 & 0.17 & $ 0$ & $ 0$ \\
\enddata
\tablecomments{Results provided for the components at $\vlsr \la -150$ \km. Errors on $v$, $b$, and $N$ are $1\sigma$ errors. Components with the velocities followed by a colon have their that velocity component fixed in the PF. Components with a Doppler parameter set to $b = 7:$ \km\ means that $b$ was fixed to 7 \km\ in the PF (otherwise the PF would lead to unreasonably small $b$-values); the column densities estimated for these components could be larger if the true $b< 7$ \km. Components with $b$-values followed by a colon and with no error bars have uncertain parameters owing to severe blending of saturated lines. Components with column densities preceded by ``$(>)$" and followed by "$:$" indicate that these components are (most likely) saturated and therefore unreliable (for a true lower limit on $N$, the reader should refer to the AOD results summarized in Table~\ref{t-results}). The flag $f_{\rm MS}$ has the following definition: $0=$ not contaminated by the MS; $-1 =$ contaminated by the MS (see \S\ref{s-ms-contamination}). The flag $f_{\rm M31}$ has the following definition: $0=$ associated with the CGM of M31 (if $f_{\rm MS} = 0$); $-1 =$ associated with the thick disk of M31 (if $f_{\rm MS} = 0$; see \S\ref{s-m31-dh}). $a$: this \oi\ is severely blended with the airglow \oi\ emission line, making these fit results more uncertain than the formal errors. $b$: this component is entirely contaminated by a Ly$\alpha$ absorber, but because of blending it is treated as \cii\ in the fit.
}
\end{deluxetable*}

%% file: tab4.tex
\begin{deluxetable}{lcccc}
\tablecaption{Velocity Components Associated with the Thick Disk of M31  \label{t-thick_disk}}
\tablehead{\colhead{Target} &\colhead{$R$} & \colhead{$X$} & \colhead{$Y$}& \colhead{$[v_{\rm min},v_{\rm max}]$} \\
\colhead{}  & \colhead{(kpc)} & \colhead{(kpc)}  & \colhead{(kpc)} &  \colhead{(\km)}
}
\startdata
PGC2304          & $10.7$ & $-10.3$& $2.8$   & $[-412.5, -338.0]$ \\
IVZw29           & $12.4$ & $-1.2$ & $-12.4$ & $[-412.5, -357.9]$ \\
QSO0043+4234     & $17.4$ & $2.8$  & $17.2$  & $[-214.1, -150.9]$ \\
RX\_J0046.9+4220 & $17.5$ & $10.2$ & $14.2$  & $[-184.6, -165.0]$ \\
RX\_J0048.3+3941 & $25.0$ & $14.1$ & $-20.7$ & $[-243.7, -172.6]$ \\
RX\_J0049.8+3931 & $29.0$ & $18.1$ & $-22.7$ & $[-195.9, -165.1]$ \\
\enddata
\tablecomments{The velocity range indicates the minimum and maximum velocities of components identified as co-rotating thick disk material. RX\_J0048.3+3941 is from the AMIGA Extended sample, while all the others are from the AMIGA Insider sample.  }
\end{deluxetable}

%% file: tab5.tex
\begin{deluxetable*}{lccccrl}
\tablecaption{Total Si column densities for Project AMIGA sightlines \label{t-tot-si}}
\tablewidth{0pt}
\tablehead{
\colhead{Target} & \colhead{$R$} & \colhead{$\log N_{\rm Si}$} & \colhead{$\sigma^{-}_{\log N}$} & \colhead{$\sigma^{+}_{\log N}$} & \colhead{Flag} & \colhead{Type}
\\
\colhead{} & \colhead{(kpc)} & \colhead{\cmm} & \colhead{(dex)} & \colhead{(dex))} & \colhead{} & \colhead{}
}
\startdata
PGC2304                   &     10.7 &    13.86 & $    0.04$ & $    0.04$& $     0 $ & Mixed detections and limits (\siiii\ not saturated)  \\
IVZw29                    &     12.4 &    13.65 & $    0.02$ & $    0.02$& $     0 $ & All Si ions detected (no saturation)  \\
QSO0043+4234              &     17.4 &    12.83 & $    0.14$ & $    0.14$& $     0 $ & Only \siiii\ detected (no saturation)  \\
RX\_J0046.9+4220          &     17.5 &    13.49 & $    0.05$ & $    0.04$& $    -2 $ & \siiii\ saturated with only non-detections  \\
RX\_J0048.3+3941          &     25.0 &    13.34 & $    0.01$ & $    0.01$& $    -2 $ & \siiii\ saturated with other detections  \\
RX\_J0049.8+3931          &     29.0 &    13.19 & $    0.02$ & $    0.03$& $     0 $ & All Si ions detected (no saturation)  \\
HS0033+4300               &     30.6 &    13.87 & $    0.06$ & $    0.07$& $     0 $ & All Si ions detected (no saturation)  \\
LAMOST003432.52+391836.1  &     32.9 &    13.26 & $    0.07$ & $    0.08$& $     0 $ & Only \siiii\ detected (no saturation)  \\
\enddata
\tablecomments{The flag column has the usual meaning for the column density: $0 =\,$detection (with no saturation issue), $-1 =\,$non-detection (upper limit), $-2 =\,$detection, but the absorption is saturated in at least one of the components (lower limit).}
\end{deluxetable*}

%% file: tab6.tex
\begin{deluxetable}{lcccc}
\tablecaption{Kendall's $\tau$ Correlation Test between $\log N$ and $R$ \label{t-corr_NvsR}}
\tablehead{
\colhead{Ion} & \colhead{Kendall's $\tau$} & \colhead{$p$-value}& \colhead{Kendall's $\tau$} & \colhead{$p$-value}
}
\startdata
\multicolumn{1}{c}{} & \multicolumn{2}{c}{Entire Sample}&\multicolumn{2}{c}{Filtered Sample} \\
\hline
\multicolumn{5}{c}{CGM Only} \\
\hline
\ion{O}{1}   & $-0.19 \pm 0.09$ &$ 0.06$       & $-0.32 \pm 0.13$ &$ 0.05$ \\ 
\ion{C}{2}   & $-0.24 \pm 0.09 $ & $0.01$      & $-0.28 \pm 0.09 $ & $0.01$ \\
\ion{Si}{2}  & $-0.35 \pm 0.08$ & $\ll 0.05$   & $-0.38 \pm 0.07$ & $\ll 0.05$ \\
\ion{Si}{3}  & $-0.47 \pm 0.06$ & $\ll 0.05$   & $-0.47 \pm 0.07$ & $\ll 0.05$ \\
\ion{Si}{4}  & $-0.16 \pm 0.10$ & $0.10$       & $-0.25 \pm 0.12$ & $0.04$ \\
\ion{C}{4}   & $-0.10 \pm 0.10$ & $0.31$       & $-0.09 \pm 0.09$ & $0.37$ \\
\hline
\multicolumn{5}{c}{CGM + Thick Disk} \\
\hline
\ion{O}{1}   & $-0.33 \pm 0.09$ &$\ll 0.05$  & $-0.54 \pm 0.10$  & $\ll 0.05$ \\
\ion{C}{2}   & $-0.33 \pm 0.09$ &$\ll 0.05$  & $-0.36 \pm 0.08$  & $\ll 0.05$ \\
\ion{Si}{2}  & $-0.48 \pm 0.06$ & $\ll 0.05$ & $-0.48 \pm 0.07$  & $\ll 0.05$ \\
\ion{Si}{3}  & $-0.56 \pm 0.06$ & $\ll 0.05$ & $-0.56 \pm 0.05$  & $\ll 0.05$ \\
\ion{Si}{4}  & $-0.38 \pm 0.10$ & $\ll 0.05$ & $-0.41 \pm 0.08$  & $\ll 0.05$ \\
\ion{C}{4}   & $-0.30 \pm 0.10$ & $\ll 0.05$ & $-0.31 \pm 0.12$  & $\ll 0.05$ \\
\enddata
\tablecomments{Kendall's $\tau$ correlation coefficients (with $1\sigma$ confidence intervals) and associated $p$-values comparing ion $\log N$ with $R$, derived using the \citet{isobe86} method for handling censored data. When $p$-values are significantly smaller than 0.05 (specifically, $p < 1\times 10^{-2}$), they are simply denoted as $\ll 0.05$. The filtered sample excludes upper limits that exceed $\langle \log N_{\rm det}\rangle - 0.5\sigma_{N_{\rm det}}$, where $N_{\rm det}$ represents detections at $>2\sigma$ significance.}
\end{deluxetable}

%% file: tab7.tex
\begin{deluxetable}{lcccc}
\tablecaption{Si Column Density Statistics Along Minor and Major Axes at $R \la  55$ kpc} \label{t-colaxis}
\tablewidth{0pt}
\tablehead{
\colhead{Ion} & \colhead{Mean ($\sigma$)} & \colhead{Median}& \colhead{Mean ($\sigma$)} & \colhead{Median}
}
\startdata
\hline
     & \multicolumn{2}{c}{Minor Axis}& \multicolumn{2}{c}{Major Axis} \\
\hline
\siii\  & 13.09 (0.35) & 12.56 & 13.12 (0.79) & 12.40 \\
\siiii\ & 13.41 (0.40) & 13.23 & 13.22 (0.20) & 13.16 \\
\siiv\  & 12.89 (0.53) & 12.56 & 12.70 (0.29) & 12.70 \\
\enddata
\tablecomments{The mean ($\log \langle N\rangle$), median, and standard deviations of the column densities are estimated using the survival analysis.}
\end{deluxetable}

%% file: tab8.tex
\begin{deluxetable}{lccc}
\tablecaption{Metal mass of the Cool CGM of M31\label{t-mass}}
\tablehead{\colhead{Range} & \colhead{$M^{\rm cool}_{\rm Z}$} & \colhead{$\sigma_{\rm stat}$} & \colhead{$\sigma_{\rm syst}$}\\
\colhead{(kpc)} & \colhead{($10^7$ M$_\sun$)} & \colhead{($10^7$ M$_\sun$)}& \colhead{($10^7$ M$_\sun$)}
}
\startdata
5--230      &  1.36       & $\pm 0.26$ & $\pm 0.56$  \\ 
5--300      &  1.88       & $\pm 0.33$ & $\pm 0.69$  \\
5--390      &  2.56       & $\pm 0.43$ & $\pm 0.83$  \\
5--150      &  0.79       & $\pm 0.17$ & $\pm 0.49$  \\
150--300    &  1.08       & $\pm 0.18$ & $\pm 0.28$  
\enddata
\tablecomments{$M^{\rm cool}_{\rm Z}$ corresponds to the metal mass traced by \siii, \siiii, and \siiv\ (see \S\ref{s-mass-main} for more detail).}
\end{deluxetable}

%% file: tab9.tex
\begin{deluxetable}{lcccc}
\tablecaption{Summary of updated \ovi\ Column Densities \label{t-novi}}
\tablewidth{0pt}
\tablehead{
\colhead{Target} & \colhead{$R$} & \colhead{$\log N_{\rm O\,VI}$} & \colhead{Flag}& \colhead{$\Delta \log N$} \\
\colhead{} & \colhead{(kpc)} & \colhead{[\cmm]} & \colhead{}& \colhead{(dex)}
}
\startdata
RX\_J0048.3+3941          &     25.0 &  $  14.30 \pm	0.06  	  $ &$ 0$ &$-0.17$\\
IRAS\_F00040+4325         &     93.0 &  $  14.67 \,^{+0.06}_{-0.10}$ &$ 0$ &$+0.01$\\
MRK352                    &    131.7 &  $ > 14.77 \pm	0.03	  $ &$-2$ &$+0.02$\\
PG0052+251                &    209.8 &  $ > 14.63 \pm	0.04      $ &$-2$ &$+0.06$\\
MRK335                    &    292.6 &  $  14.13 \pm	0.09      $ &$ 0$ &$-0.20$\\
UGC12163                  &    349.2 &  $  14.32 \pm	0.09      $ &$ 0$ &$-0.14$\\
PG0026+129                &    365.8 &  $  14.13 \,^{+0.19}_{-0.22}$ &$ 0$ &$+0.01$\\
MRK1502                   &    372.4 &  $ > 14.43 \pm	0.14      $ &$-2$ &$-0.16$\\
NGC7469                   &    503.9 &  $  14.13 \pm	0.11      $ &$ 0$ &$+0.15$\\
MRK304                    &    533.1 &  $ > 14.54 \pm	0.06      $ &$-2$ &$+0.00$\\
PG2349-014                &    562.9 &  $  13.92 \,^{+0.17}_{-0.29}$ &$ 0$ &$+0.00$\\
\enddata
\tablecomments{Flag definition: $0 =\,$detection, $-2 =\,$lower limit (detection with at least one component saturated; these are saturated owing to the low S/N of the \fuse\ spectra). The last column shows the difference in column density between \citetalias{lehner20} and the present study (see \S\ref{s-ovi-mass}).
}
\end{deluxetable}

%% file: tabA1.tex
\startlongtable
\begin{deluxetable*}{lcllccrrrr}
\tablecaption{Line List Identification\label{t-linelist}}
\tablehead{\colhead{Target} & \colhead{$\lambda_{\rm obs}$} & \colhead{Type} & \colhead{Ion} & \colhead{$\lambda_{\rm rest}$} & \colhead{Frame} & \colhead{$cz$} & \colhead{$z$} & \colhead{$W_{\rm \lambda}$} & \colhead{$\sigma_{W_{\lambda}}$}\\
\colhead{ } & \colhead{(\AA)} & \colhead{} & \colhead{} & \colhead{(\AA)} & \colhead{} & \colhead{(\km)} & \colhead{} & \colhead{(m\AA)} & \colhead{(m\AA)}}
\startdata
PGC2304 & 1139.977 & INTRINSIC & \ion{S}{4}& 1062.6 & H & $21811$ & $0.07277$ & 102.2 & 27.5 \\
PGC2304 & 1142.362 & INTERSTELLAR & \ion{Fe}{2}& 1142.3 & L & $-1$ & $-0.00000$ & 22.1 & 17.6 \\
PGC2304 & 1143.031 & INTERSTELLAR & \ion{Fe}{2}& 1143.2 & L & $-51$ & $-0.00016$ & 73.5 & 16.4 \\
PGC2304 & 1143.224 & INTERSTELLAR & \ion{Fe}{2}& 1143.2 & L & $-1$ & $-0.00000$ & 43.8 & 16.0 \\
PGC2304 & 1143.654 & INTERSTELLAR & \ion{Fe}{2}& 1144.9 & L & $-336$ & $-0.00112$ & 67.7 & 26.6 \\
PGC2304 & 1144.555 & INTERSTELLAR & \ion{Fe}{2}& 1144.9 & L & $-100$ & $-0.00034$ & 57.5 & 18.8 \\
PGC2304 & 1144.753 & INTERSTELLAR & \ion{Fe}{2}& 1144.9 & L & $-48$ & $-0.00016$ & 127.8 & 11.7 \\
PGC2304 & 1144.929 & INTERSTELLAR & \ion{Fe}{2}& 1144.9 & L & $-2$ & $-0.00000$ & 143.6 & 15.8 \\
\enddata
\tablecomments{Definition of the identified types of absorption features: INTERSTELLAR: absorption from the Local group environment (mostly from the MW, MS, and M31 CGM); IGMABS: intervening IGM/CGM absorber at $\Delta v > 3000 $ \km\ from the QSO redshift; PROXIMATE: proximate/associated absorber at $500< \Delta v < 3000 $ \km\ from the QSO redshift; INTRINSIC: intrinsic absorber  at $\Delta v < 500 $ \km\ from the QSO redshift; UNIDENTIFIED: unknown origin of the absorption; OTHER: FPN: fixed-pattern noise feature. In the frame column, ``L" stands for LSR frame (any absorption at $|\vlsr|\le 700$ \km),``H" for heliocentric frame (any absorption at $|\vlsr|> 700$ \km).
}
\end{deluxetable*}